\documentclass[12pt]{article}

\usepackage{axodraw4j,amsmath,amssymb,amsfonts,cancel,graphicx,slashed,pstricks,color,dsfont,fancyhdr,url,tabularx,multirow,comment}
\usepackage{hyperref}

\newcommand{\beq}{\begin{equation}}
\newcommand{\eeq}{\end{equation}}
\newcommand{\bea}{\begin{eqnarray}}
\newcommand{\eea}{\end{eqnarray}}
\newcommand{\ba}{\begin{array}}
\newcommand{\ea}{\end{array}}
\newcommand{\bec}{\begin{center}}
\newcommand{\eec}{\end{center}}
\newcommand{\bei}{\begin{itemize}}
\newcommand{\eei}{\end{itemize}}

\newcommand{\GeV}{\,\mathrm{GeV}}

\newcommand{\eV}{\,\mathrm{eV}}

\makeatletter
\newcommand{\BF}{\mathbin{\mathpalette\@BF\relax}}
\newcommand{\@BF}[2]{\ooalign{%
  \raisebox{.6\height}{\small{$#1B$}}\cr
  \smash{\raisebox{-.6\height}{\small{$#1F$}}}\cr}}
\makeatother

\numberwithin{equation}{section}

\begin{document}

\begin{flushright}
SACLAY-T15/025\\
\end{flushright}

\vspace{1.5cm}

\begin{center}

{\Large \bf   Flavour always matters in scalar triplet leptogenesis}
\vspace{1.5cm}

{\bf St\'ephane Lavignac and Beno\^it Schmauch}\\
\vspace{0.3cm}
\end{center}

{\sl Institut de Physique Th\'{e}orique, CEA-Saclay, 
91191 Gif-sur-Yvette Cedex, France~\footnote{Laboratoire de la Direction
des Sciences de la Mati\`ere du Commissariat \`a l'Energie Atomique
et Unit\'e de Recherche Associ\'ee au CNRS (URA 2306).}}\\
\vspace{0.5cm}

\abstract{
We present a flavour-covariant formalism for scalar triplet leptogenesis,
which takes into account the effects of the different lepton flavours in a consistent way. 
Our main finding is that flavour effects can never be neglected in scalar triplet
leptogenesis, even in the temperature regime where all charged lepton
Yukawa interactions are out of equilibrium.
This is at variance with the standard leptogenesis scenario with heavy Majorana neutrinos.
In particular, the so-called single flavour approximation leads to predictions for
the baryon asymmetry of the universe that can differ by a large amount from the
flavour-covariant computation in all temperature regimes.
We investigate numerically the impact of flavour effects and spectator processes
on the generated baryon asymmetry, and find that the region of triplet parameter
space allowed by successsful leptogenesis is significantly enlarged.
}

\vfill
\eject

\newpage

\setcounter{page}{1}
\pagestyle{plain}

\section{Introduction}  %
\label{sec:intro}           %

Leptogenesis~\cite{Fukugita:1986hr}, i.e. the generation of a lepton asymmetry through
the out-of-equilibrium decays of heavy particles before sphaleron freeze-out,
is one of the most appealing explanations for the origin of the observed
baryon asymmetry of the universe.
As such, it has been the subject of a large number of studies, especially
in its minimal version involving heavy right-handed neutrinos
(for a comprehensive review, see Ref.~\cite{Davidson:2008bu}).
Over the years, many refinements have been added to the computation of
the generated baryon asymmetry:
spectator processes~\cite{Buchmuller:2001sr,Nardi:2005hs},
finite temperature corrections~\cite{Giudice:2003jh}, lepton flavour
effects~\cite{Barbieri:1999ma,Endoh:2003mz,Abada:2006fw,Nardi:2006fx,Abada:2006ea}
and finally attempts to provide a full quantum mechanical formulation of thermal
leptogenesis~\cite{Buchmuller:2000nd,De Simone:2007rw,Garny:2009,Cirigliano:2009yt,
Beneke:2010wd,Beneke:2010dz,Anisimov:2010dk,Dev:2014wsa}.

By constrast, much less work has been devoted to the leptogenesis scenarios involving
fermionic~\cite{Hambye:2003rt,Fischler:2008xm,Strumia:2008cf,AristizabalSierra:2010mv,Zhuridov:2012hb}
or scalar~\cite{Hambye:2003ka,Antusch:2004xy,Hambye:2005tk,Chun:2006sp,
Hallgren:2007nq,Frigerio:2008ai,Felipe:2013kk,Sierra:2014tqa} electroweak triplets
(for a recent review of these scenarios and additional references, see Refs.~\cite{Hambye:2012fh}).
The CP asymmetry in scalar triplet/antitriplet decays was computed for
various models in Refs~\cite{Hambye:2003ka,Antusch:2004xy}.
A detailed quantitative study of scalar triplet leptogenesis
was performed in the single flavour approximation in Ref.~\cite{Hambye:2005tk},
and was extended to the supersymmetric case in Ref.~\cite{Chun:2006sp}.
Flavour effects were addressed in flavour non-covariant approaches in
Refs.~\cite{Felipe:2013kk,Sierra:2014tqa},
and spectator processes were also included in Ref.~\cite{Sierra:2014tqa}.
In this paper, we extend and improve previous works on flavour-dependent
scalar triplet leptogenesis by providing a complete set of flavour-covariant
Boltzmann equations using the density matrix formalism~\cite{density_matrix,Barbieri:1999ma}.
We find that flavour covariance is a crucial ingredient
of the computation of the generated baryon asymmetry,
and that flavour effects also matter in the temperature regime where
charged lepton Yukawa interactions are out of equilibrium.
We show in particular that, as opposed to the standard leptogenesis scenario
with right-handed neutrinos, the single flavour calculation does not provide
a good approximation to the full flavoured computation, even
when charged lepton Yukawa couplings can be neglected.

The paper is organized as follows. The framework and the basic ingredients
of flavour-dependent scalar triplet leptogenesis are presented in Section~\ref{sec:ingredients}.
In Section~\ref{sec:flavour}, we justify the use of a flavour-covariant formalism
and derive the Boltzmann equation for the density matrix in the closed time path formalism.
Spectator processes and chemical equilibrium are discussed in Section~\ref{sec:spectator}.
Section~\ref{sec:BE} provides the set of Boltzmann equations relevant to the different
temperature regimes, both in the flavour-covariant approach with spectator processes
included, and in various approximations neglecting flavour covariance
and/or spectator processes.
In Section~\ref{sec:numerical}, we study numerically
the impact of flavour effects and spectator processes on scalar triplet leptogenesis,
and we compare the flavour-covariant with the flavour non-covariant computations.
We find that the generated baryon asymmetry can be significantly enhanced
(up to several orders of magnitude) by the proper inclusion of flavour effects
and spectator processes, thus enlarging the region of triplet parameter
space allowed by successsful leptogenesis.
Finally, we present our conclusions in Section~\ref{sec:conclusions}.
The formulae for the space-time densities of reactions used in the paper
can be found in Appendix~\ref{app:reactions}.

\section{Basic ingredients of scalar triplet leptogenesis}         %
\label{sec:ingredients}                                                             %

\subsection{The framework}    %

We work in the framework of the type II seesaw model~\cite{typeII_seesaw},
i.e. the Standard Model augmented with a massive scalar electroweak triplet
which couples to left-handed leptons and to the Higgs boson as follows:
\begin{align}
  \mathcal{L}\, =\, -\frac{1}{2}\left(f_{\alpha\beta}\ell_\alpha^TCi\sigma^2\Delta\ell_\beta
    +\mu H^Ti\sigma^2\Delta^\dagger H + \mbox{h.c.} \right)-M_\Delta^2\, \mbox{tr} (\Delta^\dagger\Delta)\, ,
\label{eq:Lagrangian}
\end{align}
where $C$ is the charge conjugation matrix defined by $C \gamma^T_\mu C^{-1} = - \gamma_\mu$,
and
\begin{align}
  \Delta\, =\,
    \left( \begin{array}{cc}  \Delta^+ / \sqrt{2} & \Delta^{++} \\
    \Delta^0 & - \Delta^+ / \sqrt{2}
    \end{array} \right) ,  \qquad
  \Delta^\dagger\, =\,
    \left( \begin{array}{cc}  \Delta^- / \sqrt{2} & \Delta^{0*} \\
    \Delta^{--} & - \Delta^- / \sqrt{2}
    \end{array} \right) .
\end{align}
When decomposed on the components of each electroweak multiplet, this Lagrangian becomes
\begin{align}
  \mathcal{L}\, =\,  - & \frac{1}{2} \left\{ f_{\alpha\beta} \left(\nu_\alpha^T C \nu_\beta\, \Delta^0
    - \sqrt{2}\, \nu_\alpha^T C e_\beta\, \Delta^+ - e_\alpha^T C e_\beta\, \Delta^{++} \right) \right.  \nonumber \\
  + & \left. \mu \left( - H^0 H^0 \Delta^{0*} -\sqrt{2}\, H^0 H^+ \Delta^- + H^+ H^+ \Delta^{--} \right)
    + \mbox{h.c.} \right\}  \nonumber \\
  -\, & M_\Delta^2 \left( \Delta^{0*} \Delta^0 + \Delta^-\Delta^+ + \Delta^{--}\Delta^{++} \right) .
\end{align}
The triplet gives a contribution to the neutrino mass matrix:
\begin{align}
  (m_\Delta)_{\alpha\beta}\, =\, \frac{1}{2}\, \mu f_{\alpha\beta}\frac{v^2}{M_\Delta^2}\, ,
\label{eq:m_Delta}
\end{align}
where $v = \langle H^0 \rangle = 174 \GeV$ is the Higgs boson vacuum expectation value.
In addition to generating small neutrino masses, the heavy scalar triplet can also create
a lepton asymmetry through its decays, like the right-handed neutrinos of the type I
seesaw mechanism. There are however important differences between the standard
leptogenesis scenario involving heavy Majorana neutrinos and scalar triplet leptogenesis.
First $\Delta$ has 2 types of decays, into Higgs bosons and into antileptons,
with tree-level decay rates
\beq
  \Gamma(\Delta\rightarrow\bar \ell\bar \ell)\, =\, \frac{1}{32\pi}\, \mathrm{tr}(ff^\dagger)M_\Delta\, ,  \qquad
  \Gamma(\Delta\rightarrow HH)\, =\, \frac{1}{32\pi}\, \frac{|\mu|^2}{M_\Delta}\, .
\eeq
It will prove convenient in the following to introduce the quantities:
\beq
  \lambda_\ell\, \equiv\, \sqrt{\text{tr}(ff^\dagger)}\, ,  \qquad \qquad  \lambda_H\, \equiv\, \frac{|\mu|}{M_\Delta}\, ,
\eeq
which control the tree-level decay width and branching ratios of the scalar triplet:
\beq
  \Gamma_\Delta\, =\, \frac{1}{32\pi} \left( \lambda^2_\ell + \lambda^2_H \right) M_\Delta\, ,
\eeq
\beq
  B_\ell\, \equiv\, \mathrm{BR}(\Delta\rightarrow\bar \ell\bar \ell)\, =\, \frac{\lambda_\ell^2}{\lambda_\ell^2+\lambda_H^2}\ ,
  \qquad
  B_H\, \equiv\, \mathrm{BR}(\Delta\rightarrow HH)\, =\, \frac{\lambda_H^2}{\lambda_\ell^2+\lambda_H^2}\ .
\eeq
Second, in contrast to the heavy Majorana neutrinos of standard leptogenesis,
the scalar triplet is not a self-conjugate state. Hence, an asymmetry between
the triplet and antitriplet abundances will arise and affect the dynamics of leptogenesis.
Third, being charged under $SU(2)_L \times U(1)_Y$, the triplets and antitriplets
can annihilate before decaying.
In order to generate a large enough lepton asymmetry, decays have to happen 
with a rate higher or similar to gauge annihilations, which are typically in thermal equilibrium.
This requirement seems to conflict with Sakharov's third condition~\cite{Sakharov:1967dj},
but the fact that the triplet has several decay channels makes it possible for some of them
to occur out of equilibrium and resolves the contradiction~\cite{Hambye:2005tk}.

Finally, another important feature of scalar triplet leptogenesis, whose significance
has been missed so far, lies in the fact that the scalar triplet couples to pairs of
leptons from different generations rather than to a coherent superposition
of lepton flavours\footnote{Indeed, except for very specific flavour structures
of the parameters $f_{\alpha\beta}$, it is not possible to find a superposition
of lepton doublet flavours $\ell_\Delta  = \sum_\alpha c_\alpha \ell_\alpha$
such that the scalar triplet couplings to leptons $f_{\alpha\beta} \Delta \ell_\alpha \ell_\beta$
can be rewritten as $f \Delta \ell_\Delta \ell_\Delta$. By contrast, the couplings of a right-handed
neutrino $Y_{1\alpha} N_1 \ell_\alpha H$ can be rewritten $y N_1 \ell_{N_1\, }\! H$,
with $y = \sqrt{\sum_\alpha |Y_{1\alpha}|^2}$ and $\ell_{N_1} = \sum_\alpha Y_{1\alpha} \ell_\alpha / y$.}
(as opposed to a right-handed neutrino). As a consequence,
its dynamics cannot be described by Boltzmann equations involving a single
lepton asymmetry. This is at variance with the standard leptogenesis scenario,
in which neglecting the effects of the different lepton flavours is a good approximation
at high temperature. For scalar triplet leptogenesis, a flavour-covariant formalism
must be employed, as will be discussed in Section~\ref{sec:flavour}.

In order for leptogenesis to account for the observed baryon asymmetry
of the universe, a large enough asymmetry between the leptonic decay
rates of triplets and antitriplets is needed. It is a well-known fact~\cite{O'Donnell:1993am,Ma:1998dx}
that this condition is not satisfied with the minimal particle content of the type II seesaw
mechanism, since there is no CP asymmetry in $\Delta$/$\bar \Delta$ decays
at the one-loop level\footnote{We found a non-vanishing CP asymmetry
only at the 3-loop level, and only in the flavoured regime.}.
For this to happen, another heavy state (or possibly several heavy states)
with couplings to the lepton and Higgs doublets must be added to the model.
If this additional particle is significantly heavier than $\Delta$,
it will not be present in the thermal bath at the time of leptogenesis and
its effect can be parametrized by the effective dimension-5 operator~\cite{Hambye:2005tk}
\begin{align}
  \Delta \mathcal{L}_{\rm eff}\, =\, \frac{1}{4}\, \frac{\kappa_{\alpha\beta}}{\Lambda}\,
    (\ell^T_\alpha i \sigma^2 H)\, C\, (H^T i\sigma^2 \ell_\beta) + \mbox{h.c.}\, ,
\label{eq:Weinberg_operator}
\end{align}
suppressed by $\Lambda \gg M_\Delta$, where
$\Lambda=M^2_{\Delta'}/\mu' = M_{\Delta'}/\lambda'_H$ if the heavier particle is a scalar triplet
$\Delta'$, and $\Lambda=M$ if it is a right-handed neutrino or a fermionic triplet with mass $M$.
The operator~(\ref{eq:Weinberg_operator}) also gives a contribution to the neutrino mass matrix:
\begin{align}
  (m_\mathcal{H})_{\alpha\beta}=\frac{1}{2}\kappa_{\alpha\beta}\frac{v^2}{\Lambda}\, .
\label{eq:m_heavy}
\end{align}
In full generality one should also add an effective dimension-6 operator (which arises
at tree level if the heavier particle is a scalar triplet~\cite{Felipe:2013kk},
but only at the one-loop level if it is a right-handed neutrino or a fermionic triplet):
\begin{align}
   \Delta' \mathcal{L}_{\rm eff}\, \, =\, -\frac{1}{4}\, \frac{\eta_{\alpha\beta\gamma\delta}}{\Lambda^2}
    \left( \ell_\alpha^T C i \sigma^2 \vec{\sigma} \ell_\beta \right)\! \cdot\!
    \left( \bar{\ell}_\gamma \vec{\sigma} i \sigma^2 C \bar{\ell}_\delta^T \right) ,
\label{eq:4lepton_operator}
\end{align}
where $\eta_{\alpha\beta\gamma\delta}$ is symmetric under the exchanges $\alpha\leftrightarrow\beta$
and $\gamma\leftrightarrow\delta$.
This operator induces a contribution to the flavour-dependent CP asymmetries
in $\Delta$/$\bar \Delta$ decays that vanishes when summed over lepton flavours,
hence it only affects leptogenesis when the dynamics of the different flavours
is taken into account. However, it generally plays a subdominant role because
it is suppressed by an additional power of $\Lambda$ (and possibly also
by a loop factor) with respect to the operator~(\ref{eq:Weinberg_operator}).
A notable exception arises when~(\ref{eq:Weinberg_operator})
and~(\ref{eq:4lepton_operator}) are generated by a heavier scalar triplet $\Delta'$
with couplings to lepton and Higgs doublets such that
$\lambda_H \lambda'_H \ll \lambda_\ell \lambda'_\ell M_\Delta / M_{\Delta'}$.
In this case, the operator~(\ref{eq:4lepton_operator}) gives the dominant contribution
to the flavour-dependent CP asymmetries in $\Delta$ decays, while the total
CP asymmetry (to which only~(\ref{eq:Weinberg_operator}) contributes) is small.
This scenario, dubbed purely flavoured leptogenesis (PFL) in the literature,
has been studied\footnote{We disagree with the claim~\cite{Felipe:2013kk}
of a strong enhancement of the generated lepton asymmetry in PFL for low triplet masses,
which can be traced back to an erroneous term in the Boltzmann equations of Ref.~\cite{Felipe:2013kk}.
Indeed, the second term in Eq.~(23) of Ref.~\cite{Felipe:2013kk} generates a lepton
asymmetry in thermal equilibrium, thus violating the third Sakharov condition~\cite{Sakharov:1967dj}.}
in Refs.~\cite{Felipe:2013kk,Sierra:2014tqa}.
In this paper we shall stick to the less specific case of dominance of
the operator~(\ref{eq:Weinberg_operator}) and omit the operator~(\ref{eq:4lepton_operator}).

\subsection{CP Asymmetries in $\Delta$/$\bar \Delta$ decays}   %
\label{subsec:CP_asymmetry}                                                      %

%
\begin{figure}
\begin{center}
\fcolorbox{white}{white}{
  \begin{picture}(300,98) (63,-47)
    \SetWidth{1.0}
    \SetColor{Black}
    \Line[dash,dashsize=10,arrow,arrowpos=0.5,arrowlength=5,arrowwidth=2,arrowinset=0.2](64,2)(112,2)
    \Line[arrow,arrowpos=0.5,arrowlength=5,arrowwidth=2,arrowinset=0.2](160,50)(112,2)
    \Line[arrow,arrowpos=0.5,arrowlength=5,arrowwidth=2,arrowinset=0.2](160,-46)(112,2)
    \Line[dash,dashsize=10,arrow,arrowpos=0.5,arrowlength=5,arrowwidth=2,arrowinset=0.2](192,2)(240,2)
    \Arc[dash,dashsize=10,arrow,arrowpos=0.5,arrowlength=5,arrowwidth=2,arrowinset=0.2,clock](272,2)(32,-180,-360)
    \Arc[dash,dashsize=10,arrow,arrowpos=0.5,arrowlength=5,arrowwidth=2,arrowinset=0.2](272,2)(32,-180,0)
    \Line[arrow,arrowpos=0.5,arrowlength=5,arrowwidth=2,arrowinset=0.2](352,50)(304,2)
    \Line[arrow,arrowpos=0.5,arrowlength=5,arrowwidth=2,arrowinset=0.2](352,-46)(304,2)
    \Vertex(304,2){4}
    \Text(88,6)[cb]{\normalsize{\Black{$\Delta$}}}
    \Text(216,6)[cb]{\normalsize{\Black{$\Delta$}}}
    \Text(162,48)[lc]{\normalsize{\Black{$\ell_\alpha$}}}
    \Text(162,-44)[lc]{\normalsize{\Black{$\ell_\beta$}}}
    \Text(272,40)[cb]{\normalsize{\Black{$H$}}}
    \Text(272,-36)[ct]{\normalsize{\Black{$H$}}}
    \Text(354,48)[lc]{\normalsize{\Black{$\ell_\alpha$}}}
    \Text(354,-44)[lc]{\normalsize{\Black{$\ell_\beta$}}}
    \Text(118,2)[lc]{\normalsize{\Black{$f_{\alpha\beta}$}}}
    \Text(300,2)[rc]{\normalsize{\Black{$\dfrac{\kappa_{\alpha\beta}}{\Lambda}$}}}
  \end{picture}
}
\end{center}
\caption{Diagrams responsible for the flavour-dependent CP asymmetries $\epsilon_{\alpha\beta}$.}
\label{fig:epsilon_diagrams}
\end{figure}
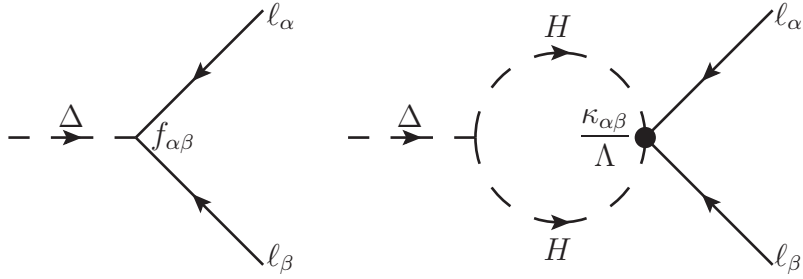

The CP asymmetries in the decays of the triplets and antitriplets are defined by
\begin{align}
  \epsilon_H\, \equiv\, 2\ \frac{\Gamma(\Delta \rightarrow H H)-\Gamma(\bar \Delta \rightarrow \bar H \bar H)}
    {\Gamma_\Delta+\Gamma_{\bar \Delta}}\, ,
\end{align}
\begin{align}
  \epsilon_{\alpha\beta}\, \equiv\,
    \frac{\Gamma(\bar \Delta\rightarrow \ell_\alpha \ell_\beta)-\Gamma(\Delta\rightarrow \bar \ell_\alpha \bar \ell_\beta)}
    {\Gamma_\Delta+\Gamma_{\bar \Delta}}\ \left( 1 + \delta_{\alpha \beta} \right) ,
\label{eq:epsilon_alphabeta_def}
\end{align}
where we included a factor 2 in the definitions of $\epsilon_H$ and $\epsilon_{\alpha \alpha}$
for later convenience. The flavour-dependent CP asymmetries $\epsilon_{\alpha \beta}$
come from the interference between the diagrams shown in Fig.~\ref{fig:epsilon_diagrams}.
With the definition~(\ref{eq:epsilon_alphabeta_def}), they can be expressed in terms of
$m_\Delta$ and $m_{\cal H}$ as
\begin{align}
  \epsilon_{\alpha\beta}\, =\, \frac{1}{4\pi}\frac{M_\Delta}{v^2}\sqrt{B_\ell B_H}\
    \frac{\mbox{Im}\left[(m^*_\Delta)_{\alpha\beta} (m_\mathcal{H})_{\alpha\beta}\right]}{\bar{m}_\Delta}\, ,
\label{eq:epsilon_alphabeta}
\end{align}
where we have introduced
\beq
  \bar{m}_\Delta\, \equiv\, \sqrt{\text{tr}(m_\Delta^\dagger m_\Delta)}\, .
\eeq
The total CP asymmetry is given by~\cite{Hambye:2005tk}
\begin{align}
  \epsilon_\Delta\, \equiv\, \sum_{\alpha, \beta} \epsilon_{\alpha \beta}\, =\,
  \frac{1}{4\pi}\frac{M_\Delta}{v^2}\sqrt{B_\ell B_H}\
 \frac{\mbox{Im} \left[ \mbox{tr} (m^\dagger_\Delta m_\mathcal{H}) \right]}{\bar{m}_\Delta}\,
  =\, \epsilon_H\, ,
\label{eq:epsilon_Delta}
\end{align}
where the last equality follows from CPT invariance.

\subsection{Washout processes}     %
\label{subsec:washout}                    %

The lepton asymmetry generated in triplet and antitriplet decays is partially washed out
by lepton number violating processes. These include the inverse decays
$\ell_\alpha \ell_\beta \rightarrow \bar \Delta$ and $\bar \ell_\alpha \bar \ell_\beta \rightarrow \Delta$,
and the $\Delta L = 2$ scatterings $\ell_\alpha\ell_\beta\leftrightarrow \bar H \bar H$
and $\ell_\alpha H\leftrightarrow\bar \ell_\beta \bar H$, mediated by $s$- and $t$-channel triplet
exchange, respectively, as well as by the higher order operator~(\ref{eq:Weinberg_operator}).
In addition, some processes that preserve total lepton number redistribute the asymmetries
between the different lepton flavours, namely
the 2 lepton--2 lepton scatterings  $\ell_\alpha\ell_\beta\leftrightarrow\ell_\gamma\ell_\delta$
and $\ell_\alpha\bar \ell_\gamma\leftrightarrow\bar \ell_\beta\ell_\delta$ (mediated by
s- and t-channel triplet exchange, respectively), and the 2--2 scatterings
involving leptons and triplets $\ell_\alpha\Delta\leftrightarrow\ell_\beta\Delta$,
$\ell_\alpha\bar \Delta\leftrightarrow\ell_\beta\bar \Delta$
and $\ell_\alpha\bar \ell_\beta\leftrightarrow\Delta\bar \Delta$ (mediated by
s-, t- and u-channel lepton exchange). Since the asymmetries stored in different lepton
flavours are subject to different washout rates, these flavour violating processes
have an impact on the erasure of the total lepton number, and we will
refer to them as washout processes, too.

We define for later reference the space-time density of triplet and antitriplet decays:
\beq
  \gamma_D\, =\, \sum_{\alpha,\beta} \left[ \gamma(\Delta \leftrightarrow \bar \ell_\alpha \bar \ell_\beta)
    + \gamma(\bar \Delta \leftrightarrow \ell_\alpha \ell_\beta) \right]
    + \gamma(\Delta \leftrightarrow H H) + \gamma(\bar \Delta \leftrightarrow \bar H \bar H)\, ,
\eeq
as well as the following combinations of flavour-dependent scattering densities:
\begin{align}
  \gamma_{\ell H}\, &=\, \sum_{\alpha,\beta} \left[ 2\gamma(\ell_\alpha\ell_\beta\leftrightarrow \bar H \bar H)
   + \gamma(\ell_\alpha H\leftrightarrow\bar \ell_\beta \bar H) \right] ,  \\
  \gamma_{4\ell}\, &=\, \sum_{\alpha,\beta,\gamma,\delta}
    \left [2\gamma(\ell_\alpha\ell_\beta\leftrightarrow \ell_\gamma\ell_\delta)
    + \gamma(\ell_\alpha\bar \ell_\gamma\leftrightarrow\bar \ell_\beta\ell_\delta) \right] ,  \\
  \gamma_{\ell\Delta}\, &=\, \sum_{\alpha,\beta} \left[ \gamma(\ell_\alpha\Delta\leftrightarrow \ell_\beta\Delta)
    + \gamma(\ell_\alpha \bar \Delta\leftrightarrow\ell_\beta\bar \Delta)
    + \gamma(\ell_\alpha\bar \ell_\beta\leftrightarrow\Delta\bar \Delta) \right] .
\end{align}
%

\section{Flavour-covariant formalism}       %
\label{sec:flavour}                                      %

\subsection{The need for a flavour-covariant formalism}  %

Leptogenesis computations are often performed in the so-called single flavour approximation,
in which a single direction in flavour space is considered.
This is a rather accurate approximation in scenarios in which a heavy Majorana
neutrino $N_1$ decaying at high temperature
is responsible for the whole lepton asymmetry. Indeed, the couplings of $N_1$
can be rewritten as (with the contraction of $SU(2)_L$ indices omitted)
\begin{equation}
  -\sum_\alpha Y_{1\alpha}\bar{N}_1 \ell_\alpha H + \mbox{h.c.}\, =\,
    -y \bar{N}_1 \ell_{N_1\, }\! H + \mbox{h.c.}\, ,
\end{equation}
where $y \equiv \sqrt{\sum_\alpha |Y_{1\alpha}|^2}$ and
$\ell_{N_1} \equiv \sum_\alpha Y_{1\alpha} \ell_\alpha / y$.
When processes mediated by the charged lepton Yukawa couplings are out of equilibrium,
i.e. when $N_1$ decays in the high temperature regime ($T > 10^{12} \GeV$), the coherence
of $\ell_{N_1}$ is effectively preserved by all interactions\footnote{Except for the $\Delta L = 2$
scatterings $\ell_\alpha H \leftrightarrow \bar \ell_\beta \bar H$ and
$\ell_\alpha \ell_\beta \leftrightarrow \bar H \bar H$ mediated by the heavier Majorana
neutrinos $N_2$ and $N_3$, which are neglected in this discussion.}
in the thermal plasma
and leptogenesis can safely be described in terms of a single lepton flavour.
When the lepton asymmetry is generated at lower temperature, on the other hand,
charged lepton Yukawa interactions enter equilibrium and destroy the coherence
of $\ell_{N_1}$. The single flavour approximation is no longer appropriate,
and the proper treatment involves a $3 \times 3$ matrix in lepton flavour space~\cite{Barbieri:1999ma},
the density matrix $(\Delta_\ell)_{\alpha\beta}$, whose diagonal entries are the
asymmetries stored in each lepton doublet $\ell_\alpha$, while the off-diagonal
entries encode the quantum correlations between the different flavours.
This matrix transforms as $\Delta_\ell \to U^* \Delta_\ell\, U^T$ under flavour
rotations $\ell \to U \ell$, and its evolution is governed by a flavour-covariant Boltzmann
equation\footnote{This formalism has been extended to the case where several heavy
Majorana neutrinos with hierarchical masses play a role in leptogenesis in Ref.~\cite{Blanchet:2011xq},
and a fully flavour-covariant formalism in which the quantum correlations between
the heavy and light neutrino flavours are taken into account has been developed in
Ref.~\cite{Dev:2014laa} (see also Refs.~\cite{Akhmedov:1998qx,Asaka:2005pn}
for an earlier use of a density matrix for the heavy neutrino flavours, in the scenario
where the baryon asymmetry is generated through CP-violating oscillations of the ``heavy''
neutrinos below the electroweak scale).}.

However, in the case of leptogenesis with right-handed neutrinos discussed above,
the density matrix formalism is only
really needed at the transition between two temperature regimes~\cite{De Simone:2006dd},
or in the case where several heavy Majorana neutrinos contribute to the generation
and washout of the lepton asymmetry
as in Refs.~\cite{Barbieri:1999ma,Vives:2005ra,Engelhard:2006yg,Blanchet:2011xq},
or in resonant leptogenesis~\cite{Pilaftsis:2005rv,Dev:2014laa}.
Otherwise there is always a natural choice of basis in which the Boltzmann equation
for $(\Delta_\ell)_{\alpha\beta}$ can be substituted for a set of Boltzmann equations
for 1, 2 or 3 flavour asymmetries.
Above $T = 10^{12} \GeV$, where all charged lepton Yukawa couplings
are out of equilibrium, the appropriate flavour basis is
$(\ell_{N_1}, \ell_{\perp 1}, \ell_{\perp 2})$,
where $\ell_{\perp 1}$ and $\ell_{\perp 2}$ are two directions in flavour space
perpendicular to $\ell_{N_1}$. In this basis, leptogenesis is well described
in terms of the sole lepton asymmetry $\Delta_{\ell_{N_1}}$.
Below $T = 10^{12} \GeV$, tau Yukawa-induced
processes like $q_3 \ell_\tau \to t_R \tau_R$ are in equilibrium and destroy
the coherence between $\ell_\tau$ and the other two lepton flavours.
However, as long as the muon Yukawa coupling is out of equilibrium
(which is the case if $T > 10^9 \GeV$), the coherence of
$\ell_0 \equiv (Y_{1e} \ell_e + Y_{1\mu} \ell_\mu) / \sqrt{|Y_{1e}|^2+|Y_{1\mu}|^2}$
is preserved, and the
dynamics of leptogenesis can be described in a 2-flavour approximation, in terms
of the asymmetries $\Delta_{\ell_0}$ and $\Delta_{\ell_\tau}$. Finally, below
$T = 10^9 \GeV$, the muon Yukawa interactions are in equilibrium as well,
and flavour coherence is completely broken. Leptogenesis is then governed
by Boltzmann equations for the three asymmetries $\Delta_{\ell_e}$,
$\Delta_{\ell_\mu}$ and $\Delta_{\ell_\tau}$.

The case of scalar triplet leptogenesis is significantly different, because
the scalar triplet does not couple to a single combination of lepton flavours
in general. This makes the use of a flavour-covariant formalism unavoidable
as long as the quantum correlations between lepton flavours
are not destroyed by charged lepton Yukawa interactions.
One can still define formally a single flavour approximation
by making the substitutions $(\Delta_\ell)_{\alpha\beta} \to \Delta_\ell$, $f_{\alpha\beta} \to \lambda_\ell$,
$\kappa_{\alpha\beta} \to \lambda_\kappa \equiv \sqrt{\mbox{tr}(\kappa \kappa^\dagger)}$
and $\mathcal{E}_{\alpha\beta} \to \epsilon_\Delta$ (where $\mathcal{E}_{\alpha\beta}$
are the flavour-covariant CP asymmetries, to be defined later), but the resulting
Boltzmann equations cannot be derived from the flavour-covariant ones by taking
the limit of vanishing Yukawa couplings. Hence, even if the scalar triplets decay
above $T = 10^{12} \GeV$, there is no guarantee that the single flavour calculation
will give a good approximation to the flavour-covariant result.
\begin{figure}[t]
\center
\includegraphics[scale=0.55]{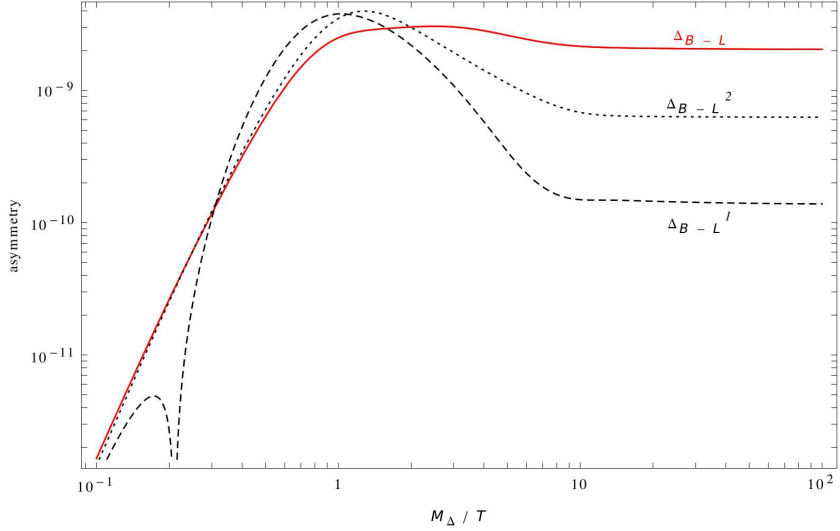}
\caption{The $B-L$ asymmetry generated in scalar triplet leptogenesis as a function of $z = m_\Delta / T$,
computed following three different prescriptions: density matrix computation
using the Boltzmann equations~(\ref{eq:BE_Sigma})--(\ref{eq:BE_Delta}),
yielding $\Delta_{B-L} = \text{tr} (\Delta_{\alpha\beta})$ {\it (red solid curve)};
naive computation using the Boltzmann equations~(\ref{eq:BE_Sigma_10^9})--(\ref{eq:BE_Delta_10^9})
written in the charged lepton mass eigenstate basis, yielding
$(\Delta_{B-L})^1 = \Delta_e + \Delta_\mu + \Delta_\tau$ {\it (black dashed curve)};
naive computation performed in the neutrino mass eigenstate basis, yielding
$(\Delta_{B-L})^2 = \Delta_1 + \Delta_2 + \Delta_3$ {\it (black dotted curve)}.
The parameters are $M_\Delta = 10^{13} \GeV$, $\lambda_H = 0.01$
and, in the notation of Eqs.~(\ref{eq:parametrization_mDelta})
and~(\ref{eq:parametrization_Uprime}), $\phi_{12} = \phi_{13} = \pi/4$, $\phi_{23} = 0$,
$\beta_1 = \pi/4$ and $\beta_2 = \pi/6$, all other phases being zero.}
\label{fig:comparaison}
\end{figure}

A less naive approach is to neglect flavour covariance but keep track of the different
lepton flavours, and write classical
Boltzmann equations for the three asymmetries $\Delta_{\ell_\alpha}$.
When the lepton asymmetry is generated below $T = 10^9 \GeV$,
this is fully justified because quantum correlations between the different
lepton flavours are destroyed
by fast interactions induced by the muon and tau Yukawa couplings.
In this case, the appropriate flavour basis is the charged lepton mass
eigenstate basis: $\Delta_{\ell_\alpha} = \{\Delta_{\ell_e}, \Delta_{\ell_\mu}, \Delta_{\ell_\tau} \}$.
When all charged lepton Yukawa interactions are out of equilibrium,
however, the choice of the flavour basis becomes a delicate issue.
In fact, it turns out that Boltzmann equations
written in different bases cannot be transformed into one another by a flavour
rotation $\ell \to U \ell$, and give different numerical results for the baryon asymmetry.
This is illustrated in Fig.~\ref{fig:comparaison}, which compares the $B-L$
asymmetry computed in the flavour-covariant forma\-lism using the density matrix
with the naive computation involving Boltzmann equations
for the flavour asymmetries $\Delta_{\ell_\alpha}$, written either
in the charged lepton or in the neutrino mass eigenstate basis.
Since physics should not depend on a choice of basis, we are led to conclude
that the density matrix formalism is the only valid approach when charged lepton
Yukawa interactions are out of equilibrium\footnote{In the limit where
scatterings involving lepton and Higgs doublets can also be neglected, there is however
a privileged basis in which scalar triplet leptogenesis can be described in terms of 
the three ``diagonal'' flavour asymmetries $(\Delta_\ell)_{\alpha\alpha} = \Delta_{\ell_\alpha}$.
Indeed, in the basis where the triplet couplings to leptons
$f_{\alpha\beta}$ are flavour diagonal, the evolution of the diagonal entries of the
density matrix becomes independent of its off-diagonal entries, as can be checked using
Eqs.~(\ref{eq:covariant_BE}), (\ref{eq:W_D}), (\ref{eq:W_4l}) and~(\ref{eq:W_ellDelta}).
This ``3-flavour approximation'', which is valid in the temperature regime where all
charged lepton Yukawa interactions are out of equilibrium, is the analog of the single
flavour approximation in the leptogenesis scenario with right-handed neutrinos.}.

In the intermediate temperature range
where the tau Yukawa coupling is in equilibrium but the muon and electron
Yukawa couplings are not, the quantum correlations between the tau and the other flavours
are destroyed. In practice, this means that the off-diagonal entries in the third line and
the third column of the density matrix are driven to zero by the fast tau Yukawa interactions.
Thus, the relevant dynamical variables in this regime are the $2\times2$ density matrix
$(\Delta^0_\ell)_{\alpha\beta}$ describing the asymmetries stored in the lepton
doublets $\ell_e$, $\ell_\mu$ and their quantum correlations, and 
the asymmetry $\Delta_{\ell_\tau}$ stored in $\ell_\tau$.

\subsection{The Boltzmann equation for the density matrix}  %
\label{subsec:density_matrix}                                                 %

The derivation of the evolution equation for the density matrix is not straightforward. 
In this paper, we shall use the closed time path (CTP) formalism~\cite{CTP},
which has been used to obtain flavoured quantum Boltzmann equations for the standard
leptogenesis scenario with heavy Majorana neutrinos~\cite{Buchmuller:2000nd,De Simone:2007rw,
Garny:2009,Cirigliano:2009yt,Beneke:2010wd,Beneke:2010dz,Anisimov:2010dk,Dev:2014wsa}
(for other approaches, also in the framework of the standard leptogenesis scenario,
see e.g. Refs.~\cite{Abada:2006fw,De Simone:2006dd} or the review~\cite{Davidson:2008bu}).
In this formalism, which is well adapted to describe non-equilibrium phenomena in quantum
field theory, particle densities are replaced by Green's functions defined on a closed path
$\mathcal{C}$ in the complex time plane going from an initial instant $t=0$ to $t=+\infty$ and back.
When applied to leptogenesis, this formalism leads to quantum Boltzmann equations involving
memory effects and off-shell corrections~\cite{Buchmuller:2000nd}; in particular,
the CP asymmetries are functions of time and their values depend on the history of the system.
Such effects can be important for resonant leptogenesis and soft
leptogenesis~\cite{De Simone:2007rw,Cirigliano:2007hb}, but they are not expected 
to play a significant role in the scenario studied in this paper, which does not involve
degenerate states. We shall therefore ignore them and make several simplifying
assumptions in order to obtain a (classical) flavour-covariant Boltzmann equation
for the density matrix. In particular, we shall ignore plasma/thermal effects and apply
the CTP formalism to quantum field theory at zero temperature.
This procedure is going to provide us with an equation of the form:

\beq
  sHz\frac{d (\Delta_\ell)_{\alpha\beta}}{dz}\, =\,
    \left(\frac{\Sigma_\Delta}{\Sigma_\Delta^\text{eq}}-1\right)\! \gamma_D\, \mathcal{E}_{\alpha\beta}
    - \mathcal{W}^D_{\alpha\beta} - \mathcal{W}^{\ell H}_{\alpha\beta}
    - \mathcal{W}^{4\ell}_{\alpha\beta} - \mathcal{W}^{\ell\Delta}_{\alpha\beta}\, ,
\label{eq:covariant_BE}
\eeq
where the right-hand side contains a source term proportional to the flavour-covariant
CP-asymmetry matrix $\mathcal{E}_{\alpha\beta}$ and washout terms 
associated with inverse triplet and antitriplet decays ($\mathcal{W}^D_{\alpha\beta}$),
$2 \to 2$ scatterings involving leptons and Higgs bosons ($\mathcal{W}^{\ell H}_{\alpha\beta}$),
2 lepton--2 lepton scatterings ($\mathcal{W}^{4\ell}_{\alpha\beta}$)
and $2 \to 2$ scatterings involving leptons and triplets ($\mathcal{W}^{\ell\Delta}_{\alpha\beta}$).
By construction, Eq.~(\ref{eq:covariant_BE}) is covariant under flavour rotations $\ell \to U \ell$,
which means that each of the matrices
$\mathcal{E}$, $\mathcal{W}^D$, $\mathcal{W}^{\ell H}$, $\mathcal{W}^{4\ell}$,
$\mathcal{W}^{\ell\Delta}$ transforms in the same way as the density matrix $\Delta_\ell$,
namely as $\mathcal{M} \to U^* \mathcal{M} U^T$.

Before proceeding with the derivation of Eq.~(\ref{eq:covariant_BE}), let us specify our notations.
For any species $X = \ell_\alpha, H, \Delta$, we define the comoving
number density $Y_X \equiv n_X / s$, where $s$ is the entropy density, and the asymmetry
stored in $X$ by $\Delta_X \equiv Y_X - Y_{\bar X} = (n_X - n_{\bar X}) / s$.
We also define the comoving number density of triplets and antitriplets
$\Sigma_\Delta \equiv (n_\Delta+n_{\bar \Delta})/s$.
The evolution of these quantities as a function of $z \equiv M_\Delta/T$ is governed by a set of
Boltzmann equations. Finally, a superscript $^{\rm eq}$ denotes an equilibrium density.

\subsubsection{Derivation of the Boltzmann equation in the CTP formalism}    %

The relevant degrees of freedom in scalar triplet leptogenesis are lepton doublets,
Higgs doublets and scalar triplets. We will therefore need the following Green's functions,
coresponding to all possible orderings of the fields along the closed time path $\mathcal{C}$
(for a review of the CTP formalism, see Ref.~\cite{Chou:1984es}):
\begin{align}
 G_{\alpha\beta}^>(x,y)\, & =\, -i\langle \ell_\alpha(x)\bar{\ell}_\beta(y)\rangle\, ,  \\
 G_{\alpha\beta}^<(x,y)\, & =\, i\langle \bar{\ell}_\beta(y)\ell_\alpha(x)\rangle\, ,  \\
 G_{\alpha\beta}^t(x,y)\, & =\, \theta(x^0-y^0)G_{\alpha\beta}^>(x,y)+\theta(y^0-x^0)G_{\alpha\beta}^<(x,y)\, ,  \\
 G_{\alpha\beta}^{\bar{t}}(x,y)\, & =\, \theta(y^0-x^0)G_{\alpha\beta}^>(x,y)+\theta(x^0-y^0)G_{\alpha\beta}^<(x,y)\, ,
\end{align}
%
%
%
where $\alpha, \beta$ are lepton flavour indices, and the $\ell$'s refer to left-handed lepton
doublets, whereas for a scalar field $\phi(x)$ (representing a Higgs doublet or a scalar triplet):
\begin{align}
 G_\phi^>(x,y)\, & =\, -i\langle \phi(x)\phi^\dagger(y)\rangle\, ,  \\
 G_\phi^<(x,y)\, & =\, -i\langle \phi^\dagger(y)\phi(x)\rangle\, ,  \\
 G_\phi^t(x,y)\, & =\, \theta(x^0-y^0)G_\phi^>(x,y)+\theta(y^0-x^0)G_\phi^<(x,y)\, ,  \\
 G_\phi^{\bar{t}}(x,y)\, & =\, \theta(y^0-x^0)G_\phi^>(x,y)+\theta(x^0-y^0)G_\phi^<(x,y)\, .
\end{align}
The brackets $\langle...\rangle$ mean that we take the average over all available states of the system.
One can write these Green's functions as a single $2 \times 2$ matrix:
\begin{equation}
 \tilde{G}_{\BF}\, = \left(\begin{matrix}G^t & \pm G^<\\G^> & -G^{\bar{t}}\end{matrix}\right) ,
\end{equation}
where the plus sign refers to bosons and the minus sign to fermions.
This matrix satisfies the following Schwinger-Dyson equation:
\begin{equation}
  \tilde{G}(x,y)\, =\, \tilde{G}^0(x,y) + \int_\mathcal{C}d^4w_1 \int_\mathcal{C}d^4w_2\,
    \tilde{G}^0(x,w_1)\tilde{\Sigma}(w_1,w_2)\tilde{G}(w_2,y)\, ,
\label{eq:SDE1}
\end{equation}
where $\tilde{\Sigma}$ is a $2 \times 2$ matrix
containing the self-energy functions $\Sigma^>$, $\Sigma^<$, $\Sigma^t$ and $\Sigma^{\bar t}$,
defined in an analogous way to the Green's functions $G^>$, $G^<$, $G^t$ and $G^{\bar t}$:
\begin{align}
 \tilde{\Sigma}_{\BF}\, = \left(\begin{matrix}\Sigma^t & \pm \Sigma^<\\ \Sigma^> & -\Sigma^{\bar{t}}\end{matrix}\right) ,
\end{align}
and $\tilde{G}^0$ is the free 2-point correlation function.
The Schwinger-Dyson equation can also be written as
\begin{equation}
  \tilde{G}(x,y)\, =\, \tilde{G}^0(x,y) + \int_\mathcal{C}d^4w_1 \int_\mathcal{C}d^4w_2\,
    \tilde{G}(x,w_1)\tilde{\Sigma}(w_1,w_2)\tilde{G}^0(w_2,y)\, .
\label{eq:SDE2}
\end{equation}
In the fermionic case, we note for later use that acting on Eqs.~(\ref{eq:SDE1})
and~(\ref{eq:SDE2}) with the operators $i\! \overrightarrow{\slashed{\partial}}_{\! x}$
and $i\! \overleftarrow{\slashed{\partial}}_{\! y}$, respectively,
gives the following equations of motion:
\begin{align}
  i\! \overrightarrow{\slashed{\partial}}_{\! x} \tilde{G}_{\alpha \beta}(x,y)\,
    & =\, \delta^{(4)}(x-y)\, \delta_{\alpha \beta}\, \tilde I\,
    +\, \sum_\gamma \int_\mathcal{C}d^4w\, \tilde{\Sigma}_{\alpha \gamma}(x,w)\tilde{G}_{\gamma \beta}(w,y)\, ,
\label{eq:dSDE1} \\
  \tilde{G}_{\alpha \beta}(x,y)\, i\! \overleftarrow{\slashed{\partial}}_{\! y}\,
    & =\, -\, \delta^{(4)}(x-y)\, \delta_{\alpha \beta}\, \tilde I\,
    +\, \sum_\gamma \int_\mathcal{C}d^4w\, \tilde{G}_{\alpha \gamma}(x,w)\tilde{\Sigma}_{\gamma \beta}(w,y)\, ,
\label{eq:dSDE2}
\end{align}
where we have restored the flavour indices and used the fact that the free Green's function
for massless fermions satisfies $i\slashed{\partial}_x \tilde{G}_{\alpha \beta}^0(x,y)
= \delta^{(4)}(x-y)\, \delta_{\alpha \beta}\, \tilde I$,
with $\tilde I$ the identity matrix in both spinor and CTP spaces.

Writing the free Dirac field as
\begin{equation}
  \psi(x)\, =\, \int\frac{d^3 \vec{p}}{(2\pi)^3\sqrt{2\omega_{\vec{p}}}} \sum_s \left(u(p,s)b(\vec{p},s)e^{-ip \cdot x}
    + v(p,s)d^\dagger(\vec{p},s)e^{ip \cdot x}\right) ,
\end{equation}
where $\omega_{\vec{p}} = \sqrt{{\vec{p}}^{\; 2} + m^2}$, we define the phase-space
distribution functions of lepton and antilepton doublets $\rho_{\alpha\beta}(\vec{p})$
and $\bar{\rho}_{\alpha\beta}(\vec{p})$ as matrices in flavour space by
\begin{align}
  \langle b_\alpha^\dagger(\vec{p})b_\beta(\vec{p'})\rangle\, =\,
    (2\pi)^3\delta^{(3)}(\vec{p}-\vec{p'})\rho_{\alpha\beta}(\vec{p})\, ,
\label{eq:rho}  \\
  \langle d_\beta^\dagger(\vec{p})d_\alpha(\vec{p'})\rangle\, =\,
    (2\pi)^3\delta^{(3)}(\vec{p}-\vec{p'})\bar{\rho}_{\alpha\beta}(\vec{p})\, .
\label{eq:rhobar}
\end{align}
The reversed order of the flavour indices $\alpha$ and $\beta$ in the definition
of $\bar {\rho}_{\alpha\beta}(\vec{p})$ ensures that the distribution functions
of lepton and antilepton doublets transform in the same way under a rotation
in flavour space, $\ell \to U \ell \,$:
\begin{align}
  \rho\ & \rightarrow\ U^* \rho\, U^T ,
\label{eq:rotation_rho}  \\
  \bar \rho\ & \rightarrow\ U^* \bar \rho\, U^T .
\label{eq:rotation_rhobar}
\end{align}
Similarly, the phase-space distribution functions of a charged scalar $\phi$ and of its antiparticle
are defined by
\begin{align}
  \langle a^\dagger_\phi(\vec{p})a_\phi(\vec{p'})\rangle\, =\,
    (2\pi)^3\delta^{(3)}(\vec{p}-\vec{p'})\rho_\phi(\vec{p})\, ,  \\
  \langle b_\phi^\dagger(\vec{p})b_\phi(\vec{p'})\rangle\, =\,
    (2\pi)^3\delta^{(3)}(\vec{p}-\vec{p'})\bar{\rho}_\phi(\vec{p})\, ,
\end{align}
where $a_\phi(\vec{p})$ and $b_\phi^\dagger(\vec{p})$ are the annihilation
and creation operators appearing in the definition of the free charged scalar field:
\begin{equation}
  \phi(x)\, =\, \int\frac{d^3 \vec{p}}{(2\pi)^3\sqrt{2\omega_{\vec{p}}}} \left( a_\phi(\vec{p}) e^{-ip \cdot x}
    + b_\phi^\dagger(\vec{p}) e^{ip \cdot x} \right) .
\end{equation}

With these definitions, the Green's functions for left-handed lepton doublets
can be written as (neglecting lepton masses)
\begin{align}
  iG_{\alpha\beta}^>(x,y)\, & =\, P_L\int \frac{d^3p}{(2\pi)^32\omega_{\vec{p}}}\ \, \slashed{p}
    \left\lbrace [\delta_{\alpha\beta}-\rho_{\beta\alpha}(\vec{p})]\, e^{-ip \cdot (x-y)}
    + \bar{\rho}_{\beta\alpha}(\vec{p})\, e^{ip \cdot (x-y)} \right\rbrace P_R\, ,  \\
  iG_{\alpha\beta}^<(x,y)\, & =\, -P_L\int \frac{d^3p}{(2\pi)^32\omega_{\vec{p}}}\ \, \slashed{p}
    \left\lbrace \rho_{\beta\alpha}(\vec{p})\, e^{-ip \cdot (x-y)}
    + [\delta_{\alpha\beta}-\bar{\rho}_{\beta\alpha}(\vec{p})]\, e^{ip \cdot (x-y)} \right\rbrace P_R\, ,
 \label{eq:G<_leptons}
\end{align}
whereas for scalars one obtains
\begin{align}
  iG_\phi^>(x,y)\, =\, \int \frac{d^3p}{(2\pi)^32\omega_{\vec{p}}} \left\lbrace
    [1+\rho_\phi(\vec{p})]\, e^{-ip \cdot (x-y)} + \bar{\rho}_{\phi}(\vec{p})\, e^{ip \cdot (x-y)} \right\rbrace\, ,  \\
  iG_\phi^<(x,y)\, =\, \int \frac{d^3p}{(2\pi)^32\omega_{\vec{p}}} \left\lbrace
    \rho_\phi(\vec{p})\, e^{-ip \cdot (x-y)} + [1+\bar{\rho}_{\phi}(\vec{p})]\, e^{ip \cdot (x-y)} \right\rbrace\, .
\end{align}
Strictly speaking, these expressions are valid for free Green's functions only,
but they will be sufficient for our purpose. Notice that, under charge conjugation,
\begin{align}
 \mathcal{C}G_{\alpha\beta}^>\mathcal{C}^{-1}(x,y)\, & =\, CG_{\beta\alpha}^{<T}(y,x)C^{-1}\,
   \equiv\, - G_{\beta\alpha_R}^{<}(y,x)\, ,  \label{eq:G_R}  \\
 \mathcal{C}G_\phi^>\mathcal{C}^{-1}(x,y)\, & =\, G_\phi^<(y,x)\, ,
\end{align}
where $C$, not to be confused with the charge conjugation operator $\mathcal{C}$,
is the charge conjugation matrix defined by $C \gamma^T_\mu C^{-1} = - \gamma_\mu$,
and the quantity $G_{\beta\alpha_R}^{<}$
in Eq.~(\ref{eq:G_R}) is defined as $G_{\beta\alpha}^{<}$ in Eq.~(\ref{eq:G<_leptons})
with $P_L$ interchanged with $P_R$.

The comoving number densities of scalars and left-handed leptons,
and of their antiparticles, are given by
\beq
  Y_\phi\, \equiv\, \frac{n_\phi}{s}\,
    =\, \frac{g_\phi}{s}\int \frac{d^3p}{(2\pi)^3}\, \rho_\phi(\vec{p})\, ,   \qquad
    Y_{\alpha\beta}\, \equiv\, \frac{n_{\alpha\beta}}{s}\,
    =\, \frac{2}{s}\int \frac{d^3p}{(2\pi)^3}\, \rho_{\alpha\beta}(\vec{p})\, ,
\label{eq:Y_Delta}
\eeq
\beq
  Y_{\bar \phi}\, \equiv\, \frac{n_{\bar \phi}}{s}\,
    =\, \frac{g_\phi}{s}\int \frac{d^3p}{(2\pi)^3}\, \bar{\rho}_\phi(\vec{p})\, ,   \qquad
    \bar{Y}_{\alpha\beta}\, \equiv\, \frac{\bar{n}_{\alpha \beta}}{s}\,
    =\, \frac{2}{s}\int \frac{d^3p}{(2\pi)^3}\, \bar{\rho}_{\alpha\beta}(\vec{p})\, ,
\label{eq:Ybar_Delta}
\eeq
where $Y_{\alpha\beta}$ and $\bar Y_{\alpha\beta}$, similarly to the phase-space distribution
functions $\rho_{\alpha\beta}(\vec{p})$ and $\bar \rho_{\alpha\beta}(\vec{p})$, are matrices
in flavour space, and $g_\phi = 2$ (resp. $g_\phi = 3$) for Higgs bosons (resp. scalar triplets).
One also defines the matrix of lepton asymmetries (hereafter called ``density matrix''):
\begin{align}
  (\Delta_\ell )_{\alpha\beta}\, \equiv\, \frac{\Delta n_{\alpha\beta}}{s}\,
    =\, Y_{\alpha\beta}-\bar{Y}_{\alpha\beta}\, .
\label{eq:Deltal_def}
\end{align}
The diagonal entries of $(\Delta_\ell )_{\alpha \beta}$ correspond to
the flavour asymmetries stored in lepton doublets, while the off-diagonal entries
encode the quantum correlations between the different flavour asymmetries.
From the definition~(\ref{eq:Deltal_def}) and from Eqs.~(\ref{eq:rotation_rho})
and~(\ref{eq:rotation_rhobar}) one can see that $\Delta_\ell$ transforms under
a rotation in flavour space $\ell \to U \ell$ as
\beq
  \Delta_\ell\, \to\, U^* \Delta_\ell\, U^T .
\eeq

One can show that $\Delta n_{\alpha\beta}$ is the zeroth component
of the current $J_{\alpha\beta}^\mu =\, :\! \bar{\ell}_\alpha\gamma^\mu \ell_\beta\! :$
(or more precisely of its average $\langle J_{\alpha\beta}^\mu \rangle$):
\begin{equation}
 \Delta n_{\alpha\beta}\, =\, \langle J_{\alpha\beta}^0 \rangle\,
  =\, \langle\, :\! \ell^\dagger_\alpha \ell_\beta\! :\, \rangle\, .
\end{equation}
We will use this fact to derive an evolution equation for $\Delta n_{\alpha\beta}$.
Noticing that
\begin{align}
 \langle \partial_\mu J_{\alpha\beta}^\mu \rangle\, =\, &-\mathrm{tr} \left[ (i\! \overrightarrow{\slashed{\partial}}_{\! x}
 + i\! \overleftarrow{\slashed{\partial}}_{\! y})G_{\beta\alpha}^>(x,y) \right]_{y\, =\, x}\ ,
\label{eq:current_divergence}
\end{align}
where the trace is taken over spinorial and $SU(2)_L$ indices, and using the Schwinger-Dyson
equations~(\ref{eq:dSDE1}) and~(\ref{eq:dSDE2}) to express the right-hand side of
Eq.~(\ref{eq:current_divergence}) in terms of the self-energy functions $\Sigma^>$ and $\Sigma^<$,
one obtains
\begin{align}
 \langle \partial_\mu J_{\alpha\beta}^\mu \rangle\, =\, &-\int_{\mathcal{C}} d^4w\ \mathrm{tr}\left[\Sigma_{\beta\gamma}^>(x,w)G_{\gamma\alpha}^t(w,x)-\Sigma_{\beta\gamma}^{\bar{t}}(x,w)G_{\gamma\alpha}^>(w,x)\right.\nonumber\\
&\left.-G_{\beta\gamma}^>(x,w)\Sigma_{\gamma\alpha}^t(w,x)+G_{\beta\gamma}^{\bar{t}}(x,w)\Sigma_{\gamma\alpha}^>(w,x)\right]  \nonumber \\
    =\, &-\int d^3w\int_0^tdt_w\ \mathrm{tr}\left[\Sigma_{\beta\gamma}^>(x,w)G_{\gamma\alpha}^<(w,x)-\Sigma_{\beta\gamma}^<(x,w)G_{\gamma\alpha}^>(w,x)\right. \nonumber\\
&\left.-G_{\beta\gamma}^>(x,w)\Sigma_{\gamma\alpha}^<(w,x)+G_{\beta\gamma}^<(x,w)\Sigma_{\gamma\alpha}^>(w,x)\right].
\end{align}
Since we consider a homogeneous and isotropic medium, the divergence of the current
reduces to $d\Delta n_{\alpha\beta} / dt$.
Finally, we incorporate the expansion of the universe by making the following
replacement in the above equation:
\begin{align}
  \langle \partial_\mu J_{\alpha\beta}^\mu \rangle\ \
    \rightarrow\ \ \frac{d\Delta n_{\alpha\beta}}{dt}+3H\Delta n_{\alpha\beta}\,
    =\, sHz\frac{d(\Delta_\ell )_{\alpha\beta}}{dz}\ .
\end{align}
We thus obtain the quantum Boltzmann equation for the density matrix $(\Delta_\ell )_{\alpha\beta}$:
\begin{align}
 sHz\frac{d(\Delta_\ell )_{\alpha\beta}}{dz}\, =\, &-\int d^3w\int_0^tdt_w\ \mathrm{tr}\left[\Sigma_{\beta\gamma}^>(x,w)G_{\gamma\alpha}^<(w,x)-\Sigma_{\beta\gamma}^<(x,w)G_{\gamma\alpha}^>(w,x)\right. \nonumber\\
&\left.-G_{\beta\gamma}^>(x,w)\Sigma_{\gamma\alpha}^<(w,x)+G_{\beta\gamma}^<(x,w)\Sigma_{\gamma\alpha}^>(w,x)\right].
\label{eq:QBE_Deltal}
\end{align}
This equation is both quantum and flavour-covariant, but we are only interested
in flavour effects. We shall therefore take the classical limit
by keeping only the contributions to $\Sigma$ involving an on-shell intermediate part.
This procedure will provide us with a flavour-covariant Boltzmann equation
of the form\footnote{Strictly speaking,
the Boltzmann equation for the $3 \times 3$ density matrix $(\Delta_\ell )_{\alpha\beta}$
is valid only in the regime
where the quantum correlations between the different lepton flavours are not affected
by the charged lepton Yukawa interactions, i.e. at $T > 10^{12} \GeV$.
When this condition is not satisfied, one must either add a term accounting for the effects
of the Yukawa-induced processes on the right-hand side of Eq.~(\ref{eq:covariant_BE}),
or impose that the appropriate entries of the density matrix $\Delta_\ell$ vanish
(see discussion later in this section and Section~\ref{sec:BE} for the relevant Boltzmann equations).
Furthermore, the effect of spectator processes such as sphalerons and Yukawa interactions,
which impose relations among the various particle asymmetries in the plasma, is not included
at this stage (it will be discussed in Section~\ref{sec:spectator}).}~(\ref{eq:covariant_BE}).

\subsubsection{Washout terms: decays and inverse decays}    %

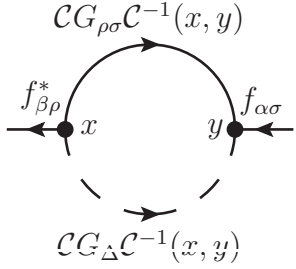
\begin{figure}
\begin{center}
\fcolorbox{white}{white}{
  \begin{picture}(144,112) (42,-171)
    \SetWidth{1.0}
    \SetColor{Black}
    \Arc[arrow,arrowpos=0.5,arrowlength=5,arrowwidth=2,arrowinset=0.2,clock](96,-128)(32,-180,-360)
    \Arc[dash,dashsize=10,arrow,arrowpos=0.5,arrowlength=5,arrowwidth=2,arrowinset=0.2](96,-128)(32,-180,0)
    \Vertex(128,-128){3.162}
    \Vertex(64,-128){3.162}
    \Line[arrow,arrowpos=0.5,arrowlength=5,arrowwidth=2,arrowinset=0.2](150,-128)(128,-128)
    \Line[arrow,arrowpos=0.5,arrowlength=5,arrowwidth=2,arrowinset=0.2](64,-128)(42,-128)
    \Text(96,-93)[cb]{\normalsize{\Black{$\mathcal{C}G_{\rho\sigma}\mathcal{C}^{-1}(x,y)$}}}
    \Text(96,-179)[cb]{\normalsize{\Black{$\mathcal{C}G_{\Delta}\mathcal{C}^{-1}(x,y)$}}}
    \Text(47,-123)[lb]{\normalsize{\Black{$f_{\beta\rho}^*$}}}
    \Text(70,-130)[lb]{\normalsize{\Black{$x$}}}
    \Text(118,-132)[lb]{\normalsize{\Black{$y$}}}
    \Text(130,-123)[lb]{\normalsize{\Black{$f_{\alpha\sigma}$}}}
\end{picture}
}
\end{center}\caption{1-loop contribution to the lepton doublet self-energy $\Sigma_{\beta\alpha}(x,y)$.}
\label{fig:Sigma_1loop}
\end{figure}

Let us first compute the flavour-covariant washout term $W^D_{\alpha \beta}$
associated with triplet/anti\-triplet decays and inverse decays. In the CTP formalism, this term arises
from the 1-loop contribution to the lepton doublet self-energy shown in Fig.~\ref{fig:Sigma_1loop}.
For the first term of the integrand on the right-hand side of Eq.~(\ref{eq:QBE_Deltal}), this gives
\begin{align}
  \mathrm{tr}\! \left[\Sigma_{\beta\gamma}^>(x,w)G_{\gamma\alpha}^<(w,x) \right]\, =\ \,
    & (-if_{\beta\rho}^*)(-if_{\gamma\sigma})\, \mathrm{tr}
    \Big[ iG_\Delta^<(w,x)(-i) G_{\sigma\rho_R}^{<}(w,x)iG_{\gamma\alpha}^<(w,x) \Big]  \nonumber \\
  =\ \, & 3 \int\frac{d^3p}{(2\pi)^32\omega_{\vec{p}}}\int\frac{d^3k}{(2\pi)^32\omega_{\vec{k}}}
    \int\frac{d^3l}{(2\pi)^32\omega_{\vec{\ell}}}\ f_{\beta\rho}^*f_{\gamma\sigma}\, 2 (k.l)\,  \nonumber \\
  & \times \left\lbrace \rho_\Delta(\vec{p})\, e^{-ip \cdot (w-x)}
    + [ 1+\bar{\rho}_\Delta(\vec{p}) ]\, e^{ip \cdot (w-x)} \right\rbrace  \nonumber \\
  & \times \left\lbrace \rho_{\rho\sigma}(\vec{k})\, e^{-ik \cdot (w-x)}
    + [ \delta_{\rho\sigma}-\bar{\rho}_{\rho\sigma}(\vec{k}) ]\, e^{ik \cdot (w-x)}\right\rbrace  \nonumber \\
  & \times \left\lbrace \rho_{\alpha\gamma}(\vec{l})\, e^{-il \cdot (w-x)}
    + [ \delta_{\alpha\gamma}-\bar{\rho}_{\alpha\gamma}(\vec{l}) ]\, e^{il \cdot (w-x)}\right\rbrace ,
\end{align}
where the factor 3 comes from the trace over $SU(2)_L$ indices.
The integration over the spatial coordinates of $w$
gives a momentum-conserving delta function.
In the integral over $t_w$, we take the classical limit by making the usual assumption that $t$ is
much smaller than the relaxation time of particle distributions, which can therefore be factorized
out of the integral, but much larger than the typical duration of a collision, so that the time integral
can be extended to infinity. This amounts to replacing the integral of oscillating exponentials
by energy-conserving delta functions, plus terms proportional to the principal values of
$1/(\omega_{\vec p}\pm\omega_{\vec k}\pm\omega_{\vec l})$, where $p$, $k$ and $l$
are the momenta of the scalar triplet and of the two leptons, respectively.
The latter terms, however, can be neglected because they arise
at second order in the CP asymmetry. Indeed, all terms involving a principal value are
proportional to $\eta_{\rho\sigma}(\vec k)+\bar{\eta}_{\rho\sigma}(\vec k )$, where
$\eta_{\rho\sigma}(\vec k)$ (resp. $\bar \eta_{\rho\sigma}(\vec k)$) parametrizes the departure
of the phase-space density $\rho_{\rho\sigma}(\vec k)$ (resp. $\bar \rho_{\rho\sigma}(\vec k)$)
from its equilibrium value:
\begin{align}
  \rho_{\rho\sigma}(\vec k)\, =\, \rho_\ell^\text{eq}(\vec k )
    \left[ \delta_{\rho\sigma}+\eta_{\rho\sigma}(\vec k) \right] , \quad
  \bar \rho_{\rho\sigma}(\vec k)\, =\, \rho_\ell^\text{eq}(\vec k )
    \left[ \delta_{\rho\sigma}+ \bar \eta_{\rho\sigma}(\vec k) \right] .
\end{align}
Since the unbalance between the lepton and antilepton densities is generated
by the asymmetries in triplet decays, which are small numbers of order $\epsilon$,
while the lepton and antilepton populations are maintained close to equilibrium
by fast electroweak interactions, one has\footnote{Eqs.~(\ref{eq:eta_minus_etabar})
and~(\ref{eq:eta_plus_etabar}) generalize the relations between equilibrium number densities
$n_{\ell_\alpha} - n_{\bar \ell_\alpha} \simeq \frac{1}{3}\, \mu_{\ell_\alpha} T^2$ and
$n_{\ell_\alpha} + n_{\bar \ell_\alpha} = 2\, n^\text{eq}_\ell (\mu_\ell = 0)
\left[ 1 + {\cal O}\, (\mu_{\ell_\alpha}/T)^2 \right]$,
where $\mu_{\ell_\alpha}$ is the chemical potential of the
lepton flavour $\alpha$ and $\mu_{\ell_\alpha} / T = {\cal O}\, (\epsilon)$.}
\begin{align}
  \eta_{\rho\sigma}(\vec k)-\bar{\eta}_{\sigma\rho}(\vec k)\, & =\, \mathcal{O} (\epsilon)\, ,
\label{eq:eta_minus_etabar}  \\
  \eta_{\rho\sigma}(\vec k)+\bar{\eta}_{\sigma\rho}(\vec{k})\, & =\, \mathcal{O} (\epsilon^2)\, .
\label{eq:eta_plus_etabar} 
\end{align}
Therefore, the terms proportional to $\eta_{\rho\sigma}(\vec k)+\bar{\eta}_{\rho\sigma}(\vec k )$
on the right-hand side of the Boltzmann equation~(\ref{eq:QBE_Deltal}) are of order $\epsilon^2$
and can safely be neglected.
By doing so one keeps only the terms that are kinematically allowed for on-shell particles.
Dropping Bose enhancement and Pauli blocking factors, we obtain
\begin{align}
 -&\int d^3w\int_0^\infty dt_w\ \mathrm{tr}\! \left[\Sigma_{\beta\gamma}^>(x,w)G_{\gamma\alpha}^<(w,x) \right]\nonumber\\
=&-\int\frac{d^3p}{(2\pi)^32\omega_{\vec{p}}}\int\frac{d^3k}{(2\pi)^32\omega_{\vec{k}}}\int\frac{d^3l}{(2\pi)^32\omega_{\vec{\ell}}}\ 3 f_{\beta\rho}^*f_{\gamma\sigma}(k.l)\, (2\pi)^4\delta^{(4)}(p-k-l)\nonumber\\
&\times\left\lbrace \rho_{\rho\sigma}(\vec{k})\rho_{\alpha\gamma}(\vec{\ell})+\rho_\Delta(\vec p)\delta_{\rho\sigma}\delta_{\alpha\gamma}\right\rbrace.
\end{align}
Proceeding in the same way to compute the other contributions from Fig.~\ref{fig:Sigma_1loop}
to the right-hand side of the Boltwmann equation~(\ref{eq:QBE_Deltal}), and introducing
the space-time density of triplet and antitriplet decays:
\begin{align}
 \gamma_D\, =\, &\int \frac{d^3p}{(2\pi)^32\omega_{\vec{p}}}\int\frac{d^3k}{(2\pi)^32\omega_{\vec{k}}}\int\frac{d^3l}{(2\pi)^32\omega_{\vec{\ell}}}\ 3 \left(\lambda_\ell^2+\lambda_H^2\right)(k.l)\nonumber\\
 &\times(2\pi)^4\delta^{(4)}(p-k-l)\left\lbrace \rho_\Delta^{\mathrm{eq}}(\vec p)+\bar{\rho}_\Delta^{\mathrm{eq}}(\vec{p})\right\rbrace,
\end{align}
we obtain the washout term $\mathcal{W}^D_{\alpha\beta}$ associated with decays and inverse decays:
\beq
 \mathcal{W}^D_{\alpha\beta}\, =\, \frac{2B_\ell}{\lambda_\ell^2}\left[(ff^\dagger)_{\alpha\beta}\frac{\Delta_\Delta}{\Sigma_\Delta^{\mathrm{eq}}}\
 +\frac{1}{4(Y_\ell^{\mathrm{eq}})^2}\left(YfY^Tf^\dagger+fY^Tf^\dagger Y - Y \leftrightarrow \bar Y \right)_{\alpha\beta}\right]\! \gamma_D\, ,
\eeq
where we remind the reader that $\Delta_\Delta \equiv (n_\Delta-n_{\bar \Delta})/s$
and $\Sigma_\Delta \equiv (n_\Delta+n_{\bar \Delta})/s$.
One can linearize this expression using again the fact that flavour-blind gauge interactions
keep the lepton densities close to their equilibrium values:
\begin{align}
  Y_{\alpha\beta}-\bar{Y}_{\alpha\beta}\, & =\, (\Delta_\ell )_{\alpha\beta}\, ,  \nonumber\\
  Y_{\alpha\beta}+\bar{Y}_{\alpha\beta}\, & =\, 2Y_\ell^{\mathrm{eq}}
    \left[ \delta_{\alpha\beta}+\mathcal{O}(\epsilon^2) \right] ,
\end{align}
which finally gives
\begin{align}
  \mathcal{W}^D_{\alpha\beta}\, =\, \frac{2B_\ell}{\lambda_\ell^2}
    \left[ (ff^\dagger)_{\alpha\beta}\frac{\Delta_\Delta}{\Sigma_\Delta^{\text{eq}}} + \frac{1}{4Y_\ell^{\text{eq}}}
    \left( 2f\Delta_\ell^Tf^\dagger + ff^\dagger\Delta_\ell + \Delta_\ell ff^\dagger \right)_{\alpha\beta}\right]\! \gamma_D\, .
\label{eq:W_D}
\end{align}
Note that the washout term~(\ref{eq:W_D}) contains a piece proportional to $\Delta_\Delta$,
which is due to decays.
This term appears on the right-hand side of the Boltzmann equation
for the $3 \times 3$ density matrix $(\Delta_\ell )_{\alpha\beta}$, Eq.~(\ref{eq:covariant_BE}).
One can easily check that it transforms as $\mathcal{W}^D \to U^* \mathcal{W}^D U^T$
under flavour rotations $\ell \to U \ell$, as required by flavour covariance.
Below $T = 10^{12} \GeV$, however, the tau Yukawa coupling is in equilibrium and drives the
$(e,\tau)$, $(\mu,\tau)$, $(\tau,e)$ and $(\tau, \mu)$ entries of the density matrix to zero.
The relevant dynamical variables in this regime
(or more precisely in the temperature range $10^{9} \GeV < T < 10^{12} \GeV$, before the muon
Yukawa coupling enters equilibrium) are $\Delta_{\ell_\tau}$, the asymmetry stored in $\ell_\tau$,
and a $2 \times 2$ matrix  $\Delta_\ell^0$ describing the flavour asymmetries stored
in the lepton doublets $\ell_e, \ell_\mu$ and their quantum correlations.
The corresponding washout terms $\tilde{\mathcal{W}}^D_{\alpha\beta}$ (where $\alpha$
and $\beta$ label any two orthogonal directions in the ($\ell_e$, $\ell_\mu$) flavour subspace)
and $\tilde{\mathcal{W}}^D_\tau$ are simply obtained by setting
$(\Delta_\ell)_{\alpha\tau} = (\Delta^*_\ell)_{\tau\alpha} = \Delta_{\ell_\tau} \delta_{\alpha\tau}$
in Eq.~(\ref{eq:W_D}), yielding
\begin{align}
  \tilde{\mathcal{W}}^D_{\alpha\beta}\, =\, & \frac{2B_\ell}{\lambda_\ell^2} \left[(ff^\dagger)_{\alpha\beta}
    \frac{\Delta_\Delta}{\Sigma_\Delta^{\text{eq}}} \right.  \nonumber \\
    & \left. +\, \frac{1}{4Y_\ell^{\text{eq}}} \left( 2f (\Delta^0_\ell)^Tf^\dagger+ff^\dagger\Delta^0_\ell
    + \Delta^0_\ell ff^\dagger \right)_{\alpha\beta}
    + \frac{1}{2Y_\ell^{\text{eq}}}\, f_{\alpha\tau}f_{\beta\tau}^*\Delta_{\ell_\tau} \right]\! \gamma_D\, ,
\label{eq:W_tilde_D}
\end{align}
and
\begin{align}
  \tilde{\mathcal{W}}^D_\tau\, =\, \frac{2B_\ell}{\lambda_\ell^2} \left[(ff^\dagger)_{\tau\tau}
    \frac{\Delta_\Delta}{\Sigma_\Delta^{\text{eq}}}
    + \frac{1}{2Y_\ell^{\text{eq}}} \left( (f (\Delta^0_\ell)^Tf^\dagger)_{\tau\tau}
    + ( (ff^\dagger)_{\tau\tau} + |f_{\tau\tau}|^2 )\, \Delta_{\ell_\tau} \right) \right]\! \gamma_D\, .
\end{align}
The resulting Boltzmann equation for $(\Delta^0_\ell)_{\alpha \beta}$
is covariant under flavour rotations in the ($\ell_e$, $\ell_\mu$) subspace.
Finally, below $T = 10^9 \GeV$, the muon Yukawa coupling enters equilibrium
and drives the $(e, \mu)$ and $(\mu, e)$ entries of the density matrix to zero. The Boltzmann
equation~(\ref{eq:covariant_BE}) then reduces to three equations for the
flavour asymmetries $\Delta_{\ell_\alpha} \equiv (\Delta_\ell)_{\alpha \alpha}$
($\alpha = e, \mu, \tau$), with a washout term $\mathcal{W}^D_\alpha$ given by
\begin{align}
  \mathcal{W}^D_\alpha\, =\, \frac{2B_\ell}{\lambda_\ell^2}\, \sum_\beta\, |f_{\alpha\beta}|^2
    \left[ \frac{\Delta_\Delta}{\Sigma_\Delta^{\text{eq}}} + \frac{\Delta_{\ell_\alpha} + \Delta_{\ell_\beta}}{2Y_\ell^{\text{eq}}}
    \right]\! \gamma_D\, .
\label{eq:W_D_alpha}
\end{align}
%

\subsubsection{Other washout terms: $2 \to 2$ scatterings}     %

The other washout terms in the Boltzmann equation~(\ref{eq:covariant_BE}) correspond
to $2 \to 2$ scatterings and arise from 2-loop contributions to the lepton doublet self-energy $\Sigma_{\beta \alpha}$.
The term $\mathcal{W}^{ \ell H}_{\alpha\beta}$ accounts for the washout of the flavour
asymmetries by the $\Delta L =2$ scatterings $\ell_\gamma \ell_\delta \leftrightarrow \bar H \bar H$
and $\ell_\gamma H \leftrightarrow \bar \ell_\delta \bar H$ and is given by
\begin{align}
 \mathcal{W}^{ \ell H}_{\alpha\beta}\, =\ & 2\left\lbrace\frac{1}{\lambda^2_\ell}\left[\frac{\left(2f\Delta_\ell^Tf^\dagger+ff^\dagger\Delta_\ell+\Delta_\ell ff^\dagger\right)_{\alpha\beta}}{4Y_\ell^\text{eq}}+\frac{\Delta_H}{Y_H^{\text{eq}}}(ff^\dagger)_{\alpha\beta}\right]\right.\! \gamma_{\ell H}^\Delta\nonumber\\
&+\frac{1}{\mbox{Re}\left[\text{tr}(f\kappa^\dagger)\right]}\left[\frac{\left(2f\Delta_\ell^T\kappa^\dagger+f\kappa^\dagger\Delta_\ell+\Delta_\ell f\kappa^\dagger\right)_{\alpha\beta}}{4Y_\ell^\text{eq}}+\frac{\Delta_H}{Y_H^{\text{eq}}}(f\kappa^\dagger)_{\alpha\beta}\right]\! \gamma_{\ell H}^\mathcal{I}\nonumber\\
&+\frac{1}{\mbox{Re}\left[\text{tr}(f\kappa^\dagger)\right]}\left[\frac{\left(2\kappa\Delta_\ell^T f^\dagger+\kappa f^\dagger\Delta_\ell+\Delta_\ell\kappa f^\dagger\right)_{\alpha\beta}}{4Y_\ell^\text{eq}}+\frac{\Delta_H}{Y_H^{\text{eq}}}(\kappa f^\dagger)_{\alpha\beta}\right]\! \gamma_{\ell H}^\mathcal{I}\nonumber\\
&\left.+\frac{1}{\lambda^2_\kappa}\left[\frac{\left(2\kappa\Delta_\ell^T\kappa^\dagger+\kappa\kappa^\dagger\Delta_\ell+\Delta_\ell\kappa\kappa^\dagger\right)_{\alpha\beta}}{4Y_\ell^\text{eq}}+\frac{\Delta_H}{Y_H^{\text{eq}}}(\kappa\kappa^\dagger)_{\alpha\beta}\right]\! \gamma_{\ell H}^\mathcal{H}\right\rbrace,
\label{eq:W_lh}
\end{align}
where $\Delta_H \equiv (n_H-n_{\bar H})/s$ is the asymmetry stored in the Higgs doublet,
$\lambda_\kappa \equiv \sqrt{\text{tr}(\kappa\kappa^\dagger)}$
and $\gamma_{\ell H}^\Delta$, $\gamma_{\ell H}^\mathcal{H}$ and $\gamma_{\ell H}^\mathcal{I}$
are the contributions (summed over lepton flavours) of different self-energy diagrams
to the space-time density of $\Delta L =2$ scatterings $\gamma_{\ell H}$. Namely, one has
$\gamma_{\ell H}=\gamma_{\ell H}^\Delta+2\gamma_{\ell H}^\mathcal{I}+\gamma_{\ell H}^\mathcal{H}$,
in which $\gamma_{\ell H}^\Delta$ is the contribution of scalar triplet exchange,
$\gamma_{\ell H}^\mathcal{H}$ is the contribution of the $D=5$ operator~(\ref{eq:Weinberg_operator})
responsible for $m_\mathcal{H}$, and $2 \gamma_{\ell H}^\mathcal{I}$ is the interference term.
Explicit expressions for the reduced cross-sections that are needed to compute numerically
these reaction densities\footnote{For some reactions
the contribution of real intermediate states must be subtracted in order
to avoid double counting in the Boltzmann equations. Although this subtraction procedure
is embedded in the CTP formalism iself~\cite{Garny:2009,Beneke:2010wd},
in practice we first take the classical limit in the quantum Boltzmann equation
in order to obtain Eqs.~(\ref{eq:W_lh}), (\ref{eq:W_4l}) and~(\ref{eq:W_ellDelta}), and then
insert the subtracted reaction densities computed in Appendix~\ref{app:reactions} into these expressions.}
(as well as the ones appearing in Eqs.~(\ref{eq:W_4l}) and~(\ref{eq:W_ellDelta}) below)
can be found in Appendix~\ref{app:reactions}.

The washout term due to the 2 lepton--2 lepton scatterings $\ell_\gamma\ell_\delta\leftrightarrow\ell_\rho\ell_\sigma$
and $\ell_\gamma\bar \ell_\rho\leftrightarrow\bar \ell_\delta\ell_\sigma$ reads
\begin{align}
\mathcal{W}^{4\ell}_{\alpha\beta}\, =\, \frac{2}{\lambda^4_\ell}&\left[ \lambda^2_\ell\, \frac{\left(2f\Delta_\ell^Tf^\dagger+ff^\dagger\Delta_\ell+\Delta_\ell ff^\dagger\right)_{\alpha\beta}}{4Y_\ell^\text{eq}}
- \frac{\text{tr}(\Delta_\ell ff^\dagger)}{Y_\ell^{\text{eq}}}(ff^\dagger)_{\alpha\beta}\right]\! \gamma_{4\ell}\, ,
\label{eq:W_4l}
\end{align}
whereas the washout term associated with the lepton-triplet scatterings $\ell_\gamma\Delta\leftrightarrow\ell_\delta\Delta$,
$\ell_\gamma\bar \Delta\leftrightarrow\ell_\delta\bar \Delta$ and
$\ell_\gamma\bar \ell_\delta\leftrightarrow\Delta\bar \Delta$ is
\begin{align}
\mathcal{W}^{\ell\Delta}_{\alpha\beta}\, =\, \frac{1}{\text{tr}(ff^\dagger ff^\dagger)}\left[\frac{1}{2Y_\ell^\text{eq}}\left(ff^\dagger ff^\dagger\Delta_\ell-2ff^\dagger\Delta_\ell ff^\dagger+\Delta_\ell ff^\dagger ff^\dagger\right)_{\alpha\beta}\right]\! \gamma_{\ell\Delta}\, .
\label{eq:W_ellDelta}
\end{align}
This completes the derivation of the washout terms appearing on the right-hand side of
Eq.~(\ref{eq:covariant_BE}). The expressions of $\mathcal{W}^{ \ell H}_{\alpha\beta}$,
$\mathcal{W}^{4\ell}_{\alpha\beta}$ and $\mathcal{W}^{\ell\Delta}_{\alpha\beta}$
relevant in the temperature ranges $10^9 \GeV < T < 10^{12} \GeV$ and $T < 10^9 \GeV$
can be obtained from Eqs.~(\ref{eq:W_lh}), (\ref{eq:W_4l}) and~(\ref{eq:W_ellDelta})
following the same procedure as the one used for $\mathcal{W}^D_{\alpha\beta}$,
see the discussion above Eqs.~(\ref{eq:W_tilde_D}) and~(\ref{eq:W_D_alpha}).
Note that $\mathcal{W}^{4\ell}_{\alpha\beta}$ and $\mathcal{W}^{\ell\Delta}_{\alpha\beta}$
vanish in the single flavour approximation where $\alpha$ and $\beta$ take a single value;
this is consistent with the fact that the 2 lepton--2 lepton and lepton-triplet scatterings
affect leptogenesis only when lepton flavour effects are taken into account.

\subsubsection{Source term}     %

%
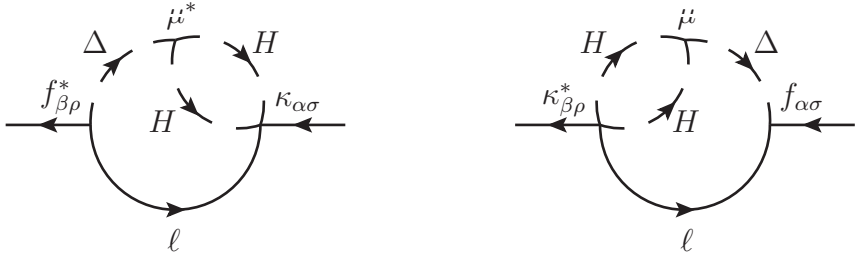
\begin{figure}
\begin{center}
\fcolorbox{white}{white}{
  \begin{picture}(322,73) (95,-78)
    \SetWidth{1.0}
    \SetColor{Black}
    \Line[arrow,arrowpos=0.5,arrowlength=5,arrowwidth=2,arrowinset=0.2](128,-43)(96,-43)
    \Line[arrow,arrowpos=0.5,arrowlength=5,arrowwidth=2,arrowinset=0.2](224,-43)(192,-43)
    \Arc[arrow,arrowpos=0.5,arrowlength=5,arrowwidth=2,arrowinset=0.2](160,-43)(32,-180,0)
    \Arc[dash,dashsize=10,arrow,arrowpos=0.5,arrowlength=5,arrowwidth=2,arrowinset=0.2,clock](158.885,-41.885)(30.905,-177.932,-272.068)
    \Arc[dash,dashsize=10,arrow,arrowpos=0.5,arrowlength=5,arrowwidth=2,arrowinset=0.2,flip](168,-35)(25.298,-18.435,108.435)
    \Arc[dash,dashsize=10,arrow,arrowpos=0.5,arrowlength=5,arrowwidth=2,arrowinset=0.2,flip,clock](184,-19)(25.298,-71.565,-198.435)
    \Line[arrow,arrowpos=0.5,arrowlength=5,arrowwidth=2,arrowinset=0.2](320,-43)(288,-43)
    \Line[arrow,arrowpos=0.5,arrowlength=5,arrowwidth=2,arrowinset=0.2](416,-43)(384,-43)
    \Arc[arrow,arrowpos=0.5,arrowlength=5,arrowwidth=2,arrowinset=0.2](352,-43)(32,-180,0)
    \Arc[dash,dashsize=10,arrow,arrowpos=0.5,arrowlength=5,arrowwidth=2,arrowinset=0.2,flip](354,-41)(30.067,-3.814,93.814)
    \Arc[dash,dashsize=10,arrow,arrowpos=0.5,arrowlength=5,arrowwidth=2,arrowinset=0.2](328,-19)(25.298,-108.435,18.435)
    \Arc[dash,dashsize=10,arrow,arrowpos=0.5,arrowlength=5,arrowwidth=2,arrowinset=0.2,clock](344,-35)(25.298,-161.565,-288.435)
    \Text(125,-39)[rb]{\normalsize{\Black{$f_{\beta\rho}^*$}}}
    \Text(198,-38)[lb]{\normalsize{\Black{$\kappa_{\alpha\sigma}$}}}
    \Text(316,-40)[rb]{\normalsize{\Black{$\kappa_{\beta\rho}^*$}}}
    \Text(389,-38)[lb]{\normalsize{\Black{$f_{\alpha\sigma}$}}}
    \Text(163,-7)[cb]{\normalsize{\Black{$\mu^*$}}}
    \Text(354,-7)[cb]{\normalsize{\Black{$\mu$}}}
    \Text(160,-91)[cb]{\normalsize{\Black{$\ell$}}}
    \Text(135,-16)[rb]{\normalsize{\Black{$\Delta$}}}
    \Text(200,-16)[rb]{\normalsize{\Black{$H$}}}
    \Text(161,-46)[rb]{\normalsize{\Black{$H$}}}
    \Text(354,-91)[cb]{\normalsize{\Black{$\ell$}}}
    \Text(389,-16)[rb]{\normalsize{\Black{$\Delta$}}}
    \Text(324,-16)[rb]{\normalsize{\Black{$H$}}}
    \Text(359,-46)[rb]{\normalsize{\Black{$H$}}}
  \end{picture}
}
\end{center}\caption{2-loop contributions to the lepton doublet self-energy $\Sigma_{\beta\alpha}$
giving rise to the CP asymmetry $\mathcal{E}_{\alpha\beta}$.}
\label{fig:CP_asymmetry}
\end{figure}

Finally, let us derive the source term of the Boltzmann equation~(\ref{eq:covariant_BE}).
In the CTP formalism, the 2-loop self-energy diagrams shown in Fig.~\ref{fig:CP_asymmetry},
when inserted in the right-hand side of Eq.~(\ref{eq:QBE_Deltal}),
give rise to the flavour-covariant source term
\begin{align}
  \mathcal{S}_{\alpha\beta}\, = \left(\frac{\Sigma_\Delta}{\Sigma_\Delta^{\mathrm{eq}}}-1\right)\!
    \gamma_D\, \mathcal{E}_{\alpha\beta}\, ,
\label{eq:source_term}
\end{align}
where $\mathcal{E}_{\alpha\beta}$ is a Hermitian matrix transforming as
$\mathcal{E} \to U^* \mathcal{E} U^T$ under a flavour rotation $\ell \to U \ell$.
The explicit form of $\mathcal{E}_{\alpha\beta}$ can be obtained without going through
the full CTP computation by noticing that, in the temperature regime $T < 10^9 \GeV$
where the quantum correlations among lepton flavours are destroyed by fast interactions
induced by the muon and tau Yukawa couplings, the source term for the flavour asymmetry
$\Delta_{\ell_\alpha} \equiv (\Delta_\ell)_{\alpha \alpha}$ ($\alpha = e, \mu, \tau$) reads
\begin{align}
 \mathcal{S}_{\alpha\alpha}\, =\, \left(\frac{\Sigma_\Delta}{\Sigma_\Delta^{\mathrm{eq}}}-1\right)\!
   \gamma_D \sum_\gamma \epsilon_{\alpha \gamma}  \, ,
\label{eq:S_alphaalpha}
\end{align}
where $\epsilon_{\alpha \gamma}$, not to be confused with $\mathcal{E}_{\alpha\gamma}$,
is the CP asymmetry in the decay $\Delta \to \bar \ell_\alpha \bar \ell_\gamma$ defined
in Eq.~(\ref{eq:epsilon_alphabeta_def}). Comparing Eq.~(\ref{eq:source_term})
with Eq.~(\ref{eq:S_alphaalpha}) and using Eq.~(\ref{eq:epsilon_alphabeta}), one obtains
the diagonal entries of $\mathcal{E}$:
\begin{align}
  \mathcal{E}_{\alpha\alpha}\, =\, \sum_{\gamma}\epsilon_{\alpha\gamma}\, =\,
    \frac{1}{4\pi}\frac{M_\Delta}{v^2}\sqrt{B_\ell B_H}\
    \frac{\mbox{Im}\left[ (m_\mathcal{H}\, m_\Delta^\dagger)_{\alpha\alpha} \right]}{\bar{m}_\Delta}\ .
\label{eq:E_alphaalpha}
\end{align}
It is then easy to derive the full flavour-covariant CP-asymmetry matrix $\mathcal{E}$ by reading
its dependence on the couplings $f_{\alpha \beta}$ and $\kappa_{\alpha \beta}$ from
Fig.~\ref{fig:CP_asymmetry}, and by using its hermiticity and covariance properties together
with Eq.~(\ref{eq:E_alphaalpha}):
\begin{align}
  \mathcal{E}_{\alpha\beta}\, =\, \frac{1}{8\pi i}\frac{M_\Delta}{v^2}\sqrt{B_\ell B_H}\
    \frac{(m_\mathcal{H\,} m_\Delta^{\dagger} - m_\Delta m_\mathcal{H}^{\dagger})_{\alpha\beta}}{\bar{m}_\Delta}\, .
\label{eq:E_alphabeta}
\end{align}
The first equality in Eq.~(\ref{eq:E_alphaalpha}) implies that the trace of this matrix is equal to
the total CP asymmetry in triplet decays:
\beq
 \mathrm{tr}\, \mathcal{E}\, =\, \sum_{\alpha, \beta} \epsilon_{\alpha \beta}\, =\, \epsilon_\Delta\, .
\eeq
This completes the derivation of the right-hand side of the flavour-covariant Boltzmann
equation~(\ref{eq:covariant_BE}).

\subsection{Full set of Boltzmann equations}  %

Finally, Eq.~(\ref{eq:covariant_BE}) must be supplemented with the Boltzmann
equations for $\Sigma_\Delta$ and $\Delta_\Delta$ (an equation for $\Delta_H$ is not
needed since, as we are going to see in Section~\ref{sec:spectator}, $\Delta_H$ can be expressed
as a function of the other asymmetries):
\begin{align}
  sHz\frac{d\Sigma_\Delta}{dz}\, & =\, -\left(\frac{\Sigma_\Delta}{\Sigma_\Delta^\text{eq}}-1\right)\! \gamma_D
    -2\left(\! \left( \frac{\Sigma_\Delta}{\Sigma_\Delta^\text{eq}} \right)^{\! 2} - 1\right)\! \gamma_A\, ,
\label{eq:BE_Sigma_Delta}  \\
  sHz\frac{d (\Delta_\ell)_{\alpha\beta}}{dz}\, & =\,
    \left(\frac{\Sigma_\Delta}{\Sigma_\Delta^\text{eq}}-1\right)\! \gamma_D\, \mathcal{E}_{\alpha\beta}
    - \mathcal{W}^D_{\alpha\beta} - \mathcal{W}^{\ell H}_{\alpha\beta}
    - \mathcal{W}^{4\ell}_{\alpha\beta} - \mathcal{W}^{\ell\Delta}_{\alpha\beta}\, ,
\label{eq:BE_Delta_l}  \\
  sHz\frac{d\Delta_\Delta}{dz}\, & =\, -\frac{1}{2}\left(\text{tr}(\mathcal{W}^D)-W^D_H\right) ,
\label{eq:BE_Delta_Delta}
\end{align}
where the first and second terms in Eq.~(\ref{eq:BE_Sigma_Delta}) come from (anti)triplet decays
and triplet-antitriplet annihilations, respectively, and $W^D_H$ in Eq.~(\ref{eq:BE_Delta_Delta})
is the washout term due to the inverse decays $H H \to \Delta$ and $\bar H \bar H \to \bar \Delta$:
\begin{align}
  W^D_H\, =\, 2B_H\left(\frac{\Delta_H}{Y_H^\text{eq}}-\frac{\Delta_\Delta}{\Sigma_\Delta^\text{eq}}\right)\! \gamma_D\, .
\label{eq:W^D_H}
\end{align}
Using Eqs.~(\ref{eq:W_D}) and~(\ref{eq:W^D_H}), the Boltzmann equation for $\Delta_\Delta$
can be rewritten as
\begin{align}
  sHz\frac{d\Delta_\Delta}{dz}\, & =\, -\left(\frac{\Delta_\Delta}{\Sigma_\Delta^\text{eq}}
    +B_\ell\frac{\text{tr}(ff^\dagger\Delta_\ell)}{\lambda_\ell^2 Y_\ell^\text{eq}}
    -B_H\frac{\Delta_H}{Y_H^\text{eq}}\right)\! \gamma_D\, .
\end{align}
Eqs.~(\ref{eq:BE_Sigma_Delta}) and~(\ref{eq:BE_Delta_Delta}) take the same form
as in the single flavour case~\cite{Hambye:2005tk}, except for the term
associated with the decays $\Delta \leftrightarrow \bar \ell_\gamma \bar \ell_\delta$,
$\bar \Delta \leftrightarrow \ell_\gamma \ell_\delta$
in the Boltzmann equation for $\Delta_\Delta$, which involves a trace with the density matrix
rather than the total lepton asymmetry.

\section{Spectator processes}          %
\label{sec:spectator}                         %

Various Standard Model (SM) reactions (strong and electroweak sphalerons, Yukawa couplings)
that contribute indirectly to the washout of the lepton flavour asymmetries have to be included
in the computation. Indeed, while these processes cannot create or erase
a lepton asymmetry by themselves (for this reason, they are usually dubbed spectator
processes), they modify the densities of the species on which the washout rates depend, 
thus affecting the final baryon asymmetry.
For instance, $\Delta_{\ell_\tau}$ can be turned into $\Delta_H$ and $\Delta_{\tau_R}$
by $y_\tau$-induced interactions, and into $\Delta_{q_3}$ by electroweak sphalerons
(where, as previously defined, we denote the asymmetry stored in the species $X$
by $\Delta_X \equiv (n_X-n_{\bar X})/s$).

In this paper, we follow the standard approach by assuming that each spectator process
is either negligible or strongly in equilibrium, in which case it imposes a relation among
the chemical potentials (hence among the asymmetries) of the species involved.
The set of relations valid in a given temperature range is then used to express the
asymmetries appearing in the Boltzmann equations in terms of asymmetries that are
conserved by all active SM processes. This was first done in the standard leptogenesis scenario
with heavy Majorana neutrinos in Refs.~\cite{Buchmuller:2001sr,Nardi:2005hs},
and in scalar triplet leptogenesis in Ref.~\cite{Sierra:2014tqa}. In the following,
we generalize the analysis of Ref.~\cite{Sierra:2014tqa} to the density matrix formalism.

The asymmetries appearing in the Boltzmann equations~(\ref{eq:BE_Sigma_Delta})--(\ref{eq:BE_Delta_Delta})
are $\Delta_\Delta$, $\Delta_H$ and the density matrix $(\Delta_\ell)_{\alpha \beta}$
(which reduces to $\Delta_{\ell_\tau}$ and the $2 \times 2$ matrix $(\Delta^0_\ell)_{\alpha \beta}$ 
in the range $10^9 \GeV < T < 10^{12} \GeV$, and to the
flavour asymmetries $\Delta_{\ell_{e,\mu,\tau}}$ below $T = 10^9 \GeV$).
Out of these, only $\Delta_\Delta$ is conserved by all SM interactions,
while $(\Delta_\ell)_{\alpha \beta}$ is affected by electroweak sphalerons
and charged lepton Yukawa couplings. It is therefore convenient to introduce
the following flavour-covariant matrix:
\beq
  \Delta_{\alpha\beta}\, =\, \frac 1 3 \Delta_B\, \delta_{\alpha\beta} - (\Delta_\ell)_{\alpha\beta}\, ,
\eeq
which is conserved by all SM processes except the charged lepton Yukawa interactions\footnote{The
diagonal entries of $\Delta_{\alpha\beta}$ reduce to $\frac 1 3 \Delta_B - \Delta_{\ell_\alpha}$,
which is conserved by electroweak sphalerons. Its off-diagonal entries are conserved
by electroweak sphalerons as well since there is no flavour off-diagonal lepton anomaly
($\partial_\mu (J^\mu_L)_{\alpha\beta} = 0$ for $\alpha \neq \beta$).}, and the asymmetries:
\beq
  \Delta_\alpha\, \equiv\, \Delta_{B/3 - L_\alpha}\, =\, \frac 1 3 \Delta_B - \Delta_{\ell_\alpha}-\Delta_{e_{R \alpha}}
    \qquad (\alpha = e, \mu, \tau)\, ,
\eeq
which are preserved by all SM interactions.
Above $T > 10^{12} \GeV$, where electroweak sphalerons and charged lepton Yukawa
couplings are out of equilibrium, $\Delta_{\alpha\beta}$ is conserved and reduces
to $- (\Delta_\ell)_{\alpha\beta}$. Below $T = 10^9 \GeV$, the tau Yukawa coupling
is in equilibrium and $\Delta_{\ell_\tau}$ is partially converted into $\Delta_{\tau_R}$.
The conserved asymmetries in the temperature range $10^9 \GeV < T < 10^{12} \GeV$
are therefore
\beq
  \Delta^0_{\alpha\beta}\, \equiv\, \frac 1 3 \Delta_B\, \delta_{\alpha\beta} - (\Delta_\ell^0)_{\alpha\beta}\, ,
    \quad \Delta_\tau \quad \mbox{and} \quad \Delta_\Delta\, ,
\eeq
where $\alpha$ and $\beta$ label any two orthogonal directions in the ($\ell_e$, $\ell_\mu$)
flavour subspace. Finally, below $T = 10^9 \GeV$, the muon Yukawa coupling is also
in equilibrium and $\Delta^0_{\alpha\beta}$ must be replaced by $\Delta_e$ and $\Delta_\mu$.

Using chemical equilibrium, one can express the Higgs and lepton flavour asymmetries
as functions of the conserved asymmetries in each temperature range.
Thus, above $T =10^{12} \GeV$:
\beq
  (\Delta_\ell)_{\alpha\beta}\, =\, - \Delta_{\alpha\beta}\, , \qquad
    \Delta_H\, =\, g^H(\Delta_{\rho\sigma},\Delta_\Delta)\, ,
\eeq
while between $T = 10^9 \GeV$ and $T =10^{12} \GeV$:
\beq
  (\Delta^0_\ell)_{\alpha\beta} = g^\ell_{\alpha\beta}(\Delta^0_{\rho\sigma}, \Delta_\tau, \Delta_\Delta)\, , \ \
    \Delta_{\ell_\tau} = g^\ell_\tau(\Delta^0_{\rho\sigma}, \Delta_\tau, \Delta_\Delta)\, , \ \
    \Delta_H = g^H(\Delta^0_{\rho\sigma}, \Delta_\tau, \Delta_\Delta)\, ,
\eeq
and below $T = 10^9 \GeV$:
\beq
  \Delta_{\ell_\alpha}\, =\, g^\ell_\alpha(\Delta_\rho,\Delta_\Delta)\, , \qquad
    \Delta_H\, =\, g^H(\Delta_\rho,\Delta_\Delta)\, ,
\eeq
where the functions $g^\ell$ and $g^H$ and their arguments depend on the temperature.

The relations among chemical potentials associated with the various spectator processes are
(we assume gauge interactions to be in equilibrium)
\begin{align}
  & \sum_{i=1,2,3}(\mu_{q_i}+2\mu_{u_i}-\mu_{d_i})\ -\! \sum_{\alpha=e,\mu,\tau}(\mu_{\ell_\alpha}+\mu_{e_\alpha})
    +2\mu_H+6\mu_\Delta\, =\, 0 \hskip .8cm (\Delta_Y = 0)
\label{eq:mu_Y}  \\
  & \sum_{i=1,2,3} (2\mu_{q_i}+\mu_{u_i}+\mu_{d_i})\, =\, 0 \hskip 6.6cm (\Delta_B = 0)
\label{eq:mu_B}  \\
  & \sum_{i=1,2,3} (2\mu_{q_i}-\mu_{u_i}-\mu_{d_i})\, =\, 0 \hskip 5.2cm \text{(QCD sphalerons)}
\label{eq:mu_QCD}  \\
  & \sum_{i=1,2,3} 3\mu_{q_i}\ +\! \sum_{\alpha=e,\mu,\tau} \mu_{\ell_\alpha}\, =\, 0 \hskip 5.4cm \text{(EW sphalerons)}
\label{eq:mu_EQ}  \\
  & \mu_{q_i}-\mu_{u_i}+\mu_H\, =\, 0 \hskip 5.65cm \text{(up-type quark Yukawa)}
\label{eq:mu_Yu}  \\
  & \mu_{q_i}-\mu_{d_i}-\mu_H\, =\, 0 \hskip 5.2cm \text{(down-type quark Yukawa)}
\label{eq:mu_Yd}  \\
  & \mu_{\ell_\alpha}-\mu_{e_\alpha}-\mu_H\, =\, 0 \hskip 5.45cm \text{(charged lepton Yukawa)}
\label{eq:mu_Ye}
\end{align}
in which we have also included the constraint due to the hypercharge neutrality of the universe,
as well as $\Delta_B = 0$, which holds as long as the electroweak sphalerons remain out of equilibrium.
These constraints on chemical potentials can be translated into relations among
particle asymmetries using the formulae, valid at leading order in $\mu/T$:
\beq
  \Delta_b\, =\, \frac{g_b}{3 s}\, \mu_b T^2 , \qquad \Delta_f\, =\, \frac{g_f}{6 s}\, \mu_f T^2 ,
\eeq
for a bosonic or fermionic species, respectively, with $g_b$ ($g_f$) the number
of internal degrees of freedom.

Taking into account the processes that are in equilibrium
in a given temperature range\footnote{Using the reaction rates available
in the literature~\cite{Campbell:1992jd,Cline:1993bd,Moore:1997im,Bento:2003jv},
one finds that the top Yukawa coupling comes into equilibrium around $T=10^{15} \GeV$,
the QCD sphalerons around $T=10^{13} \GeV$, the bottom and tau Yukawa couplings as well as
the electroweak sphalerons a little beneath $T=10^{12} \GeV$, the charm Yukawa coupling
around $T=10^{11} \GeV$, the strange and muon Yukawa couplings around $T=10^9 \GeV$,
and the electron Yukawa coupling around $T = 10^5 \GeV$.},
one arrives at the following expressions for the functions $g^\ell$ and $g^H$:
\begin{enumerate}
\item[{\it (i)}] $T>10^{15} \GeV$. In this temperature range all spectator processes
are out of equilibrium (but equality of the chemical potentials within the same gauge
multiplet is still assumed), so one simply has, from $\Delta_Y = \Delta_B = 0$:
\begin{align}
  (\Delta_\ell)_{\alpha\beta}\, & =\, -\Delta_{\alpha\beta}\, ,  \\
  \Delta_H\, & =\, -\text{tr}(\Delta_{\alpha\beta})-2\Delta_\Delta\, .
\label{sp1}
\end{align}
\item[{\it (ii)}] $10^{13} \GeV <T<10^{15} \GeV$. Only the top quark Yukawa coupling
and gauge interactions are in equilibrium. Asymmetries in $q_3$ and $t_R$ can now develop,
but baryon number is still conserved.
The constraints~(\ref{eq:mu_Y}), (\ref{eq:mu_B}) and~(\ref{eq:mu_Yu}) give
\begin{align}
  (\Delta_\ell)_{\alpha\beta}\, & =\, -\Delta_{\alpha\beta}\, ,  \\
  \Delta_H\, & =\, -\frac{2}{3}\text{tr}(\Delta_{\alpha\beta})-\frac{4}{3}\Delta_\Delta\, .
\end{align}
\item[{\it (iii)}] $10^{12} \GeV <T<10^{13} \GeV$. In this range, the QCD sphalerons
are also in equilibrium. Asymmetries in all quark species can develop, but neither
baryon number nor quark flavour are violated, hence $\Delta_{B_1} = \Delta_{B_2} = \Delta_{B_3} = 0$.
Moreover, the first two generations of quarks and $b_R$ are created
only through QCD sphalerons, implying
$\Delta_{q_1} = \Delta_{q_2} = - 2 \Delta_u = - 2 \Delta_d = - 2 \Delta_s = - 2 \Delta_c = -2 \Delta_b$.
This together with the constraints~(\ref{eq:mu_Y})--(\ref{eq:mu_QCD}) and~(\ref{eq:mu_Yu}) yields
\begin{align}
  (\Delta_\ell)_{\alpha\beta}\, & =\, - \Delta_{\alpha\beta}\, ,  \\
  \Delta_H\, & =\, -\frac{14}{23}\text{tr}(\Delta_{\alpha\beta})-\frac{28}{23}\Delta_\Delta\, .
\label{sp2}
\end{align}
\item[{\it (iv)}] $10^{9} \GeV <T<10^{12} \GeV$. In addition to the previous processes,
the electroweak sphalerons and the bottom, tau and charm Yukawa couplings
are also in equilibrium (we neglect the difference between the charm
and bottom quark equilibrium temperatures).
Quantum correlation between $\ell_\tau$ and the other lepton doublet flavours disappear,
and the relevant asymmetries in the lepton sector are $\Delta^0_{\alpha\beta}$
and $\Delta_\tau$. Due to electroweak sphalerons, equal asymmetries develop
in each quark generation: $\Delta_{B_1} = \Delta_{B_2} = \Delta_{B_3}$~($\neq 0$).
Moreover, since the right-handed quarks $u_R$, $d_R$ and $s_R$ are created only through
QCD sphalerons, one has $\Delta_u = \Delta_d = \Delta_s$. Together with
Eqs.~(\ref{eq:mu_Y}) and~(\ref{eq:mu_QCD})--(\ref{eq:mu_Ye}), this constraint gives
\begin{align}
  (\Delta_\ell^0)_{\alpha\beta}\, & =\, \left( \frac{86}{589} \text{tr}(\Delta^0_{\alpha\beta}) + \frac{60}{589} \Delta_\tau
    + \frac{8}{589} \Delta_\Delta \right) \delta_{\alpha\beta} - \Delta^0_{\alpha\beta}\, ,  \\
  \Delta_{\ell_\tau}\, & =\, \frac{30}{589}\text{tr}(\Delta^0_{\alpha\beta}) - \frac{390}{589}\Delta_\tau - \frac{52}{589}\Delta_\Delta\, ,  \\
  \Delta_H\, & =\, - \frac{164}{589} \text{tr}(\Delta^0_{\alpha\beta}) - \frac{224}{589}\Delta_\tau - \frac{344}{589}\Delta_\Delta\, .
\end{align}
\item[{\it (v)}] $10^5 \GeV < T<10^9 \GeV$. In this range, the second generation Yukawa couplings
are also in equilibrium. All three lepton flavours are now distinguishable
and the relevant asymmetries in the lepton sector are the $\Delta_\alpha$.
The QCD sphalerons lead to $\Delta_u=\Delta_d$. Together with this constraint,
Eqs.~(\ref{eq:mu_Y}) and~(\ref{eq:mu_QCD})--(\ref{eq:mu_Ye}) yield
\begin{align}
  \Delta_{\ell_e}\, & =\, - \frac{151}{179}\Delta_e + \frac{20}{179}\Delta_\mu + \frac{20}{179}\Delta_\tau
    + \frac{4}{179}\Delta_\Delta\, ,  \\
  \Delta_{\ell_\mu}\, & =\, \frac{25}{358}\Delta_e - \frac{344}{537}\Delta_\mu + \frac{14}{537}\Delta_\tau
    - \frac{11}{179}\Delta_\Delta\, ,  \\
  \Delta_{\ell_\tau}\, & =\, \frac{25}{358}\Delta_e + \frac{14}{537}\Delta_\mu - \frac{344}{537}\Delta_\tau
    - \frac{11}{179}\Delta_\Delta\, ,  \\
  \Delta_H\, & =\, - \frac{37}{179}\Delta_e - \frac{52}{179}\Delta_\mu - \frac{52}{179}\Delta_\tau
    - \frac{82}{179}\Delta_\Delta\, .
\end{align}
\item[{\it (vi)}] $T<10^5 \GeV$. All spectator processes are now in equilibrium.
Eqs.~(\ref{eq:mu_Y}) and~(\ref{eq:mu_QCD})--(\ref{eq:mu_Ye}) then give
\begin{align}
  \Delta_{\ell_e}\, & =\, - \frac{442}{711}\Delta_e + \frac{32}{711}\Delta_\mu + \frac{32}{711}\Delta_\tau
    - \frac{3}{79}\Delta_\Delta\, ,  \\
  \Delta_{\ell_\mu}\, & =\, \frac{32}{711}\Delta_e - \frac{442}{711}\Delta_\mu + \frac{32}{711}\Delta_\tau
    - \frac{3}{79}\Delta_\Delta\, ,  \\
  \Delta_{\ell_\tau}\, & =\, \frac{32}{711}\Delta_e + \frac{32}{711}\Delta_\mu - \frac{442}{711}\Delta_\tau
    - \frac{3}{79}\Delta_\Delta\, ,  \\
  \Delta_H\, & =\, - \frac{16}{79}\Delta_e - \frac{16}{79}\Delta_\mu - \frac{16}{79}\Delta_\tau
    - \frac{26}{79}\Delta_\Delta\, .
\end{align}
\end{enumerate}

\section{Boltzmann equations}   %
\label{sec:BE}                             %

Using the results of Sections~\ref{sec:flavour} and~\ref{sec:spectator}, one can write
the full set of Boltzmann equations for scalar triplet leptogenesis
 including all relevant spectator processes in a covariant way.
The dynamical variables are $\Sigma_\Delta \equiv (n_\Delta+n_{\bar \Delta})/s$,
$\Delta_\Delta \equiv (n_\Delta-n_{\bar \Delta})/s$ and, depending on the temperature
regime, the $3 \times 3$ density matrix $\Delta_{\alpha\beta}$, the $2\times 2$
density matrix $\Delta^0_{\alpha\beta}$ and the asymmetry $\Delta_\tau$,
or the three asymmetries $\Delta_e$, $\Delta_\mu$ and $\Delta_\tau$.

At temperatures higher than $10^{12} \GeV$, the system of Boltzmann equations is\footnote{One
could write the Boltzmann equations in terms of the density matrix $(\Delta_\ell)_{\alpha\beta}$
rather than $\Delta_{\alpha\beta}$, since $(\Delta_\ell)_{\alpha\beta}$ is conserved
by all SM processes above $T = 10^{12} \GeV$. However, it is more convenient to use
$\Delta_{\alpha\beta}$ as a dynamical variable, as $(\Delta_\ell)_{\alpha\beta}$ is no longer
preserved by electroweak sphalerons when the temperature drops below $10^{12} \GeV$.}
\begin{align}
  sHz\frac{d\Sigma_\Delta}{dz}\, & =\, -\left(\frac{\Sigma_\Delta}{\Sigma_\Delta^\text{eq}}-1\right)\! \gamma_D
    -2\left(\! \left( \frac{\Sigma_\Delta}{\Sigma_\Delta^\text{eq}} \right)^{\! 2} - 1\right)\! \gamma_A\, ,
\label{eq:BE_Sigma}  \\
  sHz\frac{d\Delta_{\alpha\beta}}{dz}\, & =\,
    -\left(\frac{\Sigma_\Delta}{\Sigma_\Delta^\text{eq}}-1\right)\! \gamma_D\, \mathcal{E}_{\alpha\beta}
    +\mathcal{W}^D_{\alpha\beta}+\mathcal{W}^{\ell H}_{\alpha\beta} +\mathcal{W}^{4\ell}_{\alpha\beta}
    +\mathcal{W}^{\ell\Delta}_{\alpha\beta}\, ,
\label{eq:BE_leptons}  \\
  sHz\frac{d\Delta_\Delta}{dz}\, & =\, -\frac{1}{2}\left(\text{tr}(\mathcal{W}^D)-W^D_H\right) ,
\label{eq:BE_Delta}
\end{align}
with the flavour-covariant CP-asymmetry matrix $\mathcal{E}_{\alpha\beta}$ given by Eq.~(\ref{eq:E_alphabeta}),
and the washout terms $\mathcal{W}^D_{\alpha\beta}$, $\mathcal{W}^{\ell H}_{\alpha\beta}$,
$\mathcal{W}^{4\ell}_{\alpha\beta}$, $\mathcal{W}^{\ell\Delta}_{\alpha\beta}$ and $W^D_H$
given by Eqs.~(\ref{eq:W_D}), (\ref{eq:W_lh}), (\ref{eq:W_4l}), (\ref{eq:W_ellDelta}) and
(\ref{eq:W^D_H}), respectively. In the expressions for the washout terms, the asymmetries
$(\Delta_\ell)_{\alpha\beta}$ and $\Delta_H$ should be substituted for
$- \Delta_{\alpha\beta}$ and  $g^H(\Delta_{\rho\sigma},\Delta_\Delta)$, respectively,
where the function $g^H$ depends on the temperature and can be found
in Section~\ref{sec:spectator}, Cases {\it (i)} to {\it (iii)}.
Once this is done, the system of equations~(\ref{eq:BE_Sigma})--(\ref{eq:BE_Delta})
has a closed form and can be solved numerically.

When the temperature drops below $10^{12} \GeV$, the system of Boltzmann equations becomes
\begin{align}
  sHz\frac{d\Sigma_\Delta}{dz}\, & =\, -\left(\frac{\Sigma_\Delta}{\Sigma_\Delta^\text{eq}}-1\right)\! \gamma_D
    -2\left(\! \left( \frac{\Sigma_\Delta}{\Sigma_\Delta^\text{eq}} \right)^{\! 2} - 1\right)\! \gamma_A\, ,
\label{eq:BE_Sigma_10^9_10^12}  \\
  sHz\frac{d\Delta^0_{\alpha\beta}}{dz}\, & =\,
    -\left(\frac{\Sigma_\Delta}{\Sigma_\Delta^\text{eq}}-1\right)\! \gamma_D\, \mathcal{E}_{\alpha\beta}
    +\tilde{\mathcal{W}}^D_{\alpha\beta}+\tilde{\mathcal{W}}^{\ell H}_{\alpha\beta}+\tilde{\mathcal{W}}^{4\ell}_{\alpha\beta}
    +\tilde{\mathcal{W}}^{\ell\Delta}_{\alpha\beta}  \quad (\alpha, \beta = e, \mu)\, ,
\label{eq:BE_Delta0_10^9_10^12}  \\
  sHz\frac{d\Delta_\tau}{dz}\, & =\,
    -\left(\frac{\Sigma_\Delta}{\Sigma_\Delta^\text{eq}}-1\right)\! \gamma_D\, \mathcal{E}_{\tau\tau}
    +\tilde{\mathcal{W}}^D_{\tau}+\tilde{\mathcal{W}}^{\ell H}_{\tau}+\tilde{\mathcal{W}}^{4\ell}_{\tau}
    +\tilde{\mathcal{W}}^{\ell\Delta}_{\tau} ,
\label{eq:BE_Delta_tau_10^9_10^12}  \\
  sHz\frac{d\Delta_\Delta}{dz}\, & =\, -\frac{1}{2}\left(\text{tr}(\tilde{\mathcal{W}}^D)+\tilde{\mathcal{W}}^D_\tau-W^D_H\right) ,
\label{eq:BE_Delta_10^9_10^12}
\end{align}
where the washout terms $\tilde{\mathcal{W}}^{\, \cdots}_{\alpha\beta}$
$(\alpha, \beta = e, \mu)$ and $\tilde{\mathcal{W}}^{\, \cdots}_{\tau}$ can be obtained by setting
$(\Delta_\ell)_{\alpha\tau} = (\Delta^*_\ell)_{\tau\alpha} = \Delta_{\ell_\tau} \delta_{\alpha\tau}$
in the expressions
of the corresponding $\mathcal{W}^{\, \cdots}_{\alpha\beta}$ $(\alpha, \beta = e, \mu, \tau)$,
and the asymmetries $(\Delta^0_\ell)_{\alpha\beta}$, $\Delta_{\ell_\tau}$ and $\Delta_H$
appearing in the washout terms should be replaced by the functions
$g^\ell_{\alpha\beta}(\Delta^0_{\rho\sigma}, \Delta_\tau, \Delta_\Delta)$,
$g^\ell_\tau(\Delta^0_{\rho\sigma}, \Delta_\tau, \Delta_\Delta)$ and
$g^H(\Delta^0_{\rho\sigma}, \Delta_\tau, \Delta_\Delta)$
given in Section~\ref{sec:spectator}, Case {\it (iv)}.
Finally, below $T = 10^9 \GeV$:
\begin{align}
  sHz\frac{d\Sigma_\Delta}{dz}\, & =\, -\left(\frac{\Sigma_\Delta}{\Sigma_\Delta^\text{eq}}-1\right)\! \gamma_D
    -2\left(\! \left( \frac{\Sigma_\Delta}{\Sigma_\Delta^\text{eq}} \right)^{\! 2} - 1\right)\! \gamma_A\, ,
\label{eq:BE_Sigma_10^9}  \\
  sHz\frac{d\Delta_\alpha}{dz}\, & =\,
    -\left(\frac{\Sigma_\Delta}{\Sigma_\Delta^\text{eq}}-1\right)\! \gamma_D \epsilon_\alpha
    +\mathcal{W}^D_{\alpha}+\mathcal{W}^{\ell H}_{\alpha} +\mathcal{W}^{4\ell}_{\alpha}
    +\mathcal{W}^{\ell\Delta}_{\alpha}  \quad (\alpha = e, \mu, \tau)\, ,
\label{eq:BE_leptons_10^9}   \\
  sHz\frac{d\Delta_\Delta}{dz}\, & =\, -\frac{1}{2}\left(\sum_\alpha \mathcal{W}^D_\alpha-W^D_H\right) ,
\label{eq:BE_Delta_10^9}
\end{align}
where $\epsilon_\alpha \equiv \mathcal{E}_{\alpha\alpha} = \sum_\gamma \epsilon_{\alpha\gamma}$,
the washout terms $\tilde{\mathcal{W}}^{\, \cdots}_\alpha$ can be obtained
by setting $(\Delta_\ell)_{\alpha\beta} = \Delta_{\ell_\alpha} \delta_{\alpha\beta}$
in the expressions of the corresponding $\mathcal{W}^{\, \cdots}_{\alpha\beta}$,
and the asymmetries $\Delta_{\ell_\alpha}$ and $\Delta_H$ appearing in the washout terms
should be substituted for the functions $g^\ell_\alpha(\Delta_\rho,\Delta_\Delta)$ and
$g^H(\Delta_\rho,\Delta_\Delta)$ defined
in Section~\ref{sec:spectator}, Cases {\it (v)} and {\it (vi)}.

In Section~\ref{sec:numerical}, we will compare the results obtained with the above system
of covariant Boltzmann equations (hereafter referred to as the full computation)
to various approximations in which flavour effects and/or spectator processes are neglected.
We give below the equations corresponding to these approximations.

\paragraph{Flavoured computation, no spectator processes}

In this approximation, all SM Yukawa interactions as well as the QCD and electroweak
sphalerons are neglected (rigorously, this is legitimate only above $T = 10^{15} \GeV$,
but here it is assumed for arbitrary temperatures).
The relevant Boltzmann equations are Eqs.~(\ref{eq:BE_Sigma})--(\ref{eq:BE_Delta})
with $(\Delta_\ell)_{\alpha\beta}$ and $\Delta_H$ in the washout terms replaced by
(Case {\it (i)} of Section~\ref{sec:spectator})
\beq
  (\Delta_\ell)_{\alpha \beta} = - \Delta_{\alpha \beta}\, , \qquad
  \Delta_H = - \mbox{tr} (\Delta_{\alpha\beta}) - 2 \Delta_\Delta\, .
\label{eq:no_spectator}
\eeq
%

\paragraph{Single flavour approximation, no spectator processes}

In this approximation, both flavour effects and spectator processes are neglected.
To derive the relevant Boltzmann equations, we start from
Eqs.~(\ref{eq:BE_Sigma_Delta})--(\ref{eq:BE_Delta_Delta}) and we implement
the single flavour approximation by substituting
\beq
  (\Delta_\ell)_{\alpha \beta} \rightarrow \Delta_\ell\, ,  \quad  f_{\alpha \beta} \rightarrow \lambda_\ell\, ,
  \quad \kappa_{\alpha \beta} \rightarrow \lambda_\kappa\, ,  \quad
  \mathcal{E}_{\alpha\beta} \rightarrow \mbox{tr}\, \mathcal{E} = \epsilon_\Delta\, .
\eeq
This gives:
\begin{align}
  sHz\frac{d\Sigma_\Delta}{dz}\, & =\, -\left(\frac{\Sigma_\Delta}{\Sigma_\Delta^\text{eq}}-1\right)\! \gamma_D
    -2\left(\! \left( \frac{\Sigma_\Delta}{\Sigma_\Delta^\text{eq}} \right)^{\! 2} - 1\right)\! \gamma_A\, ,  \\
  sHz\frac{d\Delta_\ell}{dz}\, & =\, \left(\frac{\Sigma_\Delta}{\Sigma_\Delta^\text{eq}}-1\right)\!
    \gamma_D \epsilon_\Delta+W^D_\ell+W^{\ell H} ,  \\
  sHz\frac{d\Delta_\Delta}{dz}\, & =\, \frac{1}{2} \left( W^D_\ell-W^D_H \right) ,
\end{align}
where $W^D_H$ and $W^D_\ell$ are the washout terms due to inverse triplet and
antitriplet decays:
\beq
  W^D_\ell\, =\, -2 B_\ell \left(\frac{\Delta_\ell}{Y_\ell^\text{eq}}
    +\frac{\Delta_\Delta}{\Sigma_\Delta^\text{eq}}\right)\! \gamma_D\, ,  \quad
  W^D_H\, =\,  -2 B_H\left(\frac{\Delta_H}{Y_H^\text{eq}}
    -\frac{\Delta_\Delta}{\Sigma_\Delta^\text{eq}}\right)\! \gamma_D\, ,
\eeq
while $W^{\ell H}$ represents the $2 \to 2$ scatterings involving two leptons and two Higgs bosons:
\beq
 W^{\ell H}\, =\, -2\left(\frac{\Delta_\ell}{Y_\ell^\text{eq}}+\frac{\Delta_H}{Y_H^\text{eq}}\right)\! \gamma_{\ell H}\, .
\eeq
Then we impose the single flavour version of the relations~(\ref{eq:no_spectator}):
\beq
  \Delta_\ell = - \Delta\, , \qquad \Delta_H = - \Delta - 2 \Delta_\Delta\, ,
\eeq
where $\Delta \equiv  \Delta_{B-L}$. This gives
\begin{align}
  sHz\frac{d\Sigma_\Delta}{dz}\, & =\, -\left(\frac{\Sigma_\Delta}{\Sigma_\Delta^\text{eq}}-1\right)\! \gamma_D
    -2\left(\! \left( \frac{\Sigma_\Delta}{\Sigma_\Delta^\text{eq}} \right)^{\! 2} - 1\right)\! \gamma_A\, ,
\label{eq:BE_Sigma_1FA}  \\
  sHz\frac{d\Delta}{dz}\, & =\, -\left(\frac{\Sigma_\Delta}{\Sigma_\Delta^\text{eq}}-1\right)\! \gamma_D \epsilon_\Delta
    -2B_\ell\left(\frac{\Delta}{Y_\ell^\text{eq}}
    -\frac{\Delta_\Delta}{\Sigma_\Delta^\text{eq}}\right)\! \gamma_D  \nonumber \\
  & \qquad\qquad\qquad\qquad -2\left(\frac{\Delta}{Y_\ell^\text{eq}}
    + \frac{\Delta+2\Delta_\Delta}{Y_H^{\text{eq}}}\right)\! \gamma_{\ell H}\, ,
\label{eq:BE_leptons_1FA}  \\
  sHz\frac{d\Delta_\Delta}{dz}\, & =\, -\left(\frac{\Delta_\Delta}{\Sigma_\Delta^\text{eq}}
    -B_\ell\frac{\Delta}{Y_\ell^\text{eq}}+B_H\frac{\Delta+2\Delta_\Delta}{Y_H^\text{eq}}\right)\! \gamma_D\, .
\label{eq:BE_Delta_1FA}
\end{align}
This case has been studied in detail in Ref.~\cite{Hambye:2005tk}.

We stress again that the single flavour approximation is not a limiting case
of the flavoured computation, in the sense that it does not become
exact when the charged lepton Yukawa couplings are sent to zero. This is why,
in some regions of the parameter space, the single flavour approximation
is actually a very bad approximation to the flavour-dependent computation,
even though charged lepton Yukawa interactions are out of equilibrium.
This behaviour, which contrasts with the standard leptogenesis scenario involving
hierarchical heavy Majorana neutrinos, will be demonstrated numerically in Section~\ref{sec:numerical}.

\paragraph{Flavour non-covariant approach, spectator processes included}
 
In this case, we take into account spectator processes but do not require
flavour-covariance of the Boltzmann equations.
We assume instead that, in each temperature regime, a basis for the $\ell_\alpha$'s
can be chosen in which scalar triplet leptogenesis is described by the evolution
of ``diagonal'' lepton flavour asymmetries. This is the approach that has been
followed so far in the literature~\cite{Felipe:2013kk,Sierra:2014tqa} (however
without taking into account spectator processes in Ref.~\cite{Felipe:2013kk}),
and it is known to be a good approximation in the standard leptogenesis scenario
with heavy Majorana neutrinos~\cite{Abada:2006fw}, except at the transition between
two temperature regimes. In practice, one works within the single flavour
approximation above $T = 10^{12} \GeV$, and in a two-flavour approximation
in the temperature range $10^9 \GeV < T < 10^{12} \GeV$, where the fast
tau-Yukawa interactions destroy the coherence between
the tau and the other two flavours. Finally, below $T = 10^9 \GeV$, all three lepton
flavours are distinguishable and the relevant dynamical variables are the
flavour asymmetries $\Delta_\alpha$, $\alpha = e, \mu, \tau$; in this case the Boltzmann
equations are given by Eqs.~(\ref{eq:BE_Sigma_10^9})--(\ref{eq:BE_Delta_10^9}).

Thus, for $T > 10^{12} \GeV$, the Boltzmann equations are
\begin{align}
  sHz\frac{d\Sigma_\Delta}{dz}\, & =\, -\left(\frac{\Sigma_\Delta}{\Sigma_\Delta^\text{eq}}-1\right)\! \gamma_D
    -2\left(\! \left( \frac{\Sigma_\Delta}{\Sigma_\Delta^\text{eq}} \right)^{\! 2} - 1\right)\! \gamma_A\, ,
\label{eq:BE_Sigma_1FA_SP}  \\
  sHz\frac{d\Delta}{dz}\, & =\, -\left(\frac{\Sigma_\Delta}{\Sigma_\Delta^\text{eq}}-1\right)\! \gamma_D \epsilon_\Delta
    -2B_\ell \left(\frac{\Delta}{Y_\ell^\text{eq}} - \frac{\Delta_\Delta}{\Sigma_\Delta^\text{eq}}\right)\! \gamma_D\, ,
    \nonumber  \\
  &\qquad\qquad\qquad\qquad
    -2\left(\frac{\Delta}{Y_\ell^\text{eq}}-\frac{g^H(\Delta,\Delta_\Delta)}{Y_H^{\text{eq}}}\right)\! \gamma_{\ell H}\, ,  \\
  sHz\frac{d\Delta_\Delta}{dz}\, & =\, -\left( \frac{\Delta_\Delta}{\Sigma_\Delta^\text{eq}}
    - B_\ell\frac{\Delta}{Y_\ell^\text{eq}} - B_H\frac{g^H(\Delta,\Delta_\Delta)}{Y_H^\text{eq}}\right)\! \gamma_D\, ,
\label{eq:BE_Delta_1FA_SP}
\end{align}
where $g^H(\Delta,\Delta_\Delta) = - (\Delta + 2 \Delta_\Delta)$, $- 2 (\Delta + 2 \Delta_\Delta)/3$ and
$- 14 (\Delta + 2 \Delta_\Delta)/23$ for $T > 10^{15} \GeV$, $10^{13} \GeV < T < 10^{15} \GeV$
and $10^{12} \GeV < T < 10^{13} \GeV$, respectively.
For $10^9 \GeV <T<10^{12} \GeV$ we work in a 2-flavour approximation, as was done
in Ref.~\cite{Sierra:2014tqa}. Namely, we consider only the asymmetry stored in $\ell_\tau$
and the overall asymmetry stored in the $\ell_e$, $\ell_\mu$ doublets, i.e. we perform
the single-flavour approximation in the ($\ell_e$, $\ell_\mu$) subspace by substituting
\bea
  & (\Delta^0_\ell)_{\alpha \beta} \rightarrow \Delta_{\ell_0}\, ,  \qquad
    \Delta^0_{\alpha \beta} \rightarrow \Delta_0\, ,  \qquad
    \mathcal{E}^0_{\alpha\beta} \rightarrow \mbox{tr}\, \mathcal{E}^0 \equiv \epsilon_0\, ,  \nonumber \\
  & f_{\alpha \beta} \rightarrow \sqrt{\mbox{tr} (\tilde f \tilde f^\dagger)}\, \equiv f_{00}\, ,  \quad
    f_{\alpha \tau} \rightarrow \sqrt{|f_{e \tau}|^2 + |f_{\mu \tau}|^2}\, \equiv f_{0\tau}\, ,  \\
  & \kappa_{\alpha \beta} \rightarrow \sqrt{\mbox{tr} (\tilde \kappa \tilde \kappa^\dagger)}\, \equiv \kappa_{00}\, ,  \quad
    \kappa_{\alpha \tau} \rightarrow \sqrt{|\kappa_{e \tau}|^2 + |\kappa_{\mu \tau}|^2}\, \equiv \kappa_{0\tau}\, ,  \nonumber
\eea
where $\alpha, \beta = e, \mu$ and
\beq
  \mathcal{E}^0\, \equiv\, \left( \begin{matrix} \mathcal{E}_{ee} & \mathcal{E}_{e\mu} \\ \mathcal{E}_{\mu e} & \mathcal{E}_{\mu\mu} \end{matrix} \right) ,  \quad
  \tilde{f}\, \equiv\, \left( \begin{matrix} f_{ee} & f_{e\mu} \\ f_{\mu e} & f_{\mu\mu} \end{matrix} \right) ,  \quad
  \tilde{\kappa}\, \equiv\, \left( \begin{matrix} \kappa_{ee} & \kappa_{e\mu}  \\
    \kappa_{\mu e} & \kappa_{\mu\mu} \end{matrix} \right) .
\eeq
We thus obtain the following system of Boltzmann equations (in which $\alpha, \gamma = 0, \tau$):
\begin{align}
  sHz & \frac{d\Sigma_\Delta}{dz}\, =\, -\left(\frac{\Sigma_\Delta}{\Sigma_\Delta^\text{eq}}-1\right)\! \gamma_D
    -2\left(\! \left( \frac{\Sigma_\Delta}{\Sigma_\Delta^\text{eq}} \right)^{\! 2} - 1\right)\! \gamma_A\, ,
\label{eq:BE_Sigma_2FA}  \\
  sHz & \frac{d\Delta_\alpha}{dz}\, =\, -\left(\frac{\Sigma_\Delta}{\Sigma_\Delta^\text{eq}}-1\right)\! \gamma_D \epsilon_\alpha
  + 2 B_\ell \sum_\gamma C^D_{\alpha\gamma}\left(\frac{g^\ell_\alpha(\Delta_\rho,\Delta_\Delta)+g^\ell_\gamma(\Delta_\rho,\Delta_\Delta)}{2Y_\ell^{\text{eq}}}+\frac{\Delta_\Delta}{\Sigma_\Delta^{\text{eq}}}\right)\! \gamma_D  \nonumber \\
&+2\sum_\gamma\left(\frac{g^\ell_\alpha(\Delta_\rho,\Delta_\Delta)+g^\ell_\gamma(\Delta_\rho,\Delta_\Delta)}{2Y_\ell^\text{eq}}+\frac{g^H(\Delta_\rho,\Delta_\Delta)}{Y_H^{\text{eq}}}\right) \left(C^\Delta_{\alpha\gamma}\gamma_{\ell H}^\Delta+C^\mathcal{I}_{\alpha\gamma}\gamma_{\ell H}^\mathcal{I}+C^\mathcal{H}_{\alpha\gamma}\gamma_{\ell H}^\mathcal{H}\right)  \nonumber\\
&+\sum_\gamma \left(\frac{g^\ell_\alpha(\Delta_\rho,\Delta_\Delta)-g^\ell_\gamma(\Delta_\rho,\Delta_\Delta)}{Y_\ell^{\text{eq}}}\right)\left(C^{4\ell}_{\alpha\gamma}\gamma_{4\ell}+C^{\ell\Delta}_{\alpha\gamma}\gamma_{\ell\Delta}\right) ,
\label{eq:BE_leptons_2FA}  \\
 sHz & \frac{d\Delta_\Delta}{dz}\, =\, -\left(\frac{\Delta_\Delta}{\Sigma_\Delta^\text{eq}}+ B_\ell \sum_{\alpha,\,\gamma} C^D_{\alpha\gamma}\frac{g^\ell_\alpha(\Delta_\rho,\Delta_\Delta)+g^\ell_\gamma(\Delta_\rho,\Delta_\Delta)}{2Y_\ell^{\text{eq}}}-B_H\frac{g^H(\Delta_\rho,\Delta_\Delta)}{Y_H^\text{eq}}\right)\! \gamma_D\, ,
\label{eq:BE_Delta_2FA}
\end{align}
where $\epsilon_\tau \equiv \mathcal{E}_{\tau\tau}$,
\begin{align}
  g^\ell_0(\Delta_\rho,\Delta_\Delta)\, & =\, - \frac{503}{589} \Delta_0 + \frac{60}{589} \Delta_\tau
    + \frac{8}{589} \Delta_\Delta\, ,  \nonumber \\
  g^\ell_\tau(\Delta_\rho,\Delta_\Delta)\, & =\, \frac{30}{589} \Delta_0 - \frac{390}{589}\Delta_\tau
    - \frac{52}{589}\Delta_\Delta\, ,  \\
  g^H(\Delta_\rho,\Delta_\Delta)\, & =\, - \frac{164}{589} \Delta_0 - \frac{224}{589}\Delta_\tau
    - \frac{344}{589}\Delta_\Delta\, ,  \nonumber
\end{align}
and the coefficients $C^X_{\alpha\gamma}$ are given by ($\alpha, \gamma = 0, \tau$)
\begin{align}
  &C^D_{00} = \frac{f^2_{00}}{\lambda^2_\ell}\, ,  \qquad
    C^D_{0\tau} = C^D_{\tau 0} = \frac{f^2_{0 \tau}}{\lambda^2_\ell}\, ,\ \qquad
    C^D_{\tau\tau} = \frac{|f_{\tau\tau}|^2}{\lambda^2_\ell}\, ,  \nonumber \\
  &C^\Delta_{\alpha\gamma} = C^D_{\alpha\gamma}\, ,  \qquad
    C^{\mathcal{I}}_{\alpha\gamma} =
    2\, \frac{f_{\alpha\gamma}\kappa_{\alpha\gamma}}{\mbox{Re}\, [\text{tr}(f^\dagger\kappa)]}\, ,
    \qquad  C^{\mathcal{H}}_{\alpha\gamma} = \frac{\kappa_{\alpha\gamma}}{\lambda_\kappa^2}\, ,  \\
  &C^{4\ell}_{0\tau} = C^{4\ell}_{\tau0} = C^D_{0\tau}\, ,  \qquad
    C^{\ell\Delta}_{0\tau} = C^{\ell\Delta}_{\tau0}
    = \frac{|f_{00}f_{0\tau}+f_{0\tau}f_{\tau\tau}|^2}{\text{tr}(ff^\dagger ff^\dagger)}\, .  \nonumber
\end{align}
%

\section{Numerical study}  %
\label{sec:numerical}         %

In this section, we investigate numerically the impact of lepton flavour effects
on scalar triplet leptogenesis using the flavour-covariant formalism
introduced previously. In order to assess quantitatively the relevance of
lepton flavour covariance, we compare the results of the density matrix
computation with the ones obtained in flavour non-covariant approximations.
We also compare the relative impacts of spectator processes and
of lepton flavour covariance on the generated baryon asymmetry.
More specifically:
\begin{itemize}
\item in the temperature regime $T > 10^{12} \GeV$, where all charged lepton Yukawa
couplings are out of equilibrium, the flavour-covariant computation involves the $3 \times 3$
density matrix $\Delta_{\alpha \beta}$, whose evolution is governed by the Boltzmann
equation~(\ref{eq:BE_leptons}).
The result of this computation is compared with the one obtained
in the single flavour approximation, for which the relevant Boltzmann equations
are Eqs.~(\ref{eq:BE_Sigma_1FA_SP})--(\ref{eq:BE_Delta_1FA_SP})
or Eqs.~(\ref{eq:BE_Sigma_1FA})--(\ref{eq:BE_Delta_1FA}),
depending on whether spectator processes are taken into account or not.
The relevance of flavour effects above $T = 10^{12} \GeV$ is a novel effect
specific to scalar triplet leptogenesis, which has been overlooked in previous works.
\item in  the regime $10^9 \GeV < T < 10^{12} \GeV$, where the tau Yukawa coupling
is in equilibrium while the muon and electron Yukawa couplings are not, the flavour-covariant
computation involves the $2 \times 2$ density matrix $\Delta^0_{\alpha \beta}$
($\alpha, \beta = e, \mu$) and the asymmetry $\Delta_\tau$, whose evolutions are governed by 
the Boltzmann equations~(\ref{eq:BE_Delta0_10^9_10^12}) and~(\ref{eq:BE_Delta_tau_10^9_10^12}).
This computation is compared with the 2-flavour approximation in which the $2 \times 2$
density matrix $\Delta^0_{\alpha \beta}$ is replaced by $\Delta_0$ (the overall asymmetry
in the charges $B/3 - L_e$ and $B/3 - L_\mu$)
and the Boltzmann equations are given by Eqs.~(\ref{eq:BE_Sigma_2FA})--(\ref{eq:BE_Delta_2FA}).
Again, the relevance of flavour effects in the ($\ell_e$, $\ell_\mu$) subspace
in the temperature range where the electron and muon Yukawa coupligs are out of equilibrium
is a new effect specific to scalar triplet leptogenesis.
\end{itemize}
Below $T = 10^9 \GeV$, flavour covariance is completely broken by fast processes
induced by the tau and muon Yukawa couplings and the relevant dynamical parameters
are $\Delta_e$, $\Delta_\mu$ and $\Delta_\tau$, whose evolution is governed
by the Boltzmann equations~(\ref{eq:BE_Sigma_10^9})--(\ref{eq:BE_Delta_10^9}).

\subsection{Parameters}               %

The parameters of scalar triplet leptogenesis are the triplet mass $M_\Delta$
and its couplings to Higgs ($\mu$ or $\lambda_H \equiv |\mu|/M_\Delta$) and lepton doublets
($f_{\alpha \beta}$), as well as the coefficients of the effective dimension-5
operators~(\ref{eq:Weinberg_operator}), $\kappa_{\alpha \beta}$. These parameters
are not all independent, however, as they are related to the neutrino mass matrix by
\begin{align}
  m_\Delta+ m_\mathcal{H}\, =\, m_\nu\, ,
\label{eq:m_nu}
\end{align}
where $m_\Delta$ and $m_\mathcal{H}$ are defined by Eqs.~(\ref{eq:m_Delta})
and~(\ref{eq:m_heavy}), respectively, and $m_\nu = U^* D_\nu U^\dagger$
in the charged lepton mass eigenstate basis, with $D_\nu = \mbox{Diag} (m_1, m_2, m_3)$,
$m_i$ ($i = 1,2,3$) the neutrino masses and $U$ the PMNS matrix.
Thus, for given neutrino parameters, the $\kappa_{\alpha \beta}$'s
are fully determined once $M_\Delta$, $\lambda_H$ and the $f_{\alpha \beta}$'s
are fixed. It is then convenient to write the triplet contribution to the neutrino mass matrix as
\begin{align}
  m_\Delta\, & =\, U^* V^* D_\Delta V^\dagger U^\dagger ,
\end{align}
where $D_\Delta$ is a diagonal matrix with real positive entries and $V$
is the unitary matrix that relates the lepton doublet basis in which $m_\nu$
is diagonal to the one in which $m_\Delta = D_\Delta$. We adopt the following
parametrization for $V$:
\beq
  V\, =\, \left( \begin{matrix}
    e^{-i\beta_1} & 0 & 0 \\ 0 & e^{-i\beta_2} & 0 \\ 0 & 0 & e^{-i\beta_3}
    \end{matrix} \right)
  U' (\phi_{23},\phi_{13},\phi_{12},\gamma)
  \left(\begin{matrix}
    e^{i\alpha_1} & 0 & 0 \\ 0 & 1 & 0 \\ 0 & 0 & e^{i\alpha_2}
    \end{matrix} \right) ,
\label{eq:parametrization_mDelta}
\eeq
\beq
  U'\, = \left(\begin{matrix}  c_{12}c_{13} & s_{12}c_{13} & s_{13}  \\
    -s_{12}c_{23}-c_{12}s_{13}s_{23}\, e^{i\gamma} & c_{12}c_{23}-s_{12}s_{13}s_{23}\, e^{i\gamma}
      & c_{13}s_{23}e^{i\gamma}  \\
    s_{12}s_{23}-c_{12}s_{13}c_{23}\, e^{i\gamma} & -c_{12}s_{23}-s_{12}s_{13}c_{23}\, e^{i\gamma}
      & c_{13}c_{23}e^{i\gamma}
    \end{matrix} \right) ,
\label{eq:parametrization_Uprime}
\eeq
in which $c_{ij} \equiv \cos\phi_{ij}$ and $s_{ij} \equiv \sin\phi_{ij}$.
Note that the phases $\beta_{1,2,3}$ are physical and cannot be removed
by rephasing the lepton doublets, since this freedom has already been used
to write the neutrino mass matrix in Eq.~(\ref{eq:m_nu}) in the standard
phase convention.

The flavour structure of the couplings $f_{\alpha \beta}$
(or equivalently, of the matrix $m_\Delta$) plays a crucial role in leptogenesis
as it determines the flavour-dependent CP asymmetries and washout rates.
Given the large number of parameters involved, we shall consider suitably
chosen ans\"atze for $m_\Delta$ in the following. Before presenting them,
let us first discuss on a qualitative basis how flavour effects depend
on the triplet couplings.

\subsection{Qualitative discussion of flavour effects}  %

In the flavour-covariant formalism used in this paper, the violation of CP in triplet
decays is described by an hermitian
CP-asymmetry matrix $ \mathcal{E}_{\alpha\beta}$ given by Eq.~(\ref{eq:E_alphabeta}).
Using the fact that $m_\nu = m_\Delta + m_\mathcal{H}$, one can rewrite it as
\beq
  \mathcal{E}_{\alpha\beta}\, =\, \frac{1}{8\pi i}\frac{M_\Delta}{v^2}\sqrt{B_\ell B_H}\
    \frac{(m_\nu m_\Delta^{\dagger} - m_\Delta m_\nu^{\dagger})_{\alpha\beta}}{\bar{m}_\Delta}\, ,
\eeq
where $\bar{m}_\Delta$ has been defined previously as
$\bar{m}_\Delta \equiv \sqrt{\text{tr}(m_\Delta^\dagger m_\Delta)}\, $.
Using the Cauchy-Schwarz theorem, one can derive an upper bound on the total
CP asymmetry $\epsilon_\Delta = \text{tr}\, \mathcal{E}$~\cite{Hambye:2005tk}:
\beq
  |\epsilon_\Delta|\ \leq\ \frac{1}{4\pi} \frac{M_\Delta}{v^2} \sqrt{B_\ell B_H}\,\bar{m}_\nu\, ,
\eeq
where $\bar{m}_\nu \equiv \sqrt{\text{tr}(m_\nu^\dagger m_\nu)}\, $.
This bound is saturated for $m_\Delta = i C m_\nu$ with $C$ real:
the total CP asymmetry in triplet decays is maximal when the $f_{\alpha \beta}$ couplings
and the neutrino mass matrix entries have the same flavour structure and an overall
phase difference $\pm \pi / 2$. For $C>0$, this gives
\begin{align}
  \mathcal{E}_{\alpha\beta}\, & =\, - \frac{1}{4\pi} \frac{M_\Delta}{v^2} \sqrt{B_\ell B_H}\
    \frac{(m_\nu m^\dagger_\nu)_{\alpha\beta}}{\bar{m}_\nu}\, ,  \qquad
    \epsilon_\Delta\, =\, - \frac{1}{4\pi} \frac{M_\Delta}{v^2} \sqrt{B_\ell B_H}\,\bar{m}_\nu\, .
\end{align}

In addition to a large enough CP asymmetry,
successful scalar triplet leptogenesis requires that decays dominate over
annihilations around $z \sim 1$ ($T \sim M_\Delta$), hence
the triplet decays are typically fast. In spite of this, a large lepton asymmetry
can develop when some decay channels are slower than the expansion of the universe.
For instance, in the single flavour approximation, it was shown in Ref.~\cite{Hambye:2005tk}
that a large efficiency is obtained when one of the two decay modes $\Delta \to H H$
or $\Delta \to \bar \ell \bar \ell$ is out of equilibrium, i.e. when $\lambda_H \ll \lambda_\ell$
or $\lambda_\ell \ll \lambda_H$. As we are going to see, when lepton flavour effects are taken into account,
it is also possible to reach an order one efficiency in the case $\lambda_\ell \sim \lambda_H$,
provided that some flavoured decay channel $\Delta \to \bar \ell_\alpha \bar \ell_\beta$
is slower than the expansion of the universe.
By contrast, when $\lambda_\ell \gg \lambda_H$,
all decay channels into antileptons are fast and
only the decays into Higgs bosons can occur out of equilibrium
(barring a strong hierarchy among the couplings $f_{\alpha \beta}$).
In this case, asymmetric triplet and antitriplet decays generate
an asymmetry in the Higgs sector in the first place,
which is accompanied by a $B-L$ asymmetry of comparable size
when the triplet abundance drops~\cite{Hambye:2005tk}
(see e.g. Eq.~(\ref{sp1}), which holds above $T = 10^{15} \GeV$,
and analogous relations valid in other temperature regimes in Section~\ref{sec:spectator}).
Lepton flavour does not play a prominent role in this mechanism,
hence no significant difference between the single flavour approximation
and the flavoured computation is expected in this case\footnote{An obvious
exception to this general statement is when the total CP asymmetry in
triplet decays vanishes, while the flavour-dependent CP asymmetries
$\mathcal{E}_{\alpha\beta}$ do not (or more generally when $\epsilon_\Delta$
is small due to strong cancellations between the diagonal entries of $\mathcal{E}$).
In particular, it does not apply to the scenario of purely flavoured
leptogenesis discussed in Refs.~\cite{Felipe:2013kk,Sierra:2014tqa},
which includes traceless contributions to $\mathcal{E}_{\alpha\beta}$ that are
not considered in this paper (see the comment in Section~\ref{sec:ingredients}).}.
A similar conclusion can be reached in the opposite case $\lambda_\ell \ll \lambda_H$.
Indeed, for $\lambda_\ell$ small enough, the washout of the flavoured lepton
asymmetries can be neglected and the Boltzmann equation for
$\Delta_{\alpha\beta}$ simplifies to
\beq
  sHz\frac{d\Delta_{\alpha\beta}}{dz}\, \simeq\, - \left(\frac{\Sigma_\Delta}{\Sigma_\Delta^{\text{eq}}}-1\right)\!
    \gamma_D\, \mathcal{E}_{\alpha\beta}\, .
\eeq
Taking the trace of this equation, one recovers the single flavour approximation
(with $\Delta \equiv \mbox{tr} (\Delta_{\alpha\beta}) = \Delta_{B-L}$):
\beq
  sHz\frac{d\Delta}{dz}\, \simeq\, - \left(\frac{\Sigma_\Delta}{\Sigma_\Delta^{\text{eq}}}-1\right)\!
  \gamma_D\, \epsilon_\Delta\, .
\eeq
Hence, in the case $\lambda_\ell \ll \lambda_H$ too, no significant effect
of lepton flavour is expected.

\subsection{Ans\"atze and inputs} %

The above discussion suggests that lepton flavour effects will not play
a prominent role when the scalar triplet gives a subdominant contribution
to neutrino masses (i.e. $\bar{m}_\Delta \ll \bar{m_\nu}$), because
in this case either $\Gamma (\Delta \to H H)$ or $\Gamma (\Delta \to \bar \ell \bar \ell)$
is smaller than the expansion rate of the universe\footnote{One can see this by noting
that $K_\ell K_H \approx 550 \left( \bar m_\Delta / 0.05 \eV \right)^2$, where
$K_a \equiv \Gamma (\Delta \to a a) / H(M_\Delta)$ and the Hubble rate is given
by $H(T) = 1.66 \sqrt{g_*}\, T^2/M_P$ with $g_* = g^{\rm SM}_* = 106.75$. The case
in which both $K_\ell < 1$ and $K_H < 1$ is not interesting for leptogenesis, because
triplets and antitriplets would annihilate before decaying.}, so that a large efficiency
is already reached in the single flavour approximation. For this reason,
we will only consider ans\"atze satisfying the condition $\bar{m}_\Delta = \bar{m_\nu}$.
Specifically, we will study the following ans\"atze (explicit expressions
for $m_\Delta$ are given in the charged lepton mass eigenstate basis):
\begin{itemize}
\item {\bf ansatz 1:} $m_\Delta = i m_\nu$ \

\item {\bf ansatz 2:} $m_\Delta = i \bar m_\nu\, U^* \left( \begin{array}{ccc} \sqrt{1-x^2}\, y & 0 & 0 \\
  0 & xy & 0 \\
  0 & 0 & \sqrt{1-y^2}
  \end{array} \right) U^\dagger\, $, \\ \\
where $x$ and $y$ can vary between $0$ and $1$. Ansatz 1 is a particular case of Ansatz 2
corresponding to $x_0 = m_2 / \sqrt{m^2_1 + m^2_2}$ and $y_0 = \sqrt{m^2_1 + m^2_2} / \bar m_\nu\, $. 
\item {\bf ansatz 3:} $m_\Delta = U^* V^* D_\Delta V^\dagger U^\dagger$ with $\phi_{23} = \phi_{13} = 0$
and $\alpha_1 = \alpha_2 + \gamma = \beta_3 = 0$, i.e.
\beq
  V\, =\, \left( \begin{matrix}
    e^{-i\beta_1} & 0 & 0 \\ 0 & e^{-i\beta_2} & 0 \\ 0 & 0 & 1
    \end{matrix} \right)
  \left(\begin{matrix}
    c_{12} & s_{12} & 0 \\ -s_{12} & c_{12} & 0 \\ 0 & 0 & 1
    \end{matrix} \right) ,
\eeq
with two different options for $D_\Delta$ and the phases $\beta_1$, $\beta_2$:
\begin{itemize}
\item {\bf ansatz 3-a:} $D_\Delta = D_\nu = \left( \begin{array}{ccc} m_1 & 0 & 0 \\
  0 & m_2 & 0 \\
  0 & 0 & m_3
  \end{array} \right)$, $\beta_2 = 0\, $, while $\beta_1$ and $\phi_{12}$ can vary;
\item {\bf ansatz 3-b:} $D_\Delta = \left( \begin{array}{ccc} x m_0 & 0 & 0 \\
  0 & \sqrt{1-x^2}\, m_0 & 0 \\
  0 & 0 & m_3
  \end{array} \right)$, $\beta_1 = \beta_2 = \pi/4\, $, \\ \\
with $m_0 \equiv \sqrt{m^2_1 + m^2_2}\, $, while $x$ and $\phi_{12}$ can vary
($0 \leq x \leq 1$). Note that $D_\Delta = D_\nu$ for $x = m_1 / \sqrt{m^2_1 + m^2_2}\, $.
\end{itemize}
\end{itemize}
Ansatz 1 depends only on 2 parameters, which can be chosen to be $M_\Delta$
and $\lambda_H$. The other ans\"atze involve 2 additional parameters that control
the flavour structure of the couplings $f_{\alpha\beta}$: $x$ and $y$ for Ansatz 2,
$\phi_{12}$ and $\beta_1$ for Ansatz 3-a, $\phi_{12}$ and $x$ for Ansatz 3-b.

We also need to specify the neutrino mass parameters, which serve as inputs
in the above ans\"atze and determine the couplings $\kappa_{\alpha\beta}$
through Eq.~(\ref{eq:m_nu}). For definiteness, we assume a normal hierarchy.
The running of the neutrino mass matrix up to the triplet mass scale $M_\Delta$
is performed using the Mathematica package REAP~\cite{Antusch:2005gp}.
For a hierarchical neutrino spectrum, the only effect of running is to multiply
the eigenvalues of the neutrino mass matrix by a common factor $r$~\cite{Chankowski:2001mx},
which for $M_\Delta$ around $10^9$--$10^{12} \GeV$ lies between $1.2$ and $1.4$.
We therefore define the squared mass differences as 
$\Delta m^2_{21} = r^2 \Delta m^2_\text{sun}$ and $\Delta m^2_{31} = r^2 \Delta m^2_\text{atm}$,
where $\Delta m^2_\text{sun}$ and $\Delta m^2_\text{atm}$ are the parameters
extracted from oscillation experiments,
and set the lightest eigenvalue $m_1=10^{-3} \eV$ at the scale $M_\Delta$
(which corresponds to a lightest neutrino mass in the range $(0.7\,$--$\, 0.8) \times 10^{-3} \eV$).
For the oscillations parameters, we take
$\Delta m^2_\text{sun} = 7.59 \times 10^{-5} \eV^2$, $\Delta m^2_\text{atm} = 2.47 \times 10^{-3} \eV^2$,
$\sin^2 \theta_{12} = 0.30$, $\sin^2 \theta_{23} = 0.42$ and $\sin^2 \theta_{13} = 0.023$,
within $1 \sigma$ of the most recent fit~\cite{Gonzalez-Garcia:2014bfa}.
Finally, we set all CP-violating phases of the PMNS matrix to zero. 

Let us discuss these ans\"atze in turn. Ansatz 1 maximizes the total CP asymmetry
in triplet decays, but due to the hierarchy among the couplings $f_{\alpha \beta}$ implied
by the relation $m_\Delta = i m_\nu$, it is not expected to lead to very strong flavour effects.
Indeed, taking into account the hierarchy $m_1, m_2 \ll m_3$, one can write,
in the neutrino mass eigenstate basis,
\beq
  f\, =\, i \lambda_\ell \left( \begin{matrix}
    \frac{m_1}{\bar{m}_\nu}&0&0  \\  0&\frac{m_2}{\bar{m}_\nu}&0  \\  0&0&\frac{m_3}{\bar{m}_\nu}
    \end{matrix}\right) \approx\, i \lambda_\ell
 \left(\begin{matrix} 0&0&0  \\  0&0&0  \\  0&0&1
   \end{matrix}\right) , \qquad
  \mathcal{E}\, \approx \left( \begin{matrix}
    0&0&0  \\  0&0&0  \\  0&0& \epsilon_\Delta  \end{matrix} \right) . 
\label{eq:f_ansatz1}
\eeq
Therefore, as a first approximation, leptogenesis can be described in terms of a single flavour
(the tau flavour). But since the hierarchy between neutrino masses is not very strong,
the difference between the single flavour approximation and the result
of the flavoured computation can still be significant, as we are going to see.

In Ansatz 2, the couplings $f_{\alpha\beta}$ are diagonal in the neutrino mass
eigenstate basis as in Ansatz 1, but the diagonal entries $f_{\alpha\alpha}$
are no longer proportional to neutrino masses. As a result $\epsilon_\Delta$
is not maximal, but since $f_{\alpha\beta}$ and $\mathcal{E}_{\alpha\beta}$
do not necessarily have the hierarchical structure of Eq.~(\ref{eq:f_ansatz1}), flavour effects
can be more important than in Ansatz 1 and eventually lead to a larger baryon asymmetry.

Ansatz 3 allows to investigate the effect of a misalignment between the neutrino mass
matrix $m_\nu$ and the triplet couplings to leptons $f_{\alpha\beta}$, i.e. of a
relative rotation between the lepton doublet basis in which $m_\nu = D_\nu$ and the
one in wich the $f_{\alpha\beta}$'s are diagonal. We consider two subcases,
$D_\Delta = D_\nu = \mbox{Diag} (m_1, m_2, m_3)$ (Ansatz 3-a)
and $D_\Delta = \mbox{Diag} (x m_0, \sqrt{1-x^2}\, m_0, m_3)$,
with $m_0 \equiv \sqrt{m^2_1 + m^2_2}$ (Ansatz 3-b).
The source terms are concentrated in the $(\ell_1, \ell_2)$ subspace
(i.e. $\mathcal{E}_{13} = \mathcal{E}_{23} = \mathcal{E}_{33} = 0$
in the neutrino mass eigenstate basis), in which the washout is weaker since
$|f_{\alpha\beta}| \ll |f_{33}|$ for $\alpha, \beta = 1,2$ and $f_{13} = f_{23} = 0$.
In this case, the unflavoured computation (in which the washout is controlled by
$\lambda_\ell \simeq |f_{33}|$) is likely to be a very bad approximation to the full,
flavoured computation.

\subsection{Numerical results}   %

After solving numerically the appropriate Boltzmann equations
with initial equilibrium abundances for the triplets\footnote{This assumption is justified
by the fact that gauge interactions bring the population of triplets
and antitriplets into equilibrium very quickly, except for very large triplet masses.},
leptons and Higgs bosons,
one is left with the final value of the asymmetries $\Delta_\alpha = \Delta_{B/3 - L_\alpha}$.
The baryon asymmetry is then given by~\cite{Harvey:1990qw}
\beq
  \Delta_B\, =\, \frac{12}{37}\, \Delta_{B-L}\, =\, \frac{12}{37}\, \sum_\alpha \Delta_\alpha\, ,
\eeq
where we have assumed that sphalerons go out of equilibrium below the electroweak
phase transition, as suggested by recent lattice computations~\cite{D'Onofrio:2014kta}.
\begin{figure}[t]
\center
\includegraphics[scale=0.55]{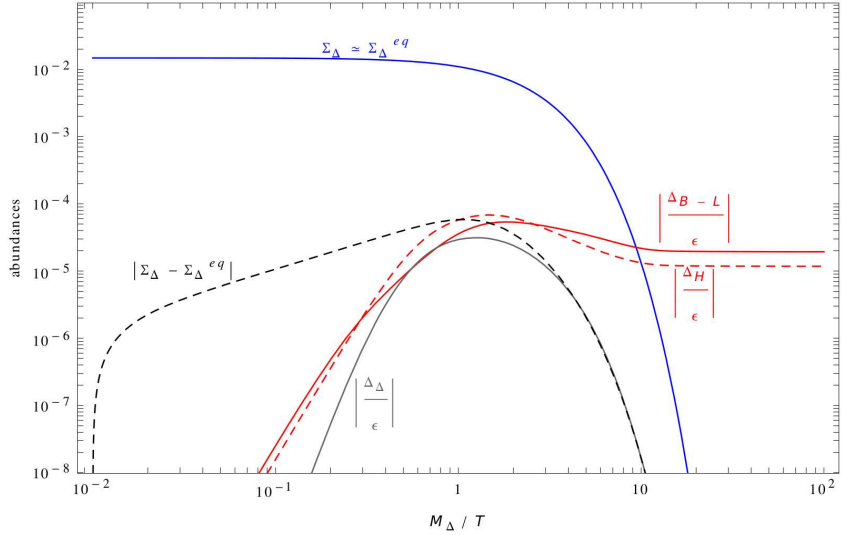}
\caption{Comoving number density of triplets and antitriplets $\Sigma_\Delta$
and asymmetries $\Delta_\Delta$, $\Delta_H$ and $\Delta_{B-L} = \mbox{Tr}\, (\Delta_{\alpha\beta})$
as a function of $z = M_\Delta / T$, for $M_\Delta = 5 \times 10^{12} \GeV$, $\lambda_H = 0.1$
and $m_\Delta = i m_\nu$. Also shown is the departure of $\Sigma_\Delta$ from its equilibrium value.
Asymmetries are plotted in units of the total CP asymmetry in triplet decays $\epsilon_\Delta$.}
\label{fig:abundances}
\end{figure}
One can also express the baryon asymmetry of the universe in terms of the more
familiar baryon-to-photon ratio $ \eta_B\ \equiv n_B / n_\gamma = 7.04\, \Delta_B$,
whose observed value from recent Planck data is~\cite{Planck:2015xua}
\beq
  \eta^{\rm obs.}_B\, =\, (6.10 \pm 0.06) \times 10^{-10} \qquad (68\%\, {\rm C.L.})\, .
\eeq
As can be seen from Fig.~\ref{fig:abundances}, the $B-L$ asymmetry is already
stabilized around $T \sim M_\Delta /100$.

\begin{figure}[t]
\center
\includegraphics[scale=0.55]{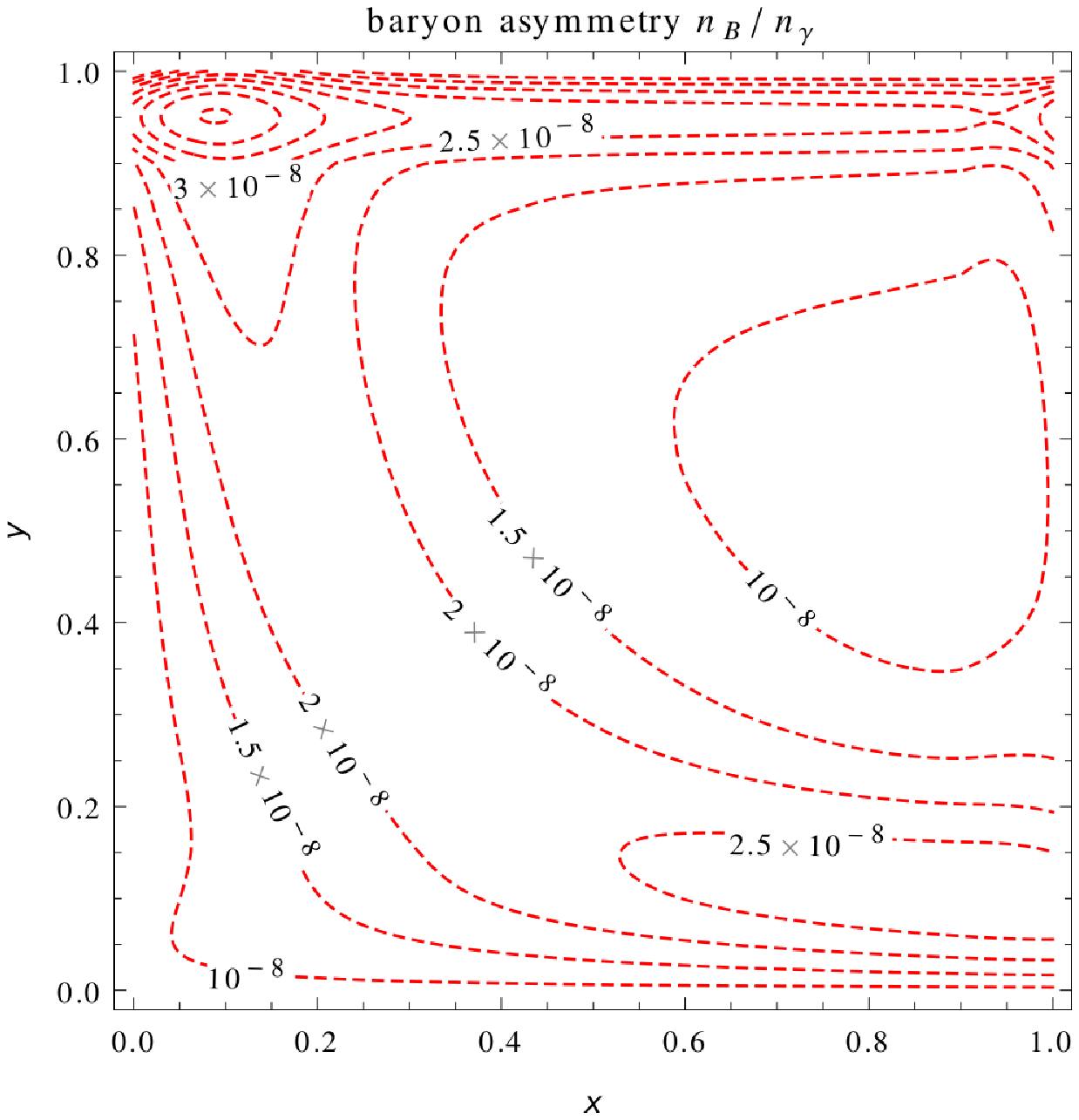}\qquad
\includegraphics[scale=0.55]{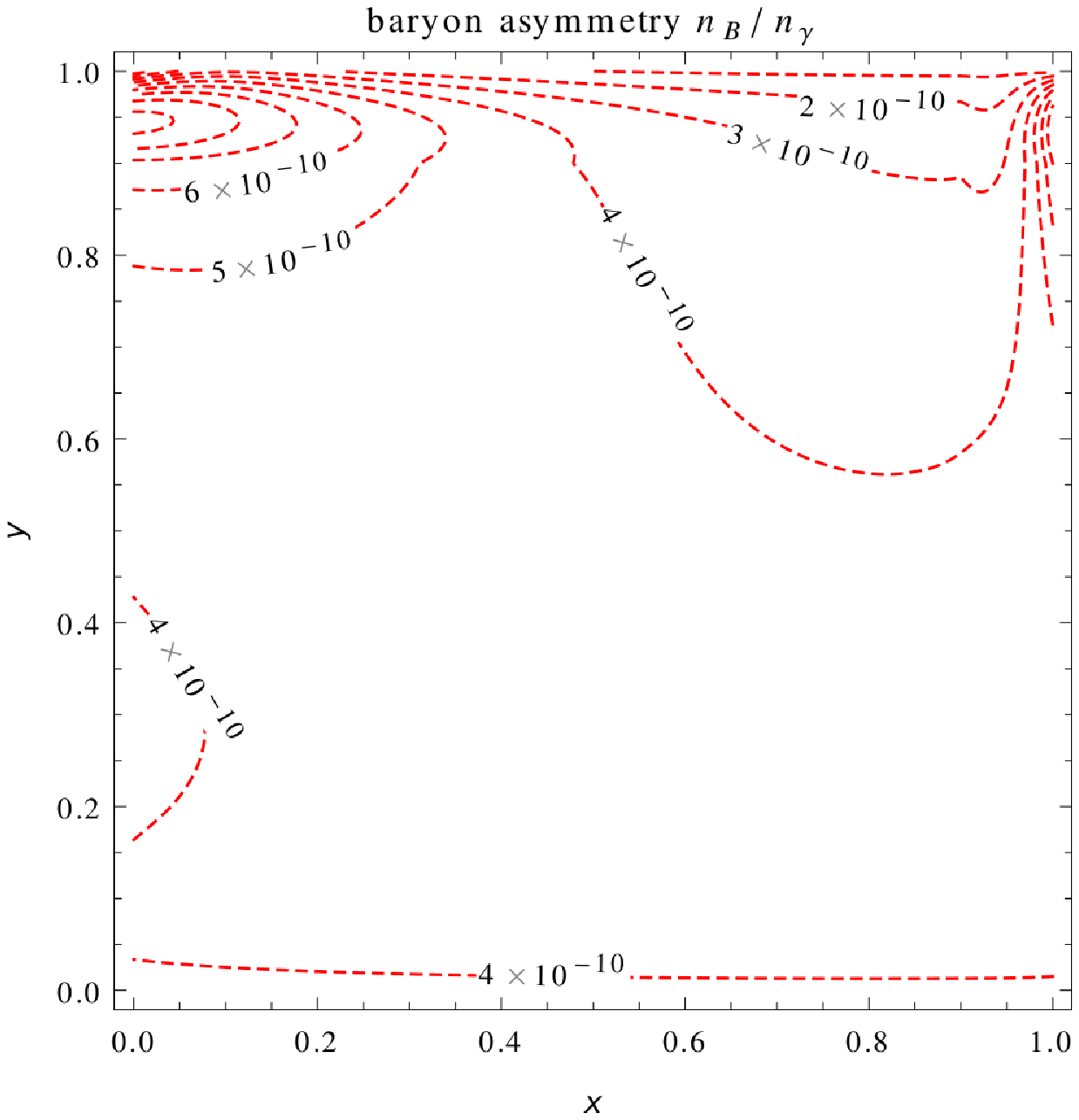}
\caption{Isocurves of the baryon-to-photon ratio $n_B / n_\gamma$ in the $(x,y)$ plane
for Ansatz 2. Left panel: $M_\Delta = 5 \times 10^{12} \GeV$, $\lambda_H = 0.4$.
Right panel: $M_\Delta = 10^{11} \GeV$, $\lambda_H = 0.025$.}
\label{fig:Ansatz_2}
\end{figure}

We are now ready to present our numerical results. To reduce the number
of free parameters, we concentrate on the ans\"atze defined in the previous
subsection and fix the neutrino parameters as specifed there.
Let us first consider Ansatz 2, which unlike Ansatz 1 does not maximize $\epsilon_\Delta$,
but allows for larger flavour effects. Since this ansatz gives same-sign source terms
in the Boltzmann equations, flavour effects are more likely to be relevant when
$B_\ell \sim B_H$; hence, for a given triplet mass, we choose
the parameter $\lambda_H$ such that $\lambda_\ell \sim \lambda_H$.
Then we
study the dependence of the baryon-to-photon ratio $n_B / n_\gamma$ on $x$ and $y$,
the parameters that control the relative sizes of the couplings $f_{\alpha\beta}$.
The result of the full computation (i.e. the computation involving the flavour-covariant
Boltzmann equations with spectator processes included) is shown
in Fig.~\ref{fig:Ansatz_2} for two different values of the triplet mass,
$M_\Delta = 5 \times 10^{12} \GeV$ and $M_\Delta = 10^{11} \GeV$.
For $M_\Delta = 5 \times 10^{12} \GeV$, $n_B / n_\gamma$ exceeds the observed
value everywhere, but this can by cured by choosing a smaller overall phase in $m_\Delta$.
We can see that the final baryon asymmetry is maximal for $x$ small and $y$ close to 1,
more precisely around $(x, y)=(0.05, 0.95)$, corresponding to
$D_\Delta = \mbox{Diag}\, (0.949, 0.0475, 0.312)\,\bar{m}_\nu$.
The baryon-to-photon ratio in this region is typically enhanced by a factor $2$ or $3$ with respect
to Ansatz 1, although the total CP asymmetry is smaller. These conclusions
are relatively stable under variations of the triplet mass as long as $B_\ell\sim B_H$.

\begin{figure}[t]
\center
\includegraphics[scale=0.55]{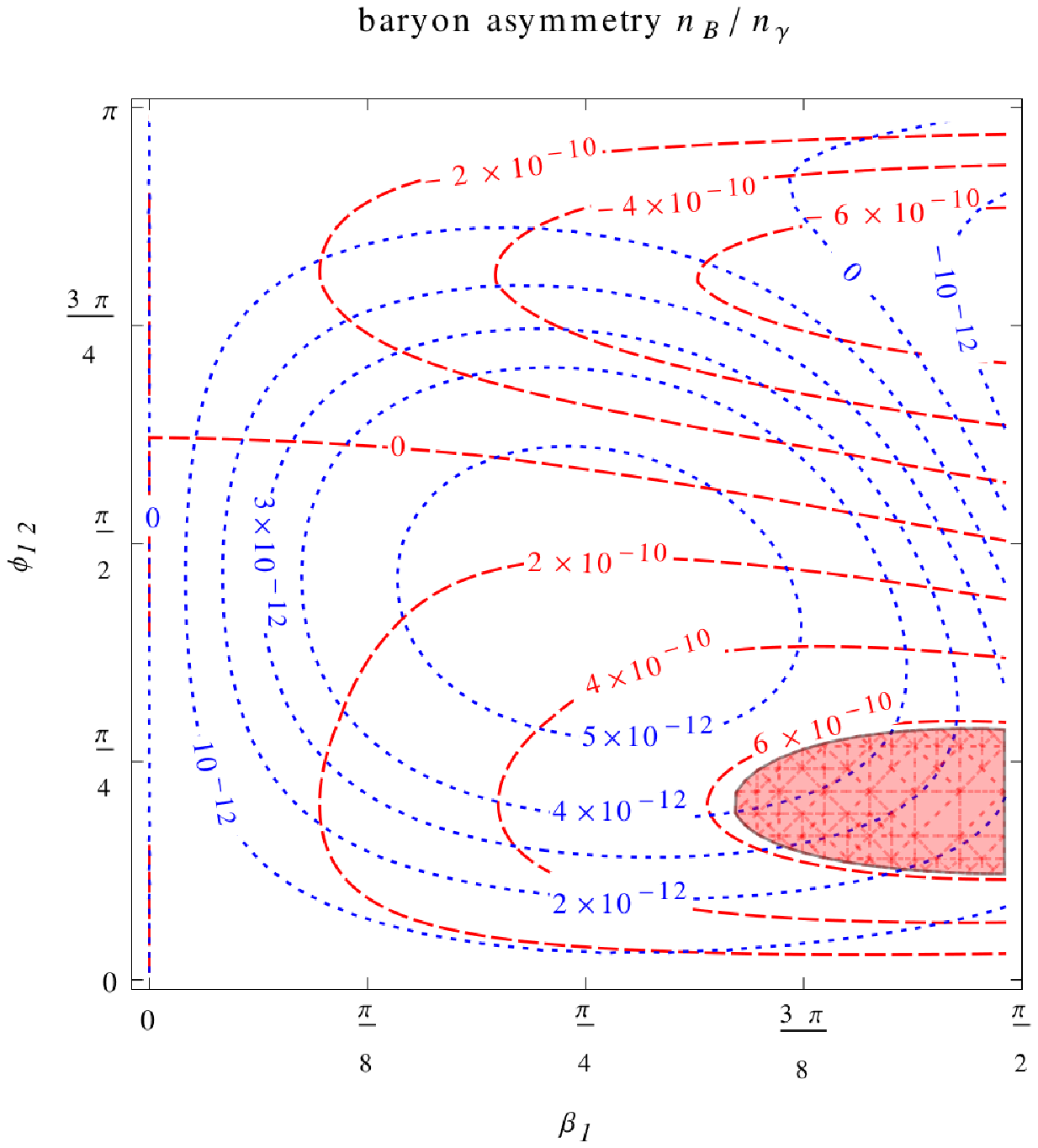}\qquad
\includegraphics[scale=0.55]{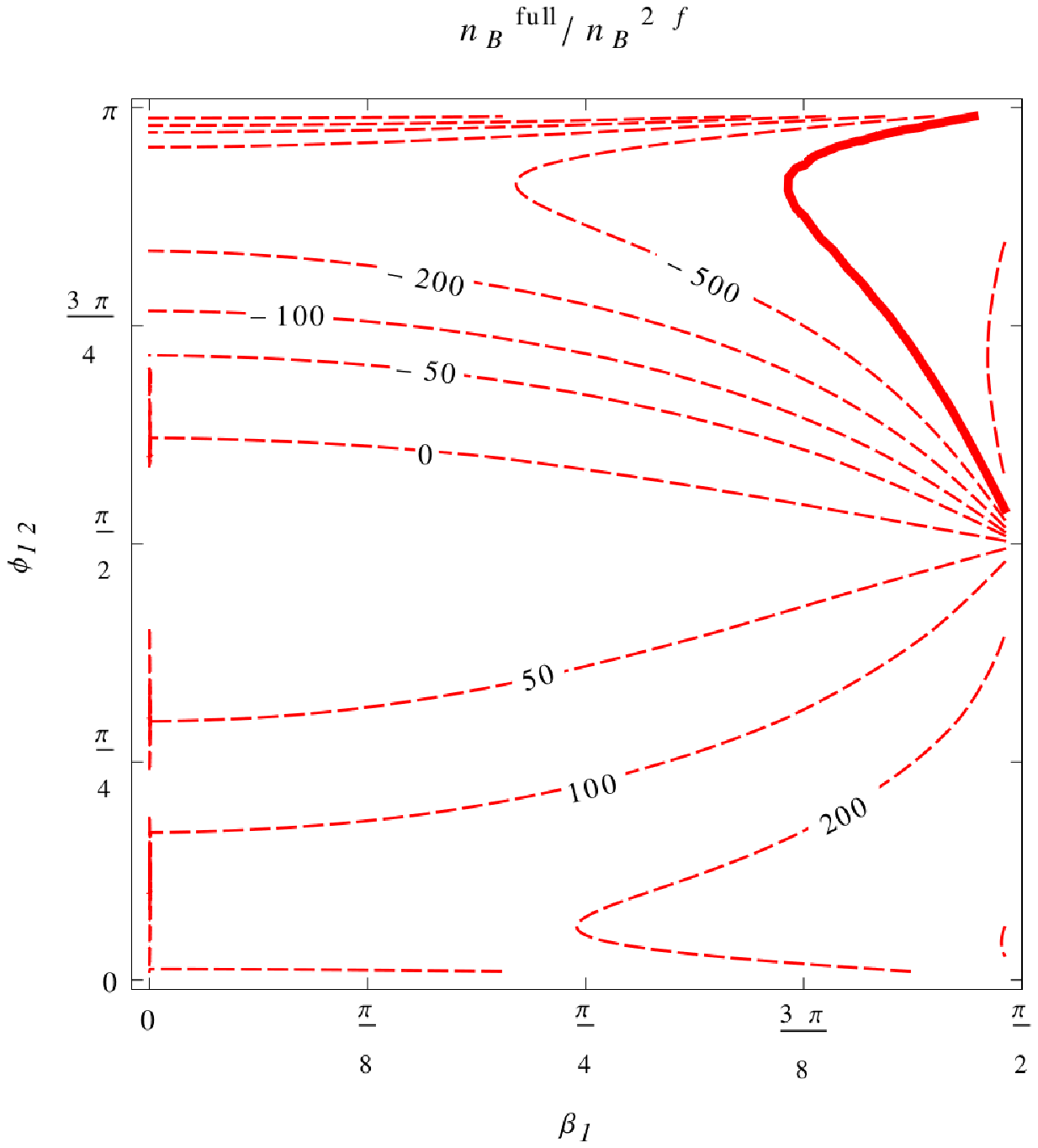}
\caption{Left panel: the baryon-to-photon ratio $n_B/n_\gamma$ predicted
by the full computation (red dashed lines) and by the 2-flavour approximation (blue dotted lines)
as a function of  $\beta_1$ and $\phi_{12}$ for Ansatz 3-a. The coloured areas indicate
where the observed baryon asymmetry can be reproduced.
Right panel: ratio of the full computation over the 2-flavour calculation of the baryon asymmetry. 
Along the solid curve, $n_B/n_\gamma$ vanishes in the 2-flavour approximation.
In both panels, $M_\Delta = 5 \times 10^{11} \GeV$ and $\lambda_H = 0.1$.}
\label{fig:Ansatz_3a}
\end{figure}
\begin{figure}[t]
\center
\includegraphics[scale=0.55]{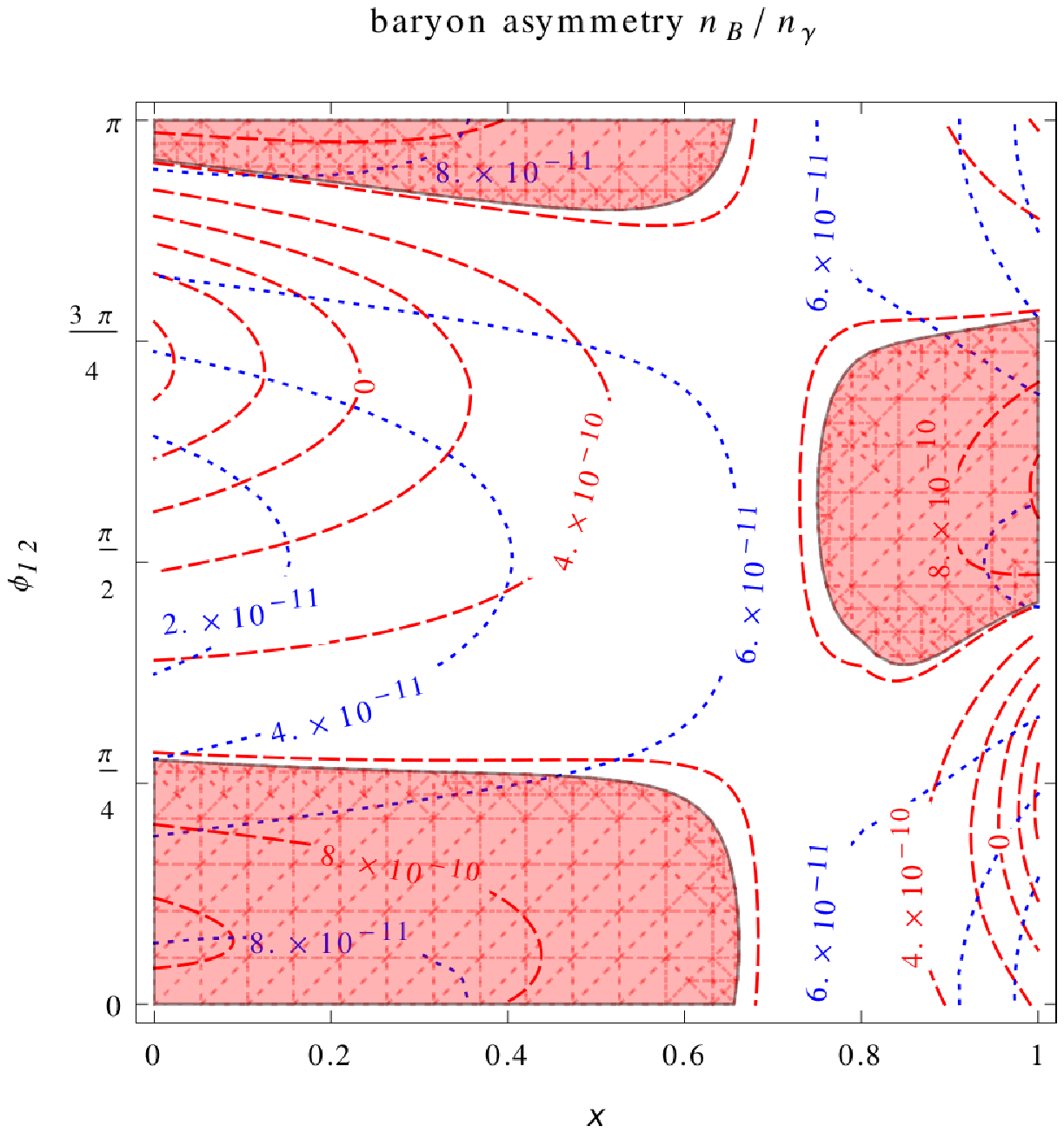}\qquad
\includegraphics[scale=0.55]{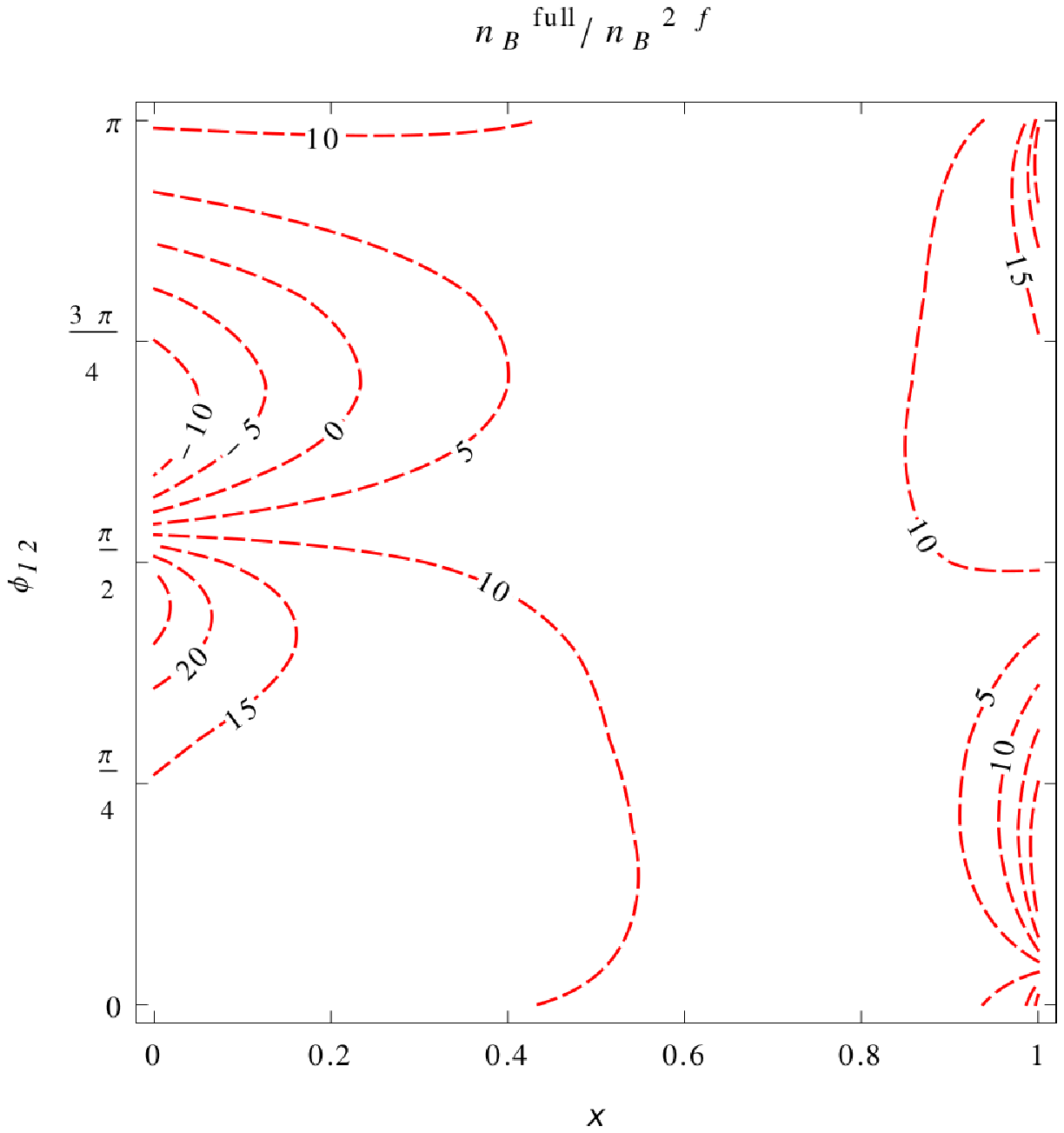}
\caption{Left panel: the baryon-to-photon ratio $n_B/n_\gamma$ predicted
by the full computation (red dashed lines) and by the 2-flavour approximation (blue dotted lines)
as a function of  $x$ and $\phi_{12}$ for Ansatz 3-b.
The coloured areas indicate where the observed baryon asymmetry can be reproduced.
Right panel: ratio of the full computation over the 2-flavour calculation of the baryon asymmetry.
In both panels, $M_\Delta = 5 \times 10^{11} \GeV$ and $\lambda_H = 0.1$.}
\label{fig:Ansatz_3b}
\end{figure}

We now proceed to study Ansatz 3, in which the CP asymmetry is concentrated
in the $(\ell_1, \ell_2)$ subspace (i.e. $\mathcal{E}_{13} = \mathcal{E}_{23} = \mathcal{E}_{33} = 0$
in the neutrino mass eigenstate basis). Let us first consider Ansatz 3-a, in which
$D_\Delta = D_\nu$ while $\beta_1$ and $\phi_{12}$ can vary, all other phases
and mixing angles in the unitary matrix $V$ being zero.
We choose a triplet mass in the range [$10^9 \GeV, 10^{12} \GeV$], and compare
the result of the full computation
with the one of the 2-flavour approximation, as described at the beginning of this section.
As can be seen from Fig.~\ref{fig:Ansatz_3a}, the baryon-to-photon ratio
computed in a flavour-covariant way can be enhanced by up to a factor
of several hundreds with respect to the 2-flavour approximation. Indeed,
important flavour effects in the $e$--$\mu$ sector are missed in this approximation.
This can be understood by examining a specific point of the parameter space,
for instance $\phi_{12}=\pi/6$ and $\beta_1=\pi/2$. In the full computation,
the sources terms in the Boltzmann equations~(\ref{eq:BE_Delta0_10^9_10^12})
and~(\ref{eq:BE_Delta_tau_10^9_10^12}) are proportional to
\begin{align}
  \mathcal{E}^0\, =\, \left(\begin{matrix}
              -10.36&-2.496\\
              -2.496&6.846
             \end{matrix}\right)\times10^{-8}\, ,   \qquad
  \mathcal{E}_{\tau\tau}\, =\, 3.513\times10^{-8}\, ,
\end{align}
where for definiteness we have written $\mathcal{E}^0$ in the $(\ell_e, \ell_\mu)$ basis,
while the flavour dependence of the washout is controlled by the triplet couplings to leptons:
\begin{align}
 \left(\begin{matrix}
              |f_{ee}| & |f_{e\mu}| & |f_{e\tau}|\\
              |f_{\mu e}| & |f_{\mu\mu}| & |f_{\mu\tau}|\\
              |f_{\tau e}| & |f_{\tau\mu}| & |f_{\tau\tau}|
             \end{matrix}\right)\,
  =\, \left(\begin{matrix}
              1.428 & 3.477 & 1.373\\
              3.477 & 9.288 & 9.394\\
              1.373 & 9.394 & 13.36
             \end{matrix}\right)\times 10^{-3}\, .
\end{align}
Thus we have a large source term for a flavour asymmetry (namely in the $e$ flavour)
that is only weakly washed out by inverse decays. As a result, a large
baryon asymmetry is generated ($n_B/n_\gamma=7.277\times10^{-10}$) even though
the total CP asymmetry vanishes (indeed, for the particular point considered
$\epsilon_\Delta = \mbox{tr}\, \mathcal{E}^0 + \mathcal{E}_{\tau\tau} = 0$).
By contrast, in the 2-flavour approximation, the CP asymmetries appearing
in the Boltzmann equations~(\ref{eq:BE_leptons_2FA}) are
\begin{align}
 \epsilon_0\, =\, -3.513\times10^{-8}\, =\, -\epsilon_\tau\, ,
\end{align}
and the washout is controlled by
\begin{align}
  \left( \begin{matrix} f_{00} & f_{0\tau}  \\  f_{0\tau} & |f_{\tau\tau}|  \end{matrix} \right)\,
  =\, \left( \begin{matrix}
              10.69 & 9.567 \\
              9.567 & 13.36
             \end{matrix} \right) \times10^{-3}\, .
\end{align}
Therefore, the asymmetries stored in the two flavours are washed out with a comparable strength,
and the resulting baryon asymmetry ($n_B/n_\gamma=1.617\times10^{-12}$) is much smaller
than in the full flavoured computation.

A similar (but numerically not as strong) enhancement of the full computation result with respect
to the 2-flavour approximation can be observed for Ansatz 3-b, in which
$D_\Delta=\text{Diag} (x\, m_0, \sqrt{1-x^2}\, m_0, m_3)$ with $m_0 = \sqrt{m^2_1 + m^2_2}$
and the phases $\beta_1$ and $\beta_2$ are set to $\pi/4$.
This is shown in Fig.~\ref{fig:Ansatz_3b}.

\begin{figure}
\center
\includegraphics[scale=0.55]{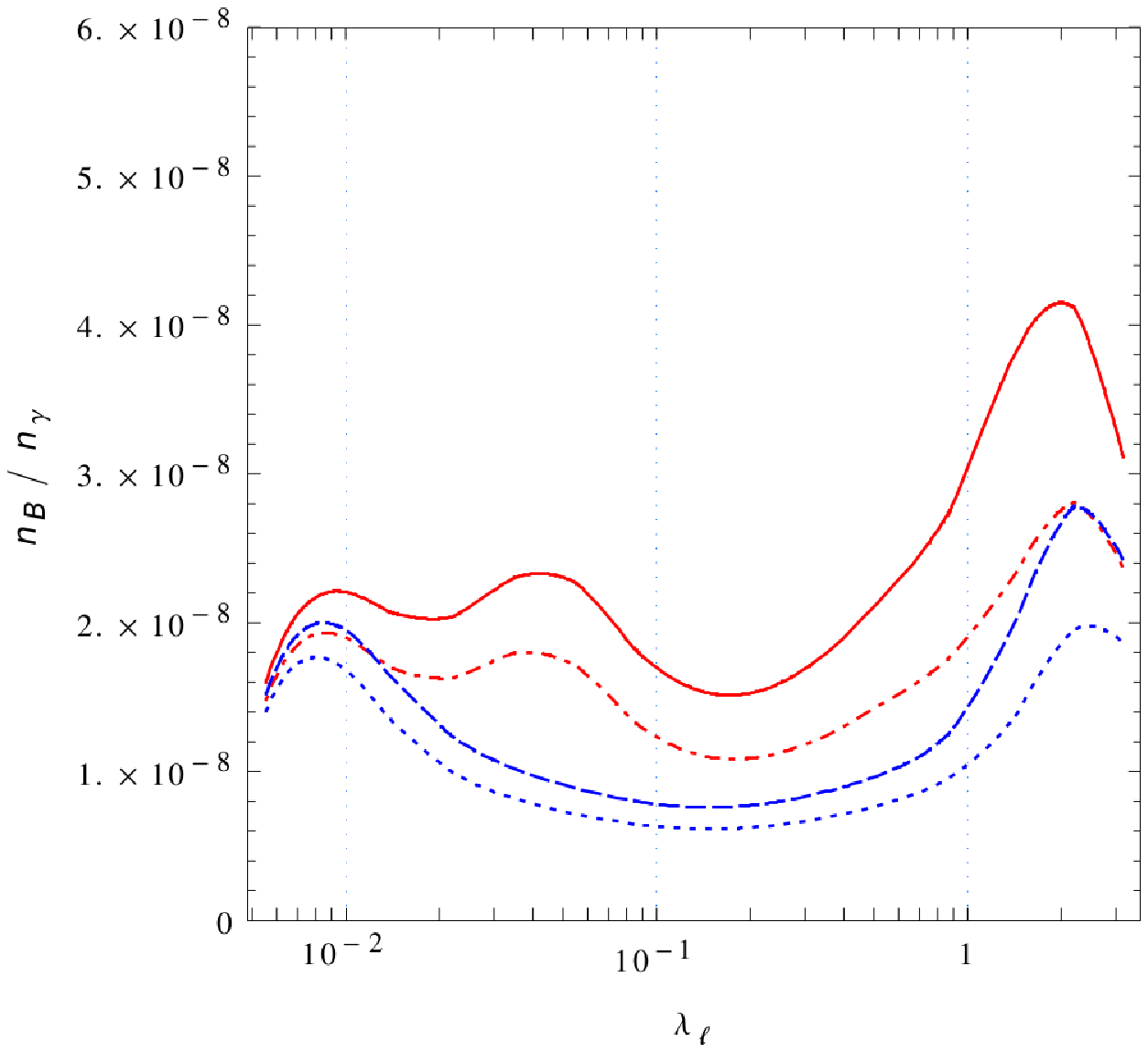}\qquad
\includegraphics[scale=0.55]{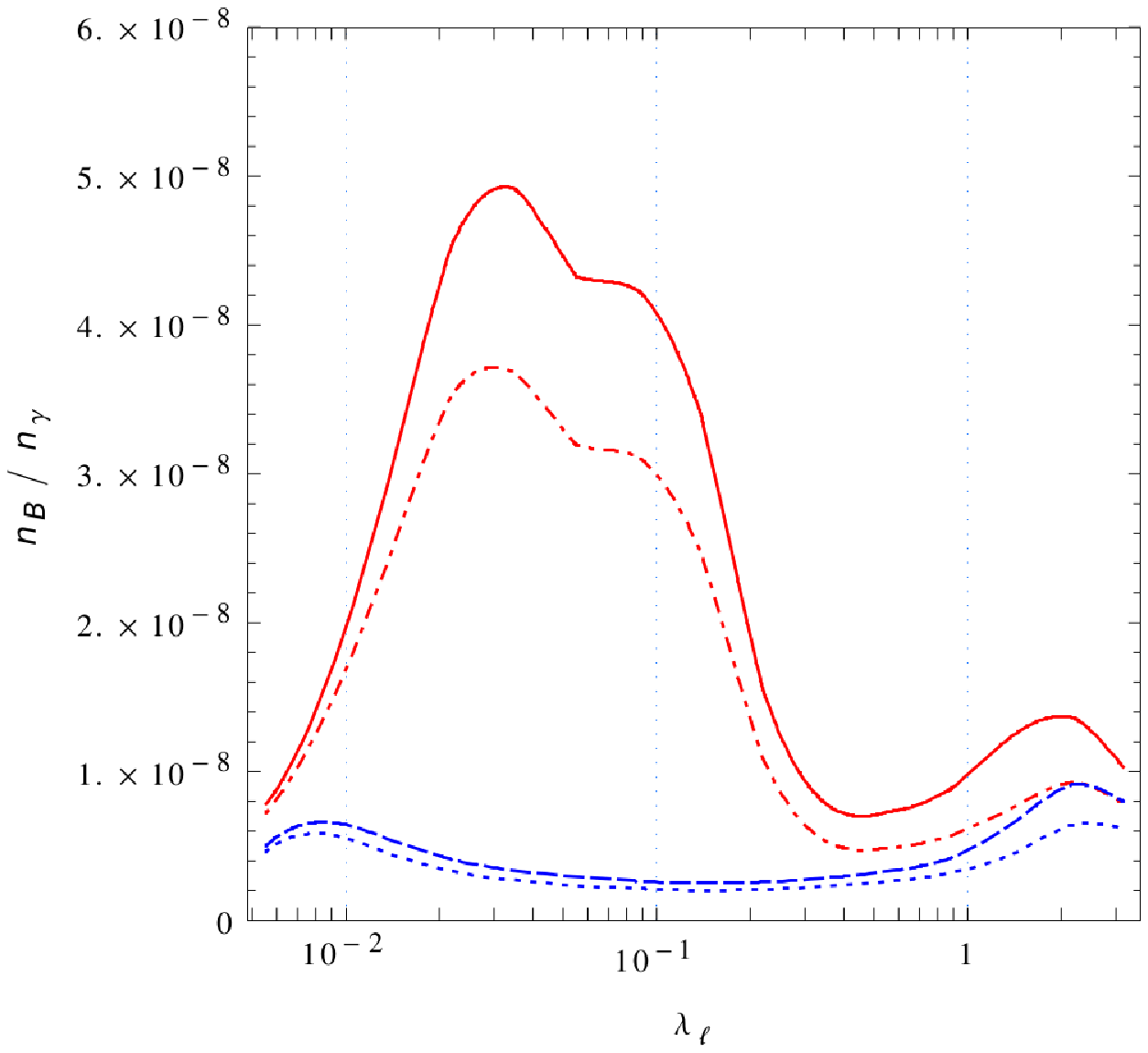}
\caption{Baryon-to-photon ratio $n_B/n_\gamma$ as a function of $\lambda_\ell$
for $M_\Delta=5\times10^{12} \GeV$, assuming Ansatz 1 (left panel) or
Ansatz 2 with $(x, y) = (0.05, 0.95)$ (right panel).
The red lines indicate the result of the flavour-covariant computation involving
the $3\times3$ density matrix $\Delta_{\alpha\beta}$,
with (solid red line) or without (dashed-dotted red line) spectator processes taken into account,
whereas the blue lines indicate the result of the single flavour approximation, taking spectator 
processes into account (blue dashed line) or not (blue dotted line).
The equality of branching ratios $B_\ell = B_H$ is realized for $\lambda_\ell \simeq 0.15$.}
\label{fig:comp1}
\vskip 0.5cm
\includegraphics[scale=0.55]{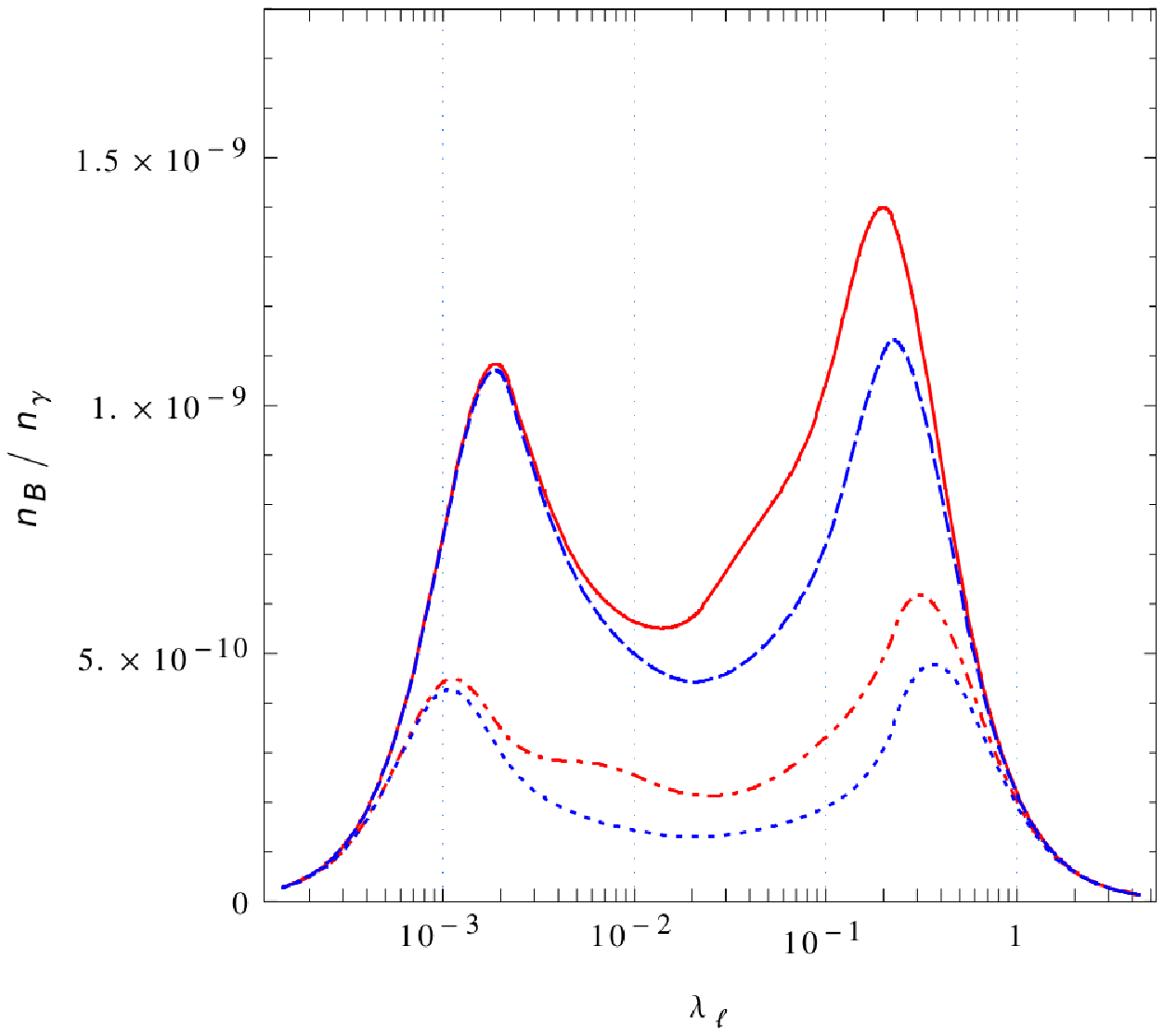}\qquad
\includegraphics[scale=0.55]{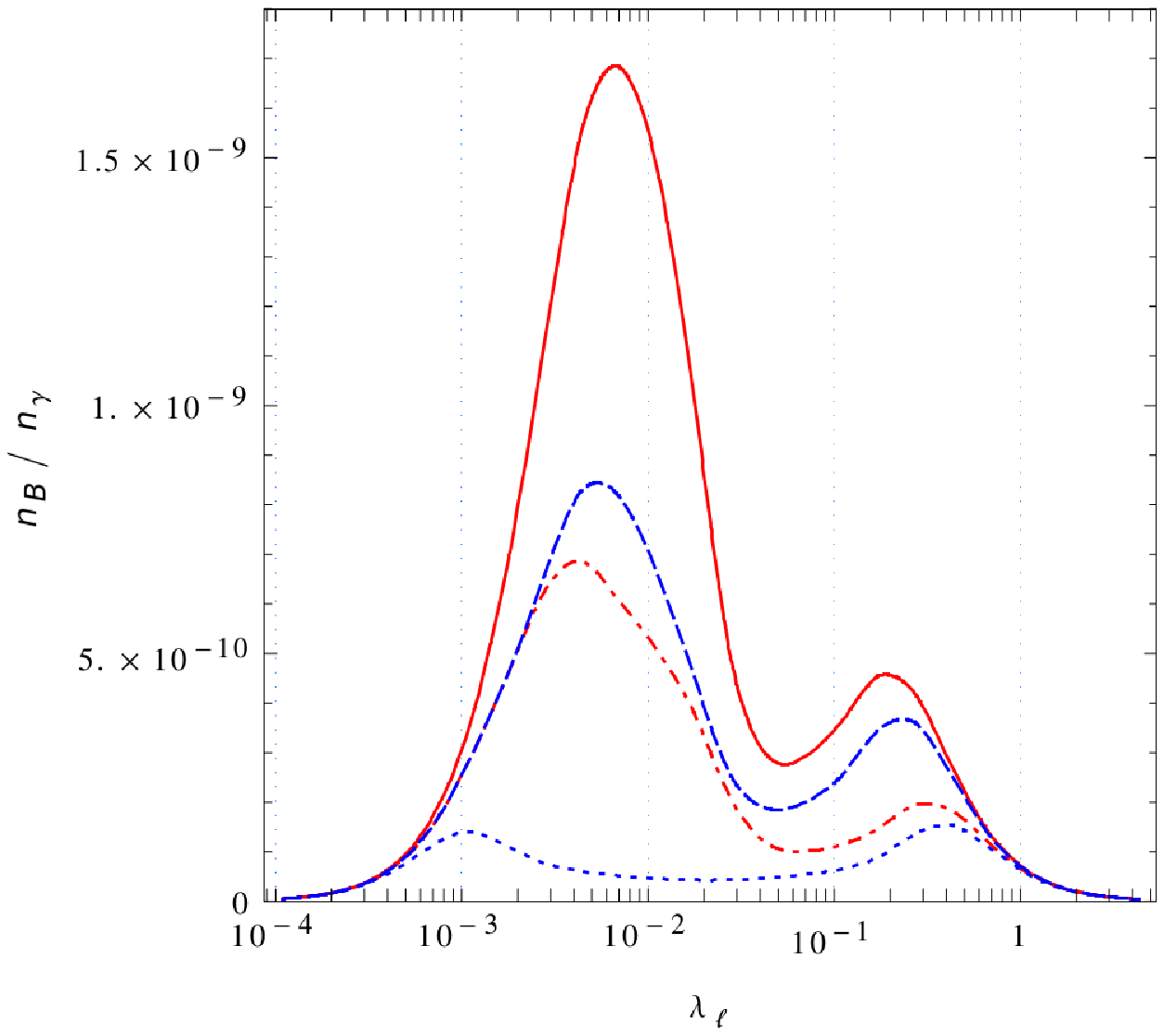}
\caption{Baryon-to-photon ratio $n_B/n_\gamma$ as a function of $\lambda_\ell$
for $M_\Delta=10^{11} \GeV$, assuming Ansatz 1 (left panel) or
Ansatz 2 with $(x, y) = (0.05, 0.95)$ (right panel).
The red lines indicate the result of the flavour-covariant computation involving
the $2\times2$ density matrix $\Delta^0_{\alpha\beta}$,
with (solid red line) or without (dashed-dotted red line) spectator processes taken into account,
whereas the blue lines indicate the result of the 2-flavour approximation, taking spectator 
processes into account (blue dashed line) or not (blue dotted line).
The equality of branching ratios $B_\ell = B_H$ is realized for $\lambda_\ell \simeq 0.021$.}
\label{fig:comp2}
\end{figure}

In Figs.~\ref{fig:comp1} and~\ref{fig:comp2}, we compare the relative impacts of spectator
processes and flavour covariance on the generated baryon asymmetry in the temperature
regimes $T > 10^{12} \GeV$ and $10^9 \GeV < T < 10^{12} \GeV$, respectively.
We consider two ans\"atze designed to produce a large baryon asymmetry
(even exceeding the observed value when the CP-violating phases are chosen
to be large, as is the case here),
namely Ansatz 1, which maximizes the total CP asymmetry, and Ansatz 2 with
$(x, y) = (0.05, 0.95)$, a parameter choice that has been shown to maximize
the final baryon asymmetry in Fig.~\ref{fig:Ansatz_2}.
In Fig.~\ref{fig:comp1}, the triplet mass has been chosen to be $M_\Delta = 5 \times 10^{12} \GeV$,
so that most of the $B-L$ asymmetry is produced at $T > 10^{12} \GeV$. For Ansatz 1
(i.e. $m_\Delta=im_\nu$), flavour effects and spectator processes have a quantitatively
similar impact on the baryon-to-photon ratio, while for Ansatz 2 flavour effects
strongly dominate over spectator processes such as QCD sphalerons
or top quark Yukawa interactions (especially for $10^{-2} \lesssim \lambda_\ell \lesssim 0.5$).
In this case, neglecting spectator processes and including flavour effects gives a much
more accurate result than doing the opposite.
In Fig.~\ref{fig:comp2}, the triplet mass is $M_\Delta=10^{11} \GeV$, hence the $B-L$
asymmetry is essentially generated in the temperature regime $10^9 \GeV < T < 10^{12} \GeV$
where the tau Yukawa coupling is in equilibrium, but the muon Yukawa coupling is not.
For Ansatz 1, the 2-flavour calculation turns out to be a rather good approximation
to the flavour-covariant computation, while neglecting spectator processes
gives a very bad estimate, except for small or large values of $\lambda_\ell$.
For Ansatz 2, on the contrary, both flavour covariance and spectator processes
have a significant impact on the baryon-to-photon ratio (except again for extreme
values of $\lambda_\ell$), and neglecting one of them underestimates the result
by up to a factor 2. The 2-flavour approximation without spectator processes
actually gives a much larger disagreement with the full computation.

\begin{figure}[t]
\center
\includegraphics[scale=0.55]{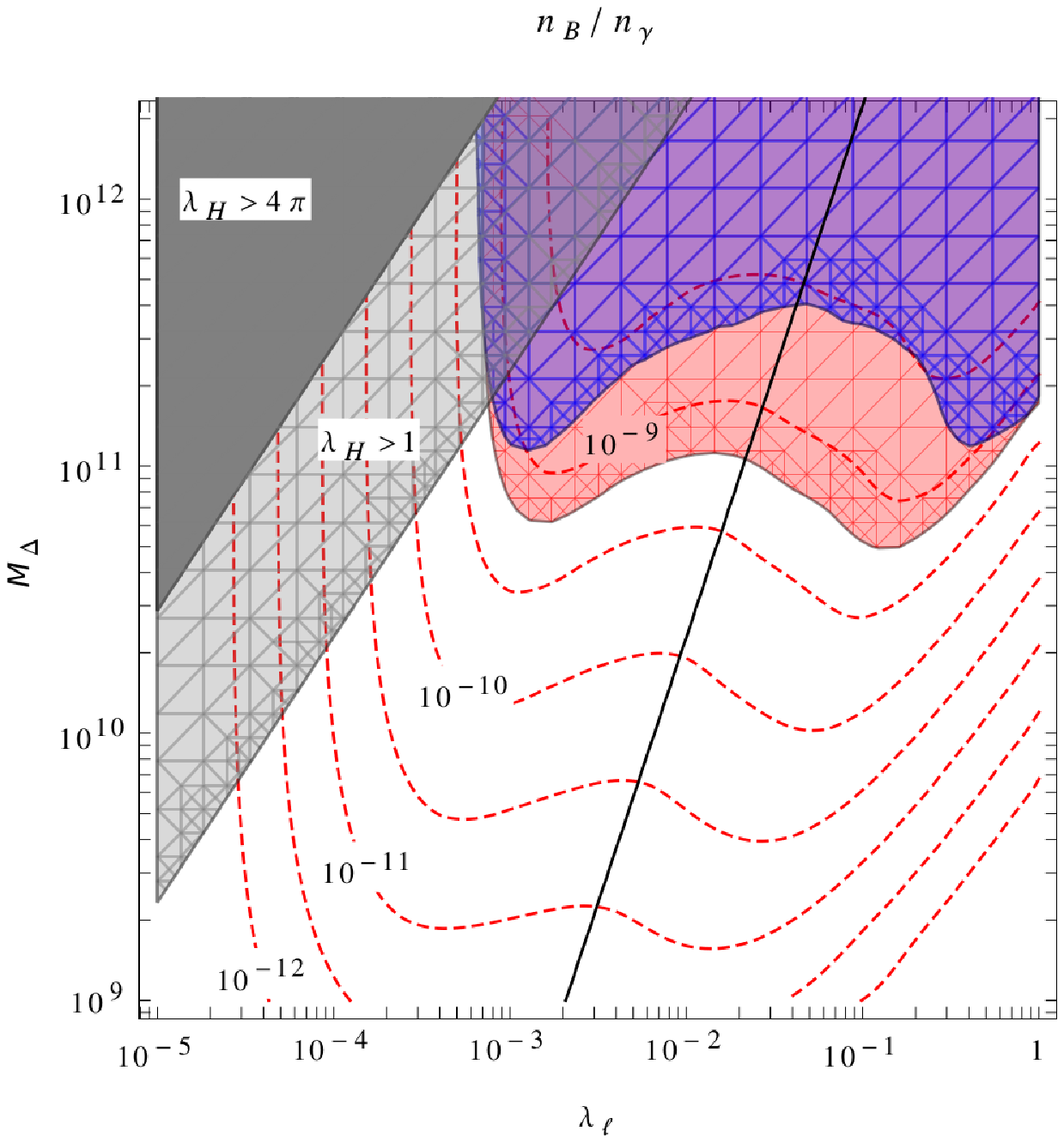}\qquad
\includegraphics[scale=0.55]{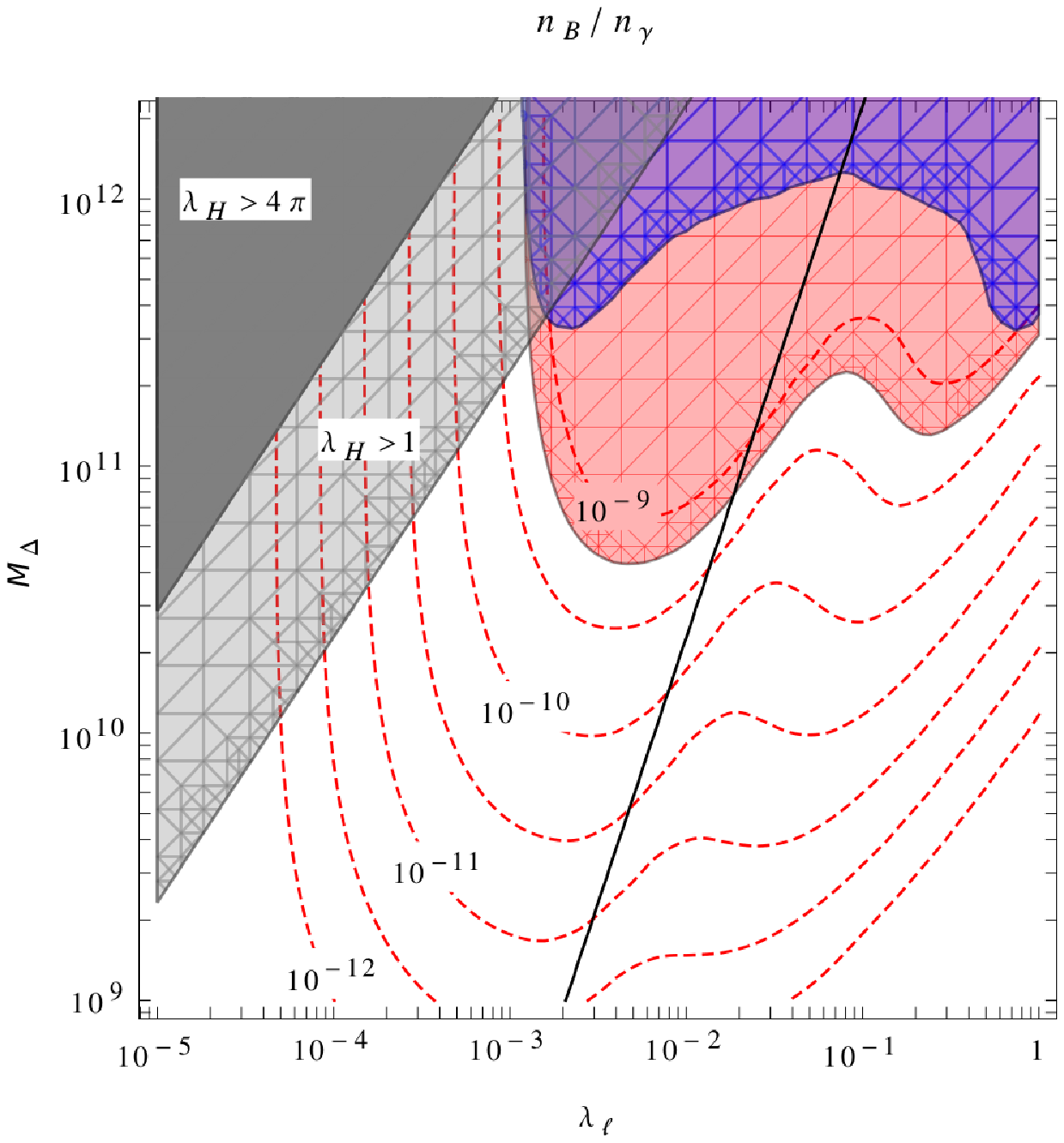}
\caption{Isocurves of the baryon-to-photon ratio $n_B/n_\gamma$ in the $(\lambda_\ell, M_\Delta)$
plane obtained performing the full computation, assuming Ansatz 1 (left panel) or
Ansatz 2 with $(x, y) = (0.05, 0.95)$ (right panel).
The coloured regions indicate where the observed baryon asymmetry can be reproduced
in the full computation (light red shading) or in the single flavour approximation
with spectator processes neglected (dark blue shading).
The solid black line corresponds to $B_\ell = B_H$.
Also shown are the regions where $\lambda_H$ is greater than $1$ or $4\pi$.}
\label{fig:M_Delta_dependence}
\end{figure}
Finally, we study in Fig.~\ref{fig:M_Delta_dependence} the dependence
of the generated baryon asymmetry on $\lambda_\ell$ and $M_\Delta$,
both for Ansatz 1 and for Ansatz 2 with $(x, y) = (0.05, 0.95)$.
The computation is performed assuming that the third generation Yukawa
couplings as well as the charm Yukawa coupling are in equilibrium, which strictly
speaking is true only in the temperature range $10^9 \GeV < T < 10^{12} \GeV$.
From Fig.~\ref{fig:M_Delta_dependence} one can conclude that
successful scalar triplet leptogenesis is possible for a triplet mass
as low as $4.4 \times 10^{10} \GeV$, to be compared with $1.2\times10^{11} \GeV$
in the approximation where flavour effects and spectator processes are neglected.
Other assumptions about the flavour structure of the triplet couplings to leptons
may allow for a lighter scalar triplet.
For comparison, we quote the lower bounds found by Ref.~\cite{Hambye:2005tk}
in the single flavour approximation with spectator processes neglected:
$M_\Delta > 1.3 \times 10^{11} \GeV$ in the case $\bar m_\Delta = 0.05 \eV$
($\approx \bar m_\nu$ for a hierarchical neutrino mass spectrum), in agreement
with our result, and $M_\Delta > 2.8 \times 10^{10} \GeV$
in the case $\bar m_\Delta = 0.001 \eV \ll \bar m_\nu$. Although we did not
consider ans\"atze satisfying $\bar m_\Delta \ll \bar m_\nu$, because flavour effects
are comparatively smaller in this case, we also expect the latter bound to be lowered
by the inclusion of flavour and spectator processes.

\section{Conclusions}     %
\label{sec:conclusions}   %

In this paper, we have shown how to consistently include the effects of
the different lepton flavours in scalar triplet leptogenesis. When charged
lepton Yukawa interactions are out of equilibrium at the time of leptogenesis,
i.e. when the lepton asymmetry is generated at high temperature,
the proper treatment of flavour effects involves a $3 \times 3$ density
matrix, whose evolution is governed by flavour-covariant Boltzmann equations.
The often-used single flavour approximation, which gives rather accurate
results in the standard leptogenesis scenario with right-handed neutrinos,
leads to predictions for the generated baryon asymmetry that can depart
by a large amount from the flavour-covariant computation.
In the intermediate temperature regime where the tau Yukawa coupling
is in equilibrium but the muon and the electron ones are not, the $3 \times 3$
density matrix can be replaced by the asymmetry stored in the tau lepton doublet
and by a $2 \times 2$ density matrix describing the flavour asymmetries in the
$(\ell_e, \ell_\mu)$ subspace and their quantum correlations. In this case too,
the 2-flavour calculation in which the $(\ell_e, \ell_\mu)$ subspace is described
by a single flavour $\ell_0$ does not in general give a good approximation of the flavour-covariant
computation. Finally, when the tau and muon Yukawa couplings are in equilibrium,
flavour covariance is completely broken and the dynamics of leptogenesis is described
in terms of the asymmetries stored in the three lepton doublets $\ell_e$, $\ell_\mu$
and $\ell_\tau$.

We performed a numerical study of the impact of flavour effects and spectator
processes on the generated baryon asymmetry
for judiciously chosen ans\"atze, and compared the flavour-covariant
computation with flavour non-covariant approximations used in the literature,
with or without spectator
processes included. We found discrepancies in the predictions for the
baryon asymmetry ranging from an order one factor to two orders of magnitude.
In particular, we showed that successful leptogenesis can easily be achieved
when the decays of the triplet into leptons and Higgs bosons occur at a similar rate,
while it would require a significantly heavier triplet in the single flavour
approximation. As a result, the minimal triplet mass allowed by successful
leptogenesis is lowered by the inclusion of flavour effects,
from $1.2\times10^{11} \GeV$ to $4.4 \times 10^{10} \GeV$ in the case
where the scalar triplet and the additional heavy states give comparable
contributions to neutrino masses.

Throughout this paper, we worked in a framework in which the contribution
of the additional heavy states to the CP asymmetries is parametrized by
an effective dimension-5 operator, but the procedure we used to derive
the flavour-covariant Boltzmann equations can also be applied
to explicit models with e.g. several scalar triplets, or with a scalar triplet and 
right-handed neutrinos. The formalism employed in this paper can also
be used to study the scenario of purely flavoured leptogenesis~\cite{Felipe:2013kk,Sierra:2014tqa}
in a flavour-covariant way. In the framework used in this paper, this
would require the addition of the effective four-lepton operator~(\ref{eq:4lepton_operator}).


\paragraph*{Acknowledgments}
We thank Thomas Hambye for useful discussions.
This work has been supported in part by the Agence Nationale de la Recherche
under contract ANR 2010 BLANC 0413 01, by the European Research Council
(ERC) Advanced Grant Higgs@LHC, and by the European Union FP7 ITN Invisibles
(Marie Curie Actions, PITN-GA-2011-289442).


\renewcommand{\theequation}{A.\arabic{equation}}
\setcounter{equation}{0}  

\begin{appendix}

\section{Reaction densities}    %
\label{app:reactions}               %

We summarize in this appendix some useful formulae for the reaction densities used
in this paper. Let us first recall the general expression for the (thermally averaged)
space-time density of a general reaction $a + b + \ldots \to i + j + \ldots\, $.
Neglecting Bose enhancement and Pauli blocking factors:
\begin{align}
  \gamma(a + b + \ldots \rightarrow i + j + \ldots)\, = \int & \frac{d^3p_a}{(2\pi)^32\omega_{\vec p_a}}\rho^{\mathrm{eq}}_a
    \frac{d^3p_b}{2\omega_{\vec p_b}(2\pi)^3}\rho^{\mathrm{eq}}_b \dots
    \frac{d^3p_i}{2\omega_{\vec p_i}(2\pi)^3}\frac{d^3p_j}{2\omega_{\vec p_j}(2\pi)^3} \dots  \nonumber \\
  & |\mathcal{M}|^2\, (2\pi)^4\delta^{(4)}(p_a+p_b+ \ldots -p_i - p_j - \ldots)\, ,
\end{align}
where $|\mathcal{M}|^2$ is the squared matrix element summed over the internal
degrees of freedom of the initial and final states, and $\rho^{\mathrm{eq}}_X (\vec p)$
is the phase-space distribution function of the particle $X$ at kinetic and chemical
equilibrium, which only depends on the bosonic or fermionic nature of $X$:
\begin{align}
  \rho^{\mathrm{eq}}_{\BF} (\vec p)\, =\, \frac{1}{e^{\beta(E-\mu)}\mp1}\, .
\end{align}
In the following all computations are done using the Maxwell-Boltzmann statistics,
i.e. neglecting the difference between bosons and fermions:
\begin{align}
  \rho^{\mathrm{eq}}_F (\vec p)\, =\, \rho^{\mathrm{eq}}_B (\vec p)\, =\, e^{-\beta(E-\mu)}\,
    \equiv\,  \rho^{\mathrm{eq}}_{\rm MB} (\vec p)\, .
\end{align}
With this approximation, the space-time density of triplet and antitriplet decays can be written as
\beq
  \gamma_D\, =\, s\Sigma_{\Delta}^{\mathrm{eq}}\, \frac{K_1(z)}{K_2(z)}\, \Gamma_{\Delta}\, ,
\eeq
where $\Sigma_\Delta \equiv (n_\Delta + n_{\bar \Delta}) / s$ is the comoving number density
of triplets and antitriplets, $K_{1, 2} (z)$ are modified Bessel functions of the second kind,
$z \equiv M_\Delta / T$ and $\Gamma_\Delta$ is the triplet decay width. For $2 \to 2$ scatterings, one has
\begin{equation}
  \gamma(a+b\rightarrow i+j)\, =\, \frac{T}{64\pi^4}\int_{s_\mathrm{min}}^\infty\! ds\, s^{1/2}\hat{\sigma}(s)\,
    K_1\! \left(\frac{\sqrt{s}}{T}\right) ,
\end{equation}
where $\hat{\sigma}(s)=2s\,\lambda(1,m_a^2/s,m_b^2/s)\,\sigma(s)$ is the reduced cross-section
summed over the internal degrees of freedom of initial and final particles.

When computing the space-time density of a $2 \to 2$ scattering,
one must take care to properly subtract the contribution
of on-shell intermediate particles, which is already taken into account in decays and inverse
decays~\cite{Kolb:1979qa}. When the resonance occurs in the $s$-channel, one can compute
the subtracted reaction density by taking away the resonant part from the squared propagator in the
narrow-width approximation~\cite{Cline:1993bd,Giudice:2003jh}:
\beq
  |D|^2\ \rightarrow\ |D|^2-\frac{\pi}{M\Gamma}\delta\left(s-M^2\right) ,
\eeq
where $M$ and $\Gamma$ are the mass and width of the intermediate particle,
and $D$ is the propagator of the intermediate state in the Breit-Wigner form:
\beq
  D\, =\, \frac{1}{s-M^2+iM\Gamma}\ .
\eeq
We have to do this for the computation of $\gamma(\ell\ell\rightarrow \bar H \bar H)$
and $\gamma(\ell\ell\rightarrow\ell\ell)$, which receive a contribution from $s$-channel
triplet exchange.
Similarly, when computing $\gamma(\ell\bar \Delta\rightarrow\ell\bar \Delta)$, one has to subtract
the contribution of  real intermediate leptons in the $u$-channel, corresponding to the
two processes $\bar \Delta\rightarrow \ell\ell$ and $\ell\ell\rightarrow \bar \Delta$.
Following Ref.~\cite{Giudice:2003jh}, we perform the subtraction in the following way:
\beq
  |D|^2\ \rightarrow\ |D|^2-\frac{\pi}{E\Gamma_{\rm th}}\delta\left(u\right) ,
\eeq
where $E$ is the energy of the intermediate lepton, $\Gamma_{\rm th}$ its thermal width, and
the propagator of the intermediate lepton is
\beq
  D\, =\, \frac{1}{u+iE\Gamma_{\rm th}}\ .
\eeq
As in Ref.~\cite{Giudice:2003jh}, we use the following representation of $\delta(x)$ in numerical computations:
\beq
  \delta(x)\, =\, \frac{2}{\pi}\frac{\epsilon^3}{(x^2+\epsilon^2)^2}\ ,
\eeq
where $\epsilon$ is a small number.
In the limit of small width, which we assume to be valid, one can simply set
$\epsilon = \Gamma/M$ for a resonance in the  $s$-channel with an intermediate
particle of mass $M$ and width $\Gamma$, and $E\Gamma_{\rm th} = \epsilon M_\Delta^2$
for a resonance in the $u$-channel, with any small value for $\epsilon$.
With this choice, one can compute the subtracted space-time densities for the various $2 \to 2$ scatterings.

In what follows, we note $x \equiv s/M_\Delta^2$. The reduced cross-sections for the scatterings
involving 2 leptons and 2 Higgs bosons are, after subtracting the contribution of the on-shell
intermediate triplet,
\begin{align}
  (\hat{\sigma}^{\ell\ell}_{\bar H \bar H})^\Delta\, & =\, \frac{3xM_\Delta^2}{8\pi v^4}\,
    \bar{m}_\Delta^2\, \frac{(1-x)^2-\epsilon^2}{[(1-x)^2+\epsilon^2]^2}\ ,  \nonumber \\
  (\hat{\sigma}^{\ell\ell}_{\bar H \bar H})^\mathcal{I}\, & =\, \frac{3xM_\Delta^2}{8\pi v^4}\,
    \mbox{Re}\! \left[\text{tr}(m_\Delta m_\mathcal{H}^\dagger)\right] \frac{1-x}{(1-x)^2+\epsilon^2}\ ,  \\
  (\hat{\sigma}^{\ell\ell}_{\bar H \bar H})^\mathcal{H}\, & =\, \frac{3xM_\Delta^2}{8\pi v^4}\,
    \bar{m}_\mathcal{H}^2\, ,  \nonumber
\end{align}
for $\ell \ell \rightarrow \bar H \bar H$, and
\begin{align}
  (\hat{\sigma}^{\ell H}_{\bar \ell \bar H})^\Delta\, & =\, \frac{3M_\Delta^2}{2\pi v^4}\, \bar{m}_\Delta^2
    \left(-\frac{1}{1+x}+\frac{\ln(1+x)}{x}\right) ,  \nonumber\\
  (\hat{\sigma}^{\ell H}_{\bar \ell \bar H})^\mathcal{I}\, & =\, \frac{3M_\Delta^2}{2\pi v^4}\,
    \mbox{Re}\! \left[\text{tr}(m_\Delta m_\mathcal{H}^\dagger)\right]\! \left(1-\frac{\ln(1+x)}{x}\right) ,  \\
  (\hat{\sigma}^{\ell H}_{\bar \ell \bar H})^\mathcal{H}\, & =\, \frac{3xM_\Delta^2}{4\pi v^4}\,
    \bar{m}_\mathcal{H}^2\, ,  \nonumber
\end{align}
for $\ell H \rightarrow \bar l \bar H$, where the superscripts $\Delta$, $\mathcal{H}$
and $\mathcal{I}$ refer to the contributions of the scalar triplet, of the $D=5$
operator~(\ref{eq:Weinberg_operator}) responsible for $m_\mathcal{H}$ and
to the interference between the two contributions, respectively.
For the 2 lepton--2 lepton scatterings $\ell \ell \to \ell \ell$ and $\ell \bar \ell \to \ell \bar \ell$, we obtain
\begin{align}
  \hat{\sigma}^{\ell\ell}_{\ell\ell}\, =\, \frac{3}{32\pi}\, \lambda^4_\ell\, x^2
    \frac{(1-x)^2-\epsilon^2}{\left[(1-x)^2+\epsilon^2\right]^2}\ ,
\end{align}
and
\begin{align}
  \hat{\sigma}^{\ell\bar \ell}_{\ell\bar \ell}\, =\, \frac{3}{16\pi}\,\lambda^4_\ell
    \left(2+\frac{2}{1+x}-\frac{4}{x}\ln(1+x)\right) .
\end{align}
We did not include the gauge contributions since they do not violate flavour, hence
they drop from the Boltzmann equations.
Finally, the reduced cross-sections for the scatterings $\ell\Delta\rightarrow\ell\Delta$,
$\ell \bar \Delta \rightarrow \ell \bar \Delta$ and $\ell \bar \ell \rightarrow \Delta \bar \Delta$
are given by, after subtracting the contribution of the on-shell intermediate lepton
in the $u$-channel,
\begin{align}
  \hat{\sigma}^{\ell\Delta}_{\ell\Delta}\, & =\, \frac{9}{32\pi}\, \text{tr}[ff^\dagger ff^\dagger]\, \frac{(x-1)^4}{x^4}\ ,  \\
  \hat{\sigma}^{\ell\bar \Delta }_{\ell\bar \Delta }\, & =\, \frac{9}{32\pi}\, \text{tr}[ff^\dagger ff^\dagger]
    \left[\, \ln\! \left( \frac{x^2 \left((x-2)^2+\epsilon ^2\right)}{x^2\epsilon ^2+1} \right)
    + \frac{(x-1)^2}{x} \frac{4-2 x}{(x-2)^2 + \epsilon ^2}\, \right] ,  \\
  \hat{\sigma}^{\Delta\bar \Delta }_{\ell\bar \ell}\, & =\, \frac{9}{32\pi}\, \text{tr}[ff^\dagger ff^\dagger]\,
    \sqrt{1-\frac{4}{x}}\, \left[\, \frac{(x-2)}{\sqrt{(x-4) x}}\,
    \ln\! \left( \frac{x-2+\sqrt{x(x-4)}}{x-2-\sqrt{x(x-4)}} \right) - 2\, \right] .  \label{eq:sigma_DDbar_llbar}
\end{align}

With the above expressions, one can compute numerically the reaction densities $\gamma_{\ell H}$,
$\gamma_{4\ell}$ and $\gamma_{\ell\Delta}$  that appear in the washout terms
$\mathcal{W}^{\ell H}_{\alpha\beta}$, $\mathcal{W}^{4\ell}_{\alpha\beta}$
and $\mathcal{W}^{\ell\Delta}_{\alpha\beta}$, respectively
(see Subsection~\ref{subsec:washout} for the definition of these reaction densities
and Eqs.~(\ref{eq:W_lh}), (\ref{eq:W_4l}) and~(\ref{eq:W_ellDelta}) for the washout terms).
We also need the space-time density of triplet-antitriplet annihilations into Standard Model particles,
$\gamma_A$, which enters the Boltzmann equation for $\Sigma_\Delta$, Eq.~(\ref{eq:BE_Sigma_Delta}).
The contribution of gauge scatterings to $\gamma_A$ has been computed in Ref.~\cite{Hambye:2005tk}.
There are also subleading contributions proportional to $\lambda^4_\ell$,
$\lambda^4_H$, $\lambda^2_\ell g^2_a$ and $\lambda^2_H g^2_a$, which we do not include in our computation.

\begin{figure}[t]
\center
\includegraphics[scale=0.6]{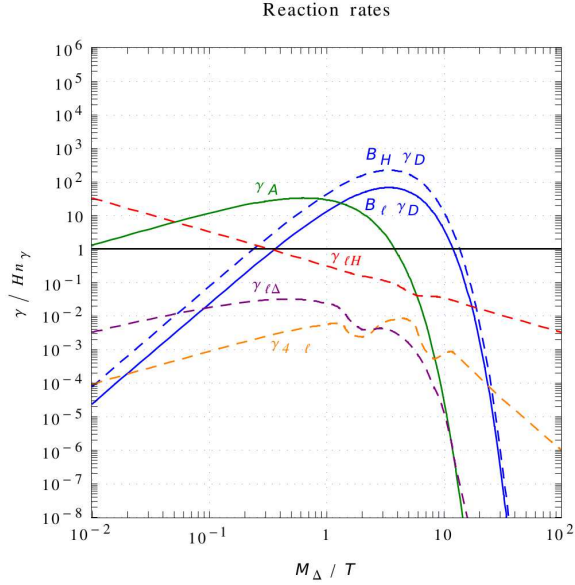}
\caption{Space-time densities of decays ($B_\ell\gamma_D$, $B_H\gamma_D$) and scatterings
($\gamma_A$, $\gamma_{\ell H}$, $\gamma_{4\ell}$ and $\gamma_{\ell\Delta}$) in units of $Hn_\gamma$,
as a function of $z = M_\Delta/T$, for the values of the parameters indicated in the text.}
\label{fig:reactions}
\end{figure}

Fig.~\ref{fig:reactions} shows the typical magnitude of the various reaction densities
and their evolution as a function of $z = M_\Delta/T$
(here for $M_\Delta = 5 \times 10^{12} \GeV$, $m_\Delta = i m_\nu$ and $\lambda_H=0.2$).
A ratio $\gamma_R / Hn_\gamma$ smaller than $1$ indicates that the reaction $R$ is slow
on a cosmological time scale.
In this example, most scatterings are negligible, except for the ones involving Higgs bosons;
however, the latter become slower than the expansion of the universe
at the time where decays start to dominate over annihilations, a necesary condition
for a large lepton asymmetry to develop~\cite{Hambye:2005tk,Hambye:2012fh}.
For the present choice of parameters, however, the third Sakharov condition would
not be satisfied in the single flavour case, as both decays into Higgs bosons and into leptons
are in equilibrium at the time of leptogenesis (i.e. at $z \lesssim 10$, before the triplet
abundance becomes strongly Boltzmann-suppressed). Taking into account flavour effects
leads to an enhanced efficiency because some ``flavoured'' decay channels
$\Delta \to \bar \ell_\alpha \bar \ell_\beta$ are significantly slower than others.


\end{appendix}



\end{document}